\documentclass[onecolumn]{aastex63}
\pdfoutput=1 
\usepackage{amsmath,amstext}
\usepackage[T1]{fontenc}
\usepackage{apjfonts, xspace, xcolor} 
\usepackage[figure,figure*]{hypcap}
\usepackage{xcolor}

\newcommand{\asec}{\arcsec\xspace}
\newcommand{\HST}{\textit{HST}\xspace}
\newcommand{\GAIA}{\textit{Gaia}\xspace}
\newcommand{\STIS}{\textit{HST}\xspace}

\newcommand{\SNIFS}{SuperNova Integral Field Spectrograph\xspace}

\newcommand{\snfactory}{SNfactory\xspace}

\newcommand{\ang}{\AA\xspace}

\newcommand{\simNalign}[2]{& \ensuremath{\sim} & \ensuremath{\mathcal{N}(#1,\ #2)} \xspace}

\newcommand{\sncosmo}{{\tt{sncosmo}}\xspace}

\newcommand{\SNRcut}{15\xspace}
\newcommand{\NLandoltAll}{27\xspace}

\newcommand{\BSpectraNominal}{between 4119 and 4289\xspace}
\newcommand{\RSpectraNominal}{between 4256 and 4261\xspace}
\newcommand{\SpectraNominal}{4289\xspace}
\newcommand{\NightsNominal}{497\xspace}

\newcommand{\NStandNominal}{46\xspace}
\newcommand{\NSecondary}{32\xspace}
\newcommand{\NPrimary}{14\xspace}

\newcommand{\NminNights}{eight\xspace}
\newcommand{\shortlong}{12\xspace}
\newcommand{\numdatapoints}{just over ten million\xspace}
\newcommand{\bdseventyfivemmag}{13\xspace}
\newcommand{\repeatibility}{13--24\,mmag\xspace}

\newcommand{\repeatlong}{14~mmag\xspace}
\newcommand{\bohlinlandoltscatter}{5--16~mmag\xspace}
\newcommand{\medatmdispersion}{7~mmag\xspace}
\newcommand{\meduncatm}{1~mmag\xspace}
\newcommand{\calspecdispersion}{6~mmag\xspace} 
\newcommand{\internalscatter}{8~mmag\xspace} 

\newcommand{\ConcludingSentence}{With our large number of observations, careful crosschecks, and \NPrimary reference stars, our results are the best calibration yet achieved with an integral-field spectrograph, and among the best calibrated surveys.\xspace}

\newcommand{\PSFMG}{\ensuremath{\mathcal{P_{MG}}}\xspace}
\newcommand{\PSFF}{\ensuremath{\mathcal{P_{\mathfrak{}{F}}}}\xspace}
\newcommand{\PSFFtilde}{\ensuremath{\mathcal{\tilde{P}_{\mathfrak{}{F}}}}\xspace}
\newcommand{\msun}{\ensuremath{{M}_\sun}\xspace}

\newcommand{\rsun}{\ensuremath{{R}_\sun}\xspace}

\newcommand{\BDSeventeen}{BD$+17^\circ$4708\xspace}
\newcommand{\NBDSeventeen}{338\xspace}
\newcommand{\BDTwentyEight}{BD$+28^\circ 4211$\xspace}
\newcommand{\BDSeventyFive}{BD$+75^\circ 325$\xspace}
\newcommand{\BDFiftyTwo}{BD$+52^\circ 913$\xspace}
\newcommand{\BDTwentyFive}{BD$+25^\circ 4655$\xspace}
\newcommand{\BDThirtyThree}{BD$+33^\circ 2642$\xspace}

\DeclareMathOperator{\sinc}{sinc}

\newcommand{\zpV}{$-21.0711 \pm 0.0016$}
\newcommand{\zpBmV}{$+0.5978 \pm  0.0023$}
\newcommand{\zpUmB}{$-0.4515 \pm 0.0027$}
\newcommand{\zpVmR}{$+0.5442 \pm 0.0009 $}
\newcommand{\zpVmI}{$+1.2444 \pm 0.0025$}

\newcommand{\ourzpV}{$-21.0749$}
\newcommand{\ourzpBmV}{$+0.5998$}
\newcommand{\ourzpVmR}{$+0.5446$}
\newcommand{\ourzpVmI}{$+1.2422$}

\newcommand{\VmRrms}{3.2}

\newcommand{\fmsV}{$+5.3\pm3.1$}
\newcommand{\fmsBmV}{$-6.2\pm3.1$}
\newcommand{\fmsVmR}{$-1.6\pm1.1$}
\newcommand{\fmsVmI}{$+7.7\pm2.7$}

\shorttitle{\snfactory Flux Calibration}
\shortauthors{D. Rubin and the Nearby Supernova Factory}

\begin{document}

\newcommand{\stsci}{\affiliation{Space Telescope Science Institute, 3700 San Martin Drive Baltimore, MD 21218, USA}}
\newcommand{\lbnl}{\affiliation{Physics Division, Lawrence Berkeley National Laboratory, 1 Cyclotron Road, Berkeley, CA, 94720}}
\newcommand{\uhawaii}{\affiliation{Department of Physics and Astronomy, University of Hawai`i at M{\=a}noa, Honolulu, Hawai`i 96822, USA}}
\newcommand{\ucberkeley}{\affiliation{Department of Physics, University of California Berkeley, Berkeley, CA 94720, USA}}
\newcommand{\humboldt}{\affiliation{Institut f\"ur Physik, Newtonstr. 15, 12489 Berlin, Humboldt-Universit\"at zu Berlin, Germany}}

\title{
Uniform Recalibration of Common Spectrophotometry Standard Stars onto the \\
CALSPEC System using the SuperNova Integral Field Spectrograph}

\author[0000-0001-5402-4647]{David Rubin}
\uhawaii
\lbnl

\author{G.~Aldering}
\affiliation{Physics Division, Lawrence Berkeley National Laboratory, 1 Cyclotron Road, Berkeley, CA, 94720}

\author[0000-0002-0389-5706]{P.~Antilogus}
\affiliation{Laboratoire de Physique Nucl\'eaire et des Hautes Energies, CNRS/IN2P3, \\ Sorbonne Universit\'e, Universit\'e de Paris, 4 place Jussieu, 75005 Paris, France}

\author[0000-0002-9502-0965]{C.~Aragon}
\affiliation{Physics Division, Lawrence Berkeley National Laboratory, 1 Cyclotron Road, Berkeley, CA, 94720}
\affiliation{College of Engineering, University of Washington 371 Loew Hall, Seattle, WA, 98195}

\author{S.~Bailey}
\affiliation{Physics Division, Lawrence Berkeley National Laboratory, 1 Cyclotron Road, Berkeley, CA, 94720}

\author[0000-0003-0424-8719]{C.~Baltay}
\affiliation{Department of Physics, Yale University, New Haven, CT, 06250-8121}

\author{S.~Bongard}
\affiliation{Laboratoire de Physique Nucl\'eaire et des Hautes Energies, CNRS/IN2P3, \\ Sorbonne Universit\'e, Universit\'e de Paris, 4 place Jussieu, 75005 Paris, France}

\author[0000-0002-5828-6211]{K.~Boone}
\affiliation{Physics Division, Lawrence Berkeley National Laboratory, 1 Cyclotron Road, Berkeley, CA, 94720}
\affiliation{Department of Physics, University of California Berkeley, 366 LeConte Hall MC 7300, Berkeley, CA, 94720-7300}
\affiliation{DIRAC Institute, Department of Astronomy, University of Washington, 3910 15th Ave NE, Seattle, WA 98195, USA}

\author[0000-0002-3780-7516]{C.~Buton}
\affiliation{Univ Lyon, Universit\'e Claude Bernard Lyon~1, CNRS/IN2P3, IP2I Lyon, F-69622, Villeurbanne, France}
    
\author[0000-0002-5317-7518]{Y.~Copin}
\affiliation{Univ Lyon, Universit\'e Claude Bernard Lyon~1, CNRS/IN2P3, IP2I Lyon, F-69622, Villeurbanne, France}

\author[0000-0003-1861-0870]{S.~Dixon}
\affiliation{Physics Division, Lawrence Berkeley National Laboratory, 1 Cyclotron Road, Berkeley, CA, 94720}
\affiliation{Department of Physics, University of California Berkeley, 366 LeConte Hall MC 7300, Berkeley, CA, 94720-7300}

\author[0000-0002-7496-3796]{D.~Fouchez}
\affiliation{Aix Marseille Univ, CNRS/IN2P3, CPPM, Marseille, France}

\author[0000-0001-6728-1423]{E.~Gangler}  
\affiliation{Univ Lyon, Universit\'e Claude Bernard Lyon~1, CNRS/IN2P3, IP2I Lyon, F-69622, Villeurbanne, France}
\affiliation{Universit\'e Clermont Auvergne, CNRS/IN2P3, Laboratoire de Physique de Clermont, F-63000 Clermont-Ferrand, France}

\author[0000-0003-1820-4696]{R.~Gupta}
\affiliation{Physics Division, Lawrence Berkeley National Laboratory, 1 Cyclotron Road, Berkeley, CA, 94720}

\author[0000-0001-9200-8699]{B.~Hayden}
\affiliation{Physics Division, Lawrence Berkeley National Laboratory, 1 Cyclotron Road, Berkeley, CA, 94720}
\affiliation{Space Telescope Science Institute, 3700 San Martin Drive Baltimore, MD, 21218}

\author{W.~Hillebrandt}
\affiliation{Max-Planck-Institut f\"ur Astrophysik,  Karl-Schwarzschild-Str. 1, D-85748 Garching, Germany}

\author[0000-0001-6315-8743]{A.~G.~Kim}
\affiliation{Physics Division, Lawrence Berkeley National Laboratory, 1 Cyclotron Road, Berkeley, CA, 94720}

\author[0000-0001-8594-8666]{M.~Kowalski}
\affiliation{Institut f\"ur Physik,  Humboldt-Universitat zu Berlin, Newtonstr. 15, 12489 Berlin}
\affiliation {DESY, D-15735 Zeuthen, Germany}

\author[0000-0002-9207-4749]{D.~K\"usters}
\affiliation {Department of Physics, University of California Berkeley, 366 LeConte Hall MC 7300, Berkeley, CA, 94720-7300}
\affiliation {DESY, D-15735 Zeuthen, Germany}

\author[0000-0002-8357-3984]{P.-F.~L\'eget}
\affiliation{Laboratoire de Physique Nucl\'eaire et des Hautes Energies, CNRS/IN2P3, \\ Sorbonne Universit\'e, Universit\'e de Paris, 4 place Jussieu, 75005 Paris, France}

\author{F.~Mondon}  
\affiliation{Universit\'e Clermont Auvergne, CNRS/IN2P3, Laboratoire de Physique de Clermont, F-63000 Clermont-Ferrand, France}

\author[0000-0001-8342-6274]{J.~Nordin}
\affiliation{Physics Division, Lawrence Berkeley National Laboratory, 1 Cyclotron Road, Berkeley, CA, 94720}
\affiliation{Institut f\"ur Physik,  Humboldt-Universitat zu Berlin, Newtonstr. 15, 12489 Berlin}

\author[0000-0003-4016-6067]{R.~Pain}
\affiliation{Laboratoire de Physique Nucl\'eaire et des Hautes Energies, CNRS/IN2P3, \\ Sorbonne Universit\'e, Universit\'e de Paris, 4 place Jussieu, 75005 Paris, France}

\author{E.~Pecontal}
\affiliation{Centre de Recherche Astronomique de Lyon, Universit\'e Lyon 1, 9 Avenue Charles Andr\'e, 69561 Saint Genis Laval Cedex, France}

\author{R.~Pereira}
\affiliation{Univ Lyon, Universit\'e Claude Bernard Lyon~1, CNRS/IN2P3, IP2I Lyon, F-69622, Villeurbanne, France}

\author[0000-0002-4436-4661]{S.~Perlmutter}
\affiliation{Physics Division, Lawrence Berkeley National Laboratory, 1 Cyclotron Road, Berkeley, CA, 94720}
\affiliation{Department of Physics, University of California Berkeley, 366 LeConte Hall MC 7300, Berkeley, CA, 94720-7300}

\author[0000-0002-8207-3304]{K.~A.~Ponder}
\affiliation{Department of Physics, University of California Berkeley, 366 LeConte Hall MC 7300, Berkeley, CA, 94720-7300}

\author{D.~Rabinowitz}
\affiliation{Department of Physics, Yale University, New Haven, CT, 06250-8121}

 \author[0000-0002-8121-2560]{M.~Rigault} 
\affiliation{Univ Lyon, Universit\'e Claude Bernard Lyon~1, CNRS/IN2P3, IP2I Lyon, F-69622, Villeurbanne, France}

\author{K.~Runge}
\affiliation{Physics Division, Lawrence Berkeley National Laboratory, 1 Cyclotron Road, Berkeley, CA, 94720}

\author[0000-0002-4094-2102]{C.~Saunders}
\affiliation{Physics Division, Lawrence Berkeley National Laboratory, 1 Cyclotron Road, Berkeley, CA, 94720}
\affiliation{Department of Physics, University of California Berkeley, 366 LeConte Hall MC 7300, Berkeley, CA, 94720-7300}
\affiliation{Princeton University, Department of Astrophysics, 4 Ivy Lane, Princeton, NJ, 08544}
\affiliation{Sorbonne Universit\'es, Institut Lagrange de Paris (ILP), 98 bis Boulevard Arago, 75014 Paris, France}

\author[0000-0002-9093-8849]{G.~Smadja}
\affiliation{Univ Lyon, Universit\'e Claude Bernard Lyon~1, CNRS/IN2P3, IP2I Lyon, F-69622, Villeurbanne, France}

\author{N.~Suzuki}   
\affiliation{Physics Division, Lawrence Berkeley National Laboratory, 1 Cyclotron Road, Berkeley, CA, 94720}
\affiliation{Kavli Institute for the Physics and Mathematics of the Universe, The University of Tokyo Institutes for Advanced Study, \\ The University of Tokyo, 5-1-5 Kashiwanoha, Kashiwa, Chiba 277-8583, Japan}

\author{C.~Tao}
\affiliation{Tsinghua Center for Astrophysics, Tsinghua University, Beijing 100084, China}
\affiliation{Aix Marseille Univ, CNRS/IN2P3, CPPM, Marseille, France}

\author[0000-0002-4265-1958]{S.~Taubenberger}
\affiliation{Max-Planck-Institut f\"ur Astrophysik, Karl-Schwarzschild-Str. 1, D-85748 Garching, Germany}

\author{R.~C.~Thomas}
\affiliation{Physics Division, Lawrence Berkeley National Laboratory, 1 Cyclotron Road, Berkeley, CA, 94720}
\affiliation{Computational Cosmology Center, Computational Research Division, \\ Lawrence Berkeley National Laboratory, 1 Cyclotron Road, Berkeley, CA, 94720}

\author{M.~Vincenzi}
\affiliation{Physics Division, Lawrence Berkeley National Laboratory, 1 Cyclotron Road, Berkeley, CA, 94720}
\affiliation{Institute of Cosmology and Gravitation, University of Portsmouth, Portsmouth, PO1 3FX, UK}

\collaboration{50}{The Nearby Supernova Factory}

\begin{abstract}

We calibrate spectrophotometric optical spectra of \NSecondary stars commonly used as standard stars, referenced to \NPrimary stars already on the \HST-based CALSPEC flux system. Observations of CALSPEC and non-CALSPEC stars were obtained with the \SNIFS over the wavelength range 3300\,\ang to 9400\,\ang as calibration for the Nearby Supernova Factory cosmology experiment. In total, this analysis used \SpectraNominal standard-star spectra taken on photometric nights. As a modern cosmology analysis, all pre-submission methodological decisions were made with the flux scale and external comparison results blinded. The large number of spectra per star allows us to treat the wavelength-by-wavelength calibration for all nights simultaneously with a Bayesian hierarchical model, thereby enabling a consistent treatment of the Type~Ia supernova cosmology analysis and the calibration on which it critically relies.  We determine the typical per-observation repeatability (median \repeatlong for exposures $\gtrsim 5$~s), the Maunakea atmospheric transmission distribution (median dispersion of \medatmdispersion with uncertainty \meduncatm),
and the scatter internal to our CALSPEC reference stars (median of \internalscatter). We also check our standards against literature filter photometry, finding generally good agreement over the full 12-magnitude range. Overall, the mean of our system is calibrated to the mean of CALSPEC at the level of $\sim3$~mmag.
\ConcludingSentence
\end{abstract}

\keywords{Flux Calibration, Spectrophotometry, Spectrophotometric Standards}

\section{Introduction}

Cosmological distance measurements through Type~Ia supernovae (SNe~Ia) rely on precise relative flux calibration across a large range of distances. The accuracy requirements are especially stringent for inferring the dark energy equation of state parameter $w$. For example, a 10~milli-magnitude (mmag) calibration offset in the distance moduli between nearby ($z \lesssim 0.1$) and mid-redshift ($z~\sim~0.5$) SNe~Ia introduces an offset of $\Delta w \sim 0.02 \mathrm{ - } 0.03$ (depending on external data constraints), comparable to the entire uncertainty budget \citep{Abbott2019}. Standard stars have long served as the basis for establishing internally consistent flux calibration systems; digital photometry has enabled mmag (i.e., $\sim 0.1\%$) flux calibration relative to such standards across the sky within a night \citep[e.g,][]{young1974, mann2011}.
Examples of optical flux calibration systems having good relative calibration include the Landolt $UBVRI$ system of filtered standard stars \citep{landolt1992, landolt2009}, the filter systems established by SDSS \citep{fukugita1996} and Pan-STARRS1 \citep{tonry12}, as well as spectrophotometric standard systems such as CALSPEC \citep{bohlin14}. Flux calibration standards are also used to separate absorption by the Earth's atmosphere from the instrumental sensitivity; this allows the resulting calibration to be extended to science targets observed at different airmasses than the standards.

In order to place those systems on a physical scale, these internally consistent systems need to be referenced to either a laboratory standard or a robust stellar model. \citet{bohlin16} provides a comprehensive review of efforts on these two fronts up to 2016. 
For SN cosmology, the most critical aspect of flux calibration is that the reference system of standard stars be wavelength-neutral, that is, possessing the same zeropoint in physical units at all wavelengths, within a wavelength-independent scale factor.

The currently predominating system for spectrophotometric calibration is based on stellar atmosphere models for three ``fundamental'' white dwarfs (WDs): GD~71, G~191B2B, and GD~153. Each of these stars is within the Local Bubble in the interstellar medium \citep{frisch2011} and thus has essentially no reddening ($<1$~mmag $E(B-V)$) due to dust along the line of sight, thereby avoiding the questions of luminosity and temperature degeneracy with dust extinction and reddening.\footnote{The external constraints pointing to the lack of dust extinction towards these particular standard stars is important for supernova cosmology because the shape and consistency of the dust extinction curve towards SNe~Ia is an important source of systematic uncertainty; there is no basis for ignoring that same source of systematic uncertainty when using stellar atmosphere models for flux calibration.}  The level to which the models for these stars corresponds to the physical calibration essential for cosmology is an area of active study, but they are believed to be closer than the laboratory-referenced calibrations that currently exist. Therefore, they constitute the reference system most commonly relied upon for the flux calibration of current supernova cosmology experiments.

These fundamental stars are, generally, too bright for large telescopes to observe with broadband photometry in typical $\sim 1$-minute exposures, and are bluer than many astronomical objects on average (e.g., galaxies, field stars, high-redshift SNe), thereby introducing calibration uncertainty when the filter bandpasses being used have uncertainty. Thus, the calibration must be transferred to fainter and redder stars. The CALSPEC network \citep{bohlin07, bohlin14, bohlin2020} has met this need, providing a practical intermediary between the fundamental WDs and the stars used to flux calibrate essentially all astronomical surveys \citep{bohlin11, betoule13, scolnic15, rubin15a, bohlin16, currie2020, Brout2021}. Despite its success, CALSPEC is observationally expensive to expand, with the highest-quality optical observations coming only from the \HST Space Telescope Imaging Spectrograph (STIS). CALSPEC also does not include many of the standards in common use \citep[e.g.,][]{hamuy92,hamuy94,oke1990}. In this work, we present an extended optical spectrophotometric standard-star network, which we are able to tightly tie to CALSPEC.

The spectrophotometric data that will be discussed here were taken with the SuperNova Integral-Field Spectrograph (SNIFS; \citealt{aldering2002, lantz2004}) as part of the Nearby Supernova Factory (SNfactory; \citealt{aldering2002}). SNIFS was built by the SNfactory collaboration to observe nearby Type~Ia supernovae for cosmological measurements, such as the dark energy equation of state and galaxy peculiar velocities. SNIFS spectroscopy covers the full optical range simultaneously using two channels separated by a dichroic. At present, the B-channel reductions span 3300\,\ang to 5200\,\ang while the R-channel reductions span 5100\,\ang to 9400\,\ang.

SNIFS was constructed and observations obtained keeping in mind the likely need to improve the flux calibration reference system in the future. For instance, parasitic light paths into SNIFS are strongly suppressed, and the $6\farcs 6\times6 \farcs 6$ field of view encloses essentially all the light from standard stars for normal ranges of atmospheric seeing and atmospheric differential refraction. This field of view is divided across a $15\times15$ element microlens array (MLA), resulting in scale of $0\farcs 43$ per lenslet. The incoming beam is $f/306$, so there are essentially no gaps or shadowing in the spatial coverage of the field. The typical delivered image quality (including atmospheric seeing, dome and telescope seeing, and guiding errors) has a median of $\sim1$\asec, so the point spread function (PSF) is well-sampled. Further details of the instrument can be found in \citet{aldering2002, lantz2004, snfinst}. To further improve the calibration, we note that the SNIFS CALibration Apparatus (SCALA; \citealt{kusters2016, lombardo2017, kusters2019}) has been constructed and installed so that eventually the SNIFS calibration can be referenced to a NIST-calibrated detector.

For the purposes of extending the CALSPEC system employing ground-based observations, establishing the relative flux above the atmosphere is critical. While conceptually straightforward, as presented for the case of the SNfactory in \citet{buton13}, this extension requires observations of many standard stars over a range of airmasses each night. Here we will go beyond the analysis presented in \citet{buton13}, which focused on characterizing the atmospheric extinction above Maunakea using the then published spectrophotometric flux tables for our stars, by putting this heterogeneous mix of stars onto the CALSPEC system. This will involve deriving new spectrophotometry having the 3300--9400\,\ang wavelength coverage of SNIFS for stars not already included in the CALSPEC sample. We depart from the usual nightly linear least squares approach to flux calibration by building a Bayesian hierarchical model to simultaneously calibrate all stars on all nights (as a function of wavelength) while deriving global parameters such as the per-observation repeatability, distribution of atmospheric extinction, and the internal consistency of CALSPEC, among others, and allowing for both inlier and outlier populations of observations.

In \S\ref{sec:network} we present the standard stars we use for this analysis, then \S\ref{sec:data} discusses the observational data and selection for this paper. \S\ref{sec:model} discusses our Bayesian hierarchical model for performing the calibration, while \S\ref{sec:internalcompare} presents the decisions and internal checks performed with the external results still blinded that led us to implement the model as we do. In \S\ref{sec:compare} we present a number of comparisons with external data, both spectrophotometry and filter photometry. We summarize and conclude in \S\ref{sec:conclusion}. Appendices~\ref{app:psf}, \ref{app:bdseventeen}, and \ref{app:MKatm} discuss our PSF model, the status of \BDSeventeen as a standard star, and a physical model for the Maunakea atmosphere, respectively.

\section{Our Standard Star Network}\label{sec:network}

\begin{figure*}[htbp]
\centering
\includegraphics[width=0.8\textwidth]{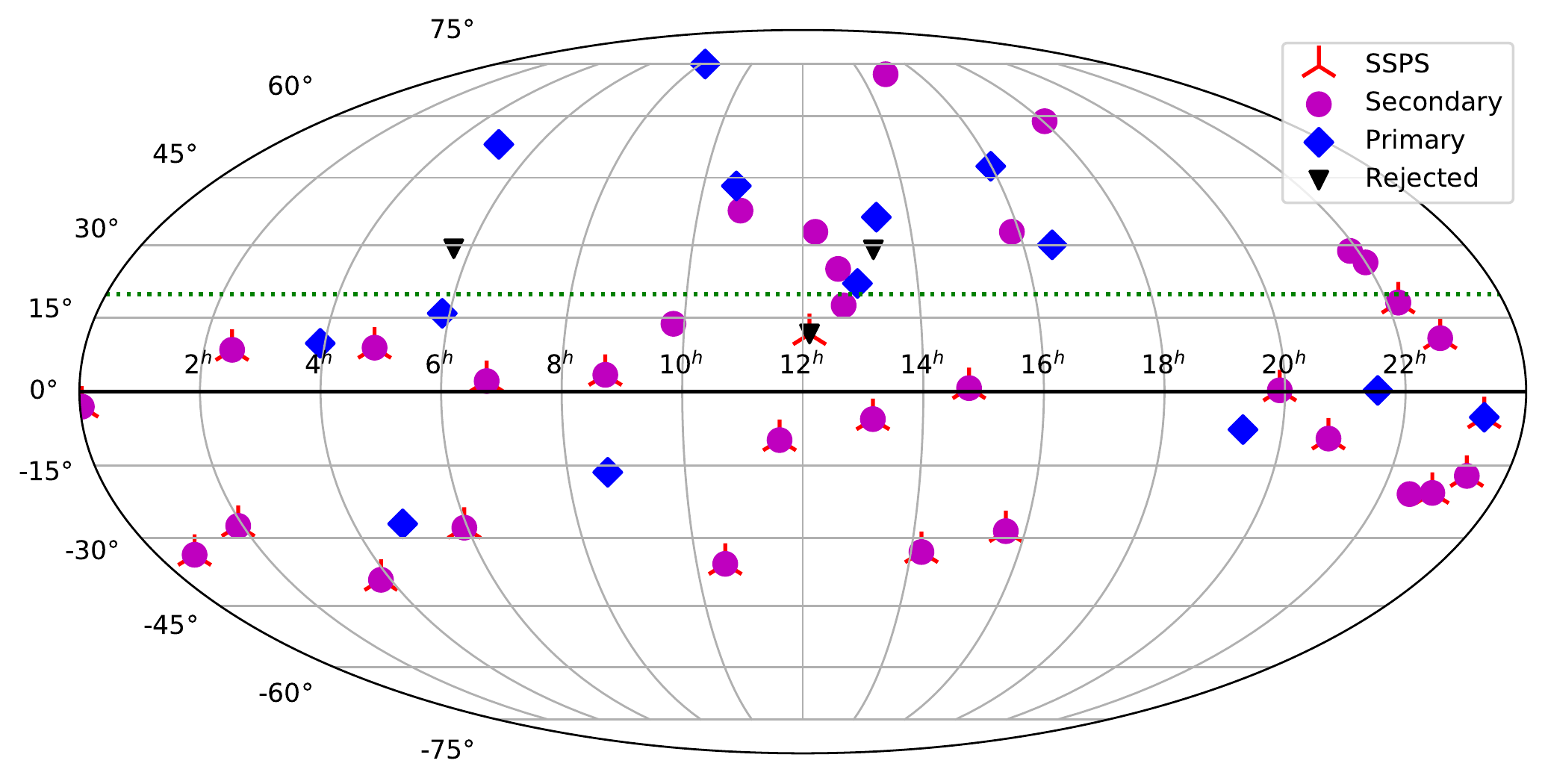}
\caption{The distribution for our standard star network on the sky. Stars are categorized by whether they are included in our set of primary CALSPEC standards or are treated as secondary standards to be recalibrated. Additionally, stars that entered our sample via the Southern Spectrophotometric Standards (SSPS) sample are highlighted; these were originally on the \citet{hayes1985} system and so are likely to change the most when transformed to the CALSPEC system. The standard stars that we ultimately rejected (see Table~\ref{tab:stdstars}) are also shown. The green dashed line indicates the declination corresponding to zenith for Maunakea. The primary and secondary standard stars are both well distributed on the sky.}
\label{fig:stds_on_sky}
\end{figure*}

When SNfactory observations with SNIFS began, there was a considerable mixture of different sets of spectrophotometric standard stars available, with no system demonstrably better than others. We also desired stars with stellar absorption lines that were weak and/or differed between stars, in order to cleanly disentangle stellar features from the instrumental response and absorption by the atmosphere. This was especially important given that spectrophotometry was often reported only in wavelength bins much broader than stellar features, and the spectrophotometric standard stars were often observed through wide slits or apertures, leading to wavelength shifts due to miscentering in the spatial direction parallel to the dispersion direction on the detector. In order to increase the number of standard stars observed each night we also desired some
bright ($V \sim 5$) stars that could be observed with 1~s exposures during nautical twilight. To construct our initial list, we examined stars from the space-based (HST+STIS) CALSPEC set of spectrophotometric standard stars (circa 2004), ground-based spectrophotometry from the set of equatorial and southern spectrophotometric standards of \citet[][hereafter SSPS\footnote{Not to be confused with the \GAIA SPSS spectrophotometric standard stars compilation}]{hamuy92,hamuy94} observable from Maunakea, a few from \citet{oke1990}, and the featureless DC white dwarf EG~131 (Lawd~74), originally presented as a standard star in \citet{bessell1999}. From these we excluded stars with very broad lines or poor wavelength coverage. Subsequent to their initial inclusion in our set of standards, some have become members of the space-based CALSPEC set of spectrophotometric standard stars. In particular, EG~131, Feige~34, HZ~4, HZ~44, HD~93521, and HR~718 ($\xi^2$~Ceti) are now part of CALSPEC.\footnote{These stars were added to the space-based part of CALSPEC in the course of our investigation. Comparison with our pre-existing models for these stars showed exceptional agreement, adding confidence in our results. See also the leave-one-out consistency check in \S\ref{sec:LOO}.}
Some stars initially in our core list of standard stars have been abandoned due to suspected variability or the presence of a nearby companion, as we discuss below. The main list of standard stars used for the \snfactory was originally presented in \citet{buton13}. An updated list with several parameters of interest is given in Table~\ref{tab:stdstars} and the distribution of these stars on the sky is presented in Figure~\ref{fig:stds_on_sky}. Figure~\ref{fig:stds_on_sky} shows that our standard star network has very good sky coverage as seen from Maunakea. Importantly for the current study, the stars that will constitute our primary calibrators, i.e., those on the space-based CALSPEC system, are well-mixed on the sky with the secondary stars that we will be recalibrating here.

\begin{longrotatetable}
\begin{deluxetable*}{llccccclcc}
\tablecaption{Table of spectrophotometric standard stars used or considered in this analysis.\label{tab:stdstars}}
\tablehead{
\colhead{Our}  & \colhead{Alternative} & \colhead{Source\tablenotemark{a}} & \colhead{Sample\tablenotemark{b}} & \colhead{Right Ascension} & \colhead{Declination} & \colhead{V  } & \colhead{MK  } & \colhead{Nights} & \colhead{Median}\\[-1.0em]
\colhead{Name} & \colhead{Name}        & \colhead{}       & \colhead{}       & \colhead{J2000          } & \colhead{J2000      } & \colhead{Mag} & \colhead{Type} & \colhead{}& \colhead{Exposure (s)}}
\startdata
\BDSeventeen                 &                 &  SC    & S & 22:11:31.375 & $+$18:05:34.16  &  9.46  & sdF8         &  208   & 180    \\
\BDSeventyFive               &                 &  SC    & P & 08:10:49.490 & $+$74:57:57.94  &  9.50  & sdO5         &  43   & 180   \\
EG~131                       & LAWD 74         &  SC    & P & 19:20:34.923 & $-$07:40:00.07  & 12.290 & DBQA5        &  200  & 300   \\
Feige 110                    &                 &  SC    & P & 23:19:58.400 & $-$05:09:56.17  & 11.50  & sdO8VIIIHe5  &  97  & 300  \\
Feige 34                     &                 &  SC    & P & 10:39:36.738 & $+$43:06:09.21  & 11.14  & sdOp         &  103  & 300   \\
G191 B2B                     & \BDFiftyTwo     &  SC    & FWD & 05:05:30.618 & $+$52:49:51.92  & 11.69  & DA.8         &  89  & 300   \\
GD 153                       &                 &  SC    & FWD & 12:57:02.322 & $+$22:01:52.63  & 13.349 & DA1.2        &  190  & 600   \\
GD 71                        &                 &  SC    & FWD & 05:52:27.620 & $+$15:53:13.23  & 13.032 & DA1.5        &  136 & 600  \\
HD 31128                     &                 &  SC    & P & 04:52:09.910 & $-$27:03:50.94  &  9.14  & F3/5Vw       &  1  & 100   \\
HD 74000                     &                 &  SC    & P & 08:40:50.804 & $-$16:20:42.51  &  9.66  & F2           &  1  & 101  \\
HD 84937                     &                 &  SC    & S & 09:48:56.098 & $+$13:44:39.32  &  8.32  & F8Vm-5     &  2   & 20  \\
HD 93521                     &                 &  SC    & S & 10:48:23.512 & $+$37:34:13.09  &  7.03  & O9.5IIInn    &  178  & 1   \\
HD 165459                    &                 &  SC    & S & 18:02:30.741 & $+$58:37:38.16  &  6.86  & A1V        &  2    & 1   \\
HZ 4                         &                 &  SC    & P & 03:55:21.988 & $+$09:47:18.13  & 14.506 & DA3.4        &  12    & 601   \\
HZ 44                        &                 &  SC    & P & 13:23:35.263 & $+$36:07:59.55  & 11.65  & sdBN0VIIHe28 &  24 & 300 \\
LDS 749B                     & LAWD 87         &  SC    & P & 21:32:16.233 & $+$00:15:14.40  & 14.674 & DB4          &  8   & 500   \\
P041C\tablenotemark{c}       & GSPC P 41-C     &  SC    & S & 14:51:57.980 & $+$71:43:17.39  & 12.16  & G0V          &  32  & 300     \\
P177D                        & GSPC P177-D     &  SC    & P & 15:59:13.579 & $+$47:36:41.91  & 13.52  & G0           &  109 & 600       \\
P330E                        & GSC 02581-02323 &  SC    & P & 16:31:33.813 & $+$30:08:46.40  & 12.917 & G2V          &  6    & 500   \\
CD-32 9927                   &                 &  SSPS  & S & 14:11:46.324 & $-$33:03:14.38  & 10.444 & A4           &  28   & 180    \\
CD-34 241                    & ``[sic] LTT~377\tablenotemark{d}''      &  SSPS  & S & 00:41:46.921 & $-$33:39:08.43  & 11.208 & F            &  47   & 300   \\
Hiltner 600\tablenotemark{c} & HD 289002       &  SSPS  & R & 06:45:13.373 & $+$02:08:14.69  & 10.44  & B1       &  25    & 180   \\
HR 718                       & $\xi^2$ Ceti    &  SC, SSPS  & S & 02:28:09.557 & $+$08:27:36.22  &  4.30  & B9III        &  167   & 1    \\
HR 1544                      & $\pi^2$ Ori     &  SSPS  & S & 04:50:36.723 & $+$08:54:00.65  &  4.35  & A1Vn         &  143      & 1   \\
HR 3454                      & $\eta$ Hya      &  SSPS  & S & 08:43:13.475 & $+$03:23:55.19  &  4.300 & B3V          &  72     & 1    \\
HR 4468                      & $\theta$ Crt    &  SSPS  & S & 11:36:40.913 & $-$09:48:08.09  &  4.673 & B9.5V        &  83      & 1 \\
HR 4963\tablenotemark{c}     & $\theta$ Vir    &  SSPS  & S & 13:09:56.984 & $-$05:32:20.47  &  4.397 & A1IVs        &  107    & 1   \\
HR 5501                      & 108 Vir         &  SSPS  & S & 14:45:30.206 & $+$00:43:02.18  &  5.665 & B9.5V        &  157     & 1  \\
HR 7596                      & 58 Aql          &  SSPS  & S & 19:54:44.795 & $+$00:16:25.05  &  5.631 & B9IV         &  228     & 1   \\
HR 7950                      & $\epsilon$ Aqr  &  SSPS  & S & 20:47:40.553 & $-$09:29:44.79  &  3.77  & B9.5V        &  146     & 1   \\
HR 8634                      & 42 Peg          &  SSPS  & S & 22:41:27.721 & $+$10:49:52.91  &  3.41  & B8V          &  141     & 1   \\
HR 9087                      & 29 Psc          &  SSPS  & S & 00:01:49.447 & $-$03:01:39.02  &  5.10  & B7III-IV     &  127     & 1  \\
LTT 1020                     & CD-28 595       &  SSPS  & S & 01:54:50.270 & $-$27:28:35.74  & 11.51  & ~            &  44     & 300   \\
LTT 1788                     & LP 995-86       &  SSPS  & S & 03:48:22.613 & $-$39:08:37.01  & 13.15  & F            &  35   & 600  \\
LTT 2415                     & L 595-22        &  SSPS  & S & 05:56:24.742 & $-$27:51:32.36  & 12.38  & sdG          &  44   & 300  \\
LTT 3864                     & CD-34 6792      &  SSPS  & S & 10:32:13.619 & $-$35:37:41.71  & 11.84  & ~            &  16  & 300    \\
LTT 6248                     & LP 916-15       &  SSPS  & S & 15:38:59.648 & $-$28:35:36.97  & 11.62  & A            &  29  & 300   \\
LTT 9239                     & LP 877-23       &  SSPS  & S & 22:52:41.035 & $-$20:35:33.00  & 11.90  & ~            &  28  & 300   \\
LTT 9491                     & EGGR 264        &  SSPS  & S & 23:19:35.388 & $-$17:05:28.47  & 14.111 & DB3          &  43  & 600    \\
\BDTwentyFive         &                 &  O90   & S & 21:59:41.975 & $+$26:25:57.40  &  9.68  & sdO6         &  61  & 180    \\
\BDTwentyEight\tablenotemark{c}  &                 &  O90   & S & 21:51:11.022 & $+$28 51 50.37  & 10.58  & sdO2VIIIHe5  &  56  & 180 \\
\BDThirtyThree        &                 &  O90   & S & 15:51:59.886 & $+$32:56:54.33  & 10.73  & O7p          &  41   & 150  \\
Feige 66                     & BD$+25^\circ 2534$  &  O90   & S & 12:37:23.516 & $+$25:03:59.87  & 10.59  & sdB1(k)      &  14  & 180    \\
Feige 67                     & BD$+18^\circ 2647$  &  O90   & S & 12:41:51.790 & $+$17:31:19.75  & 11.63  & sdOpec       &  25  & 300   \\
HZ 21                        &                 &  090   & S & 12:13:56.264 & $+$32:56:31.36  & 14.688 & DO1          &   56 & 600 \\
NGC 7293                     &                 &  O90   & S & 22:29:38.545 & $-$20:50:13.75 & 13.524 & DAO.5        &  28   & 600   \\
\cutinhead{Excluded Stars}
Feige~56\tablenotemark{e}    & HD 105183       &  SSPS  & R & 12:06:47.235 & $+$11:40:12.66  & 11.06  & sdB8IIIHe2   &  27 & 300   \\
HD 37725\tablenotemark{f}    &                 &  SC    & R & 05:41:54.370 & $+$29:17:50.96  &  8.31  & A3V        &   3  & 20 \\
HZ 43\tablenotemark{c}       &                 &  SC    & R & 13:16:21.853 & $+$29:05:55.38  & 12.66  & DAwk+M3.5Ve  &  23 & 300 \\
\enddata
\tablenotetext{a}{SC = \citet{bohlin2020}; SSPS = \citet{hamuy92, hamuy94}; 090 = \citet{oke1990}}
\tablenotetext{b}{FWD = Fundamental white dwarf, P = Primary CALSPEC star; S = Secondary star; R = Rejected.}
\tablenotetext{c}{Has companion; see \S\ref{sec:companions}}
\tablenotetext{d}{\citet{Pancino2012} showed that \citet{hamuy92} misidentified this star as LTT~377.}
\tablenotetext{e}{Suspected variable star; see \S\ref{sec:variables}}
\tablenotetext{f}{Variable star; see \S\ref{sec:variables}}
\end{deluxetable*}
\end{longrotatetable}

\subsection{Companion Stars}
\label{sec:companions}
Over the course of time, nearby companion stars have been discovered for a few of these standards, which could result in differences that depend on spatial resolution and/or orbit phase, i.e, between measurements with STIS, SNIFS, and reference photoelectric photometry. In principle, we could model the presence of these companions and then include or exclude them as needed, but we do not do that here. In particular, one of the original CALSPEC white dwarfs, HZ~43, has a companion 2.33\asec away, and has therefore been dropped from CALSPEC \citep{bohlin2001,gaia_edr3}. For this reason, we have dropped it as a calibrator as well. Another CALSPEC standard, P041C, has a red companion 0.57\asec away \citep{gilliland2011}; this is inside the 2\asec-wide \STIS slit employed for CALSPEC, unresolvable with SNIFS, and the companion is very faint over most of the optical. Therefore, we use P041C for nightly calibration but do not include it as a primary CALSPEC reference star. \GAIA finds a companion 4.3\asec away from, and $\sim8$~mag dimmer than the CALSPEC star EG~131; this level of contamination is much too small to be of consequence so we retain EG~131 as a primary CALSPEC calibrator. Feige~34 has an IR excess that \citet{latour2018} model as due to a M0 companion. However, there is no radial velocity or astrometric evidence for variations in this system so we retain it as a CALSPEC standard. 

The \citet{oke1990} standard star \BDTwentyEight has a companion \citep{massey1990, landolt2007}. However, the \GAIA EDR3 positions and proper motions \citep{gaia_edr3} indicate that over the period of our SNIFS observations the separation ranged from 3.5 -- 4.3\asec, which is outside the SNIFS spectroscopic field. As the \GAIA parallaxes indicate that this pair is not physical, their separation will continue to increase. Moreover, \GAIA finds that the fractional brightness of the companion is only 15~mmag in $G$-band.\footnote{The fractional brightnesses in the \GAIA $B_p$ and $R_p$ bands are 12 and 39~mmag, respectively.} Given its separation and faintness, there is no need to eliminate \BDTwentyEight from our sample based on the presence of this nearby star. Furthermore,  we have discovered a companion to the SPSS standard star Hiltner~600 that is 1.95\asec away and $\sim4$~mag fainter, confirmed by \GAIA. These two stars are a physical system, based on their common \GAIA proper motion. Since the configuration is stable and the combination of angular separation and relative brightness is large, the net impact of the companion on SNIFS observations is small enough that we retain Hiltner~600.

\begin{figure*}[htbp]
\centering
\includegraphics[width=0.8\textwidth]{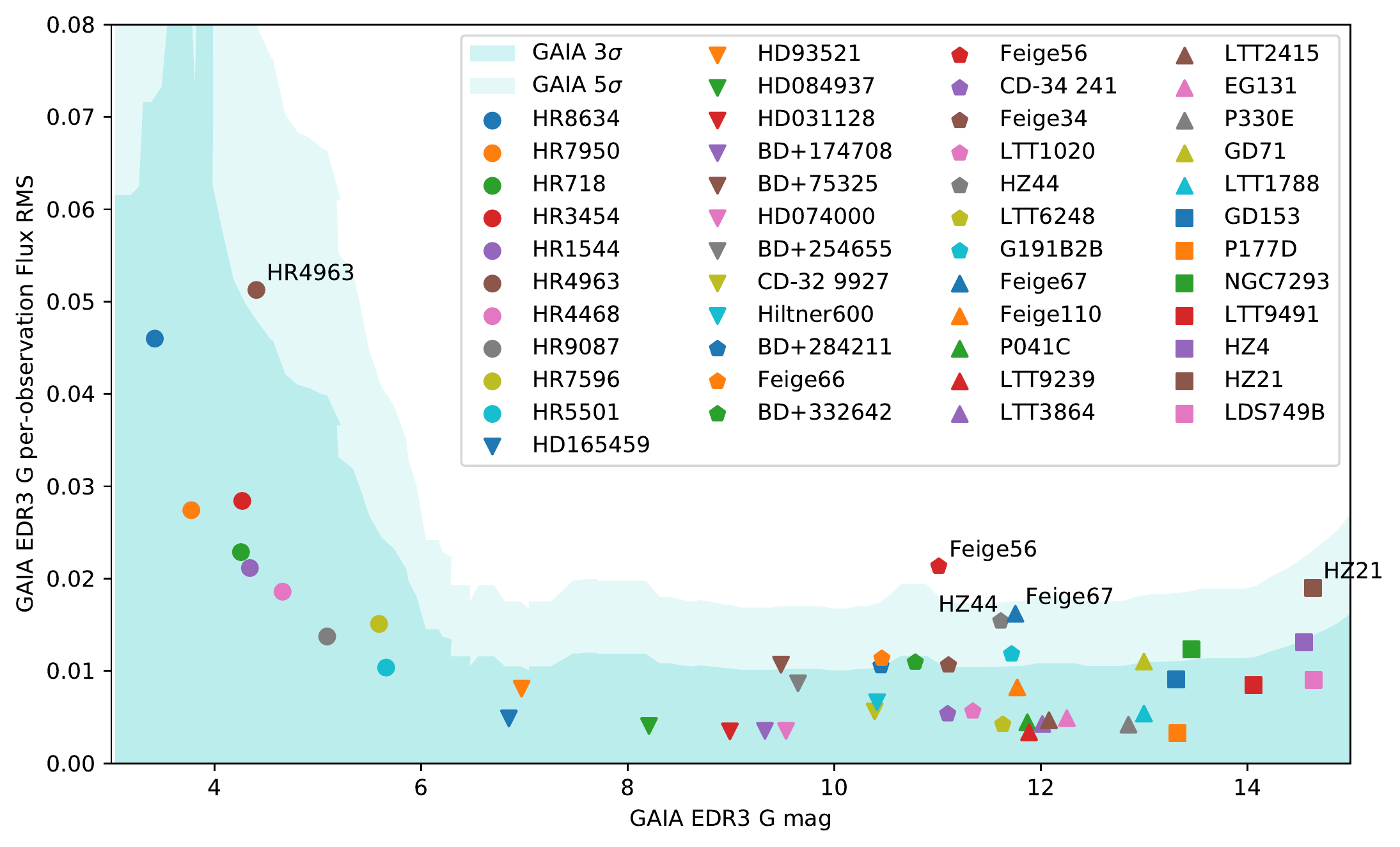}
\caption{The per-transit RMS of the \GAIA $G$ magnitude versus the $G$ magnitude. The RMS includes \GAIA measurement errors and any stellar variability that may be present. The legend identifies stars in order of their $G$-band magnitude. The cyan shaded regions indicate the $3\,\sigma$ and $5\,\sigma$ measurement uncertainty ranges expected from the typical \GAIA measurement accuracy (taken from \citealt{gaia_edr3_phot}). Stars with significant RMS --- larger than 0.015~mag and $3\,\sigma$ larger than the expected measurement uncertainty --- are labeled.}
\label{fig:gaia_rms}
\end{figure*}

\subsection{Potential Variability}
\label{sec:variables}
Additional stars in our network have been identified as variable or suspected variables in the literature. Here we examine the literature evidence for variability, signs of variability from \GAIA, and the scatter found within our own observations. 

Throughout this section we will consult the \GAIA variability results shown in 
Figure~\ref{fig:gaia_rms}. Plotted is the per-epoch RMS for all of our standard stars, inferred from the mean $G$-band flux and uncertainty and the number of transits from \GAIA EDR3 \citep{gaia_edr3, gaia_edr3_phot}. The \GAIA EDR3 observations span a period of 34~months from 25~July~2014 to 28~May~2017, and the number of observations for each of our stars ranges from 153 to 847. Also shown are the $3\sigma$ and $5\,\sigma$ per-transit measurement uncertainties inferred from Figure~14 of \citet{gaia_edr3_phot}. These indicate a number of our standards that might be variable over a period of $\sim3$~yrs at the 3\,$\sigma$ level according to this metric.

{\it HD~37725}: Late in our program we began to include observations of the newer CALSPEC star HD~37725. But \citet{marinoni16} subsequently showed that it is a $\delta$-Scuti variable star, so we no longer include it.

{\it BD$+75^\circ 325$}: \citet{bartolini1982} examined BD$+75^\circ 325$, detecting possible periodicity of 67~min and amplitude of 30~mmag, but they do not consider the result convincing. \citet{landolt2007b} also discuss the variability of BD$+75^\circ 325$, noting a rather high dispersion of 11~mmag. 
In \GAIA EDR3, BD$+75^\circ 325$ is not exceptional relative to the entire network or the expected \GAIA error bands, though its RMS of 12~mmag is consistent with \citet{landolt2007b}. As we do not detect a long-term trend over our several years of observations, we continue to include BD$+75^\circ325$ among our set of primary calibrators. In \S\ref{sec:dispersion}, we find a $\sim$ \bdseventyfivemmag~mmag offset between our observations and CALSPEC, among the worst of our primary standards.

{\it Feige~56:} \citet{marinoni16} measured variation of Feige~56 with amplitude $33\pm11$~mmag, but included it among stars having observations with drawbacks. This star shows significantly worse repeatability in \GAIA (21~mmag) than other standards of similar magnitude.\footnote{In \GAIA DR2 the uncertainties for stars with magnitudes similar to Feige~56 were much larger, such that Feige~56 has only become a clear outlier since \GAIA EDR3.} In our SNIFS observations we also see worse repeatability for this star (an extra $\sim 24$~mmag added in quadrature). Thus, we do not recommend the continued use of Feige~56 as a standard.

{\it HR4963:} In Figure~\ref{fig:gaia_rms}, this star stands out as a possible outlier. At such bright magnitudes \GAIA suffers saturation \citep{evans2018}, so measurment uncertainties increase substantially. HR4963 is a well-known close double star \citep{HR4963_binary}, with a current separation of 0.4\asec and period of 695~yrs \citep{zirm2015}. It is listed as a possible $\delta$-Scuti star in \citep{liakos2017}, but our review of the 4~yrs of monitoring performed by \citet{adelman1997} shows only 6--9~mmag of variation --- consistent with that of their comparison star. Our SNIFS observation do not show unusual variation. As \GAIA does not report the components of HR4963, we conclude that it was not resolved by \GAIA. Thus, we suspect that the binary nature of this star plus \GAIA saturation has led to larger than usual scatter in the brightness measurements. We conclude that HR4963 remains a useful standard star.

{\it HZ~21, HZ~44} and {\it Feige~67:} These three stars seem to have higher-than-expected dispersion measured by \GAIA, as seen in Figure~\ref{fig:gaia_rms}. They are only slightly fainter than the range of magnitude where \GAIA DR2 exhibited substantial uncertainty \citep{evans2018}, which EDR3 is thought to have improved \citep{gaia_edr3_phot}. Their scatter of 17~mmag is within the repeatability of SNIFS (see \S\ref{sec:dispersion}), so we are unable to provide an independent constraint on their variablility. Given the weakness of the evidence for variability, we have not excluded these three stars as standards. However, we have not observed them extensively so they carry little weight in the analysis here.

{\it \BDSeventeen:} We explicitly exclude \BDSeventeen as a primary standard, as it is suspected of being mildly variable \citep{bohlin15, marinoni16}. In Appendix~\ref{app:bdseventeen}, we show that it has a small but detectable long-term drift of $0.9 \pm 0.3$~mmag/yr. No short-term variability was found, despite our large number of observations (\NBDSeventeen), indicating that any such variability is much smaller than the repeatability of SNIFS (see \S\ref{sec:dispersion}). So we do include \BDSeventeen as a secondary standard in our primary analysis, but recalibrate it to the primary standards over the time period of our data.

{\it BD$+25^\circ 4655$:} \citet{bartolini1982} find BD$+25^\circ 4655$ to be variable, with a period of 13.5~minutes and amplitude of 70~mmag. But \GAIA EDR3 shows variation of only 9~mmag and our SNIFS observations also rule out the \citet{bartolini1982} level of variability. Therefore, we retain BD$+25^\circ 4655$.

{\it BD$+28^\circ 4211$:} Noted above for the presence of a non-physical companion, BD$+28^\circ 4211$ is also suspected of variability in \citet{marinoni16}. Their 75 observations show a linear brightness trend of $212\pm27$~mmag/day over a span of 1~hr. However, both of the other stars monitored on that same night --- neither considered variable --- also show clear brightness changes that are linear in time, albeit only about half as large. \GAIA EDR3 finds an RMS of $12\pm4$~mmag based on 245 observations over almost 3~yrs. Thus, we consider  BD$+28^\circ 4211$ sufficiently stable, and so retain it as a secondary standard star.

More sensitive tests of variability will become possible using \GAIA epoch data, although we note that \citet{marinoni16} provides tighter limits than \GAIA on the very short-term (few hour) variability of many of our standards, while  \citet{Mullally2022} checks on timescales of minutes to days.

\section{Our Dataset}\label{sec:data}

Our dataset of spectrophotometric standard stars has been obtained by the SNfactory using the SNIFS integral-field spectrograph in the course of obtaining spectrophotometric time series of nearby Type~Ia supernovae in order to improve constraints on the dark energy equation of state. SNfactory typically observed stars during evening and morning twilight, at midnight, and 2--3 times in between. During bright twilight, bright standard stars are chosen, and at midnight a CALSPEC star was given highest priority in the selection. Priority is first given to a star at low airmass, then a star at high airmass. Thereafter priority is given according to which star can best improve the calibration solution. In this calculation, bright stars (requiring $\sim$ 1~sec exposures) are given lower weight since their PSFs can exhibit more structure because few atmospheric turbulence phase distortion cells pass over the telescope for short exposures. These 1~sec exposures also experience scintillation noise, but this is estimated to be less than $\sim5$~mmag per observation, subdominant to PSF variability.\footnote{This estimate uses the Maunakea turbulence value determined by \citet{osborn2015} along with their Equation~7.} The program {\tt{stdstar\_factory}} automatically selects standard stars using these rules to select which standard star observation would provide the best flux calibration at any given time of the night given the standards already observed.

\begin{figure*}[htbp]
\centering
\includegraphics[width=\textwidth]{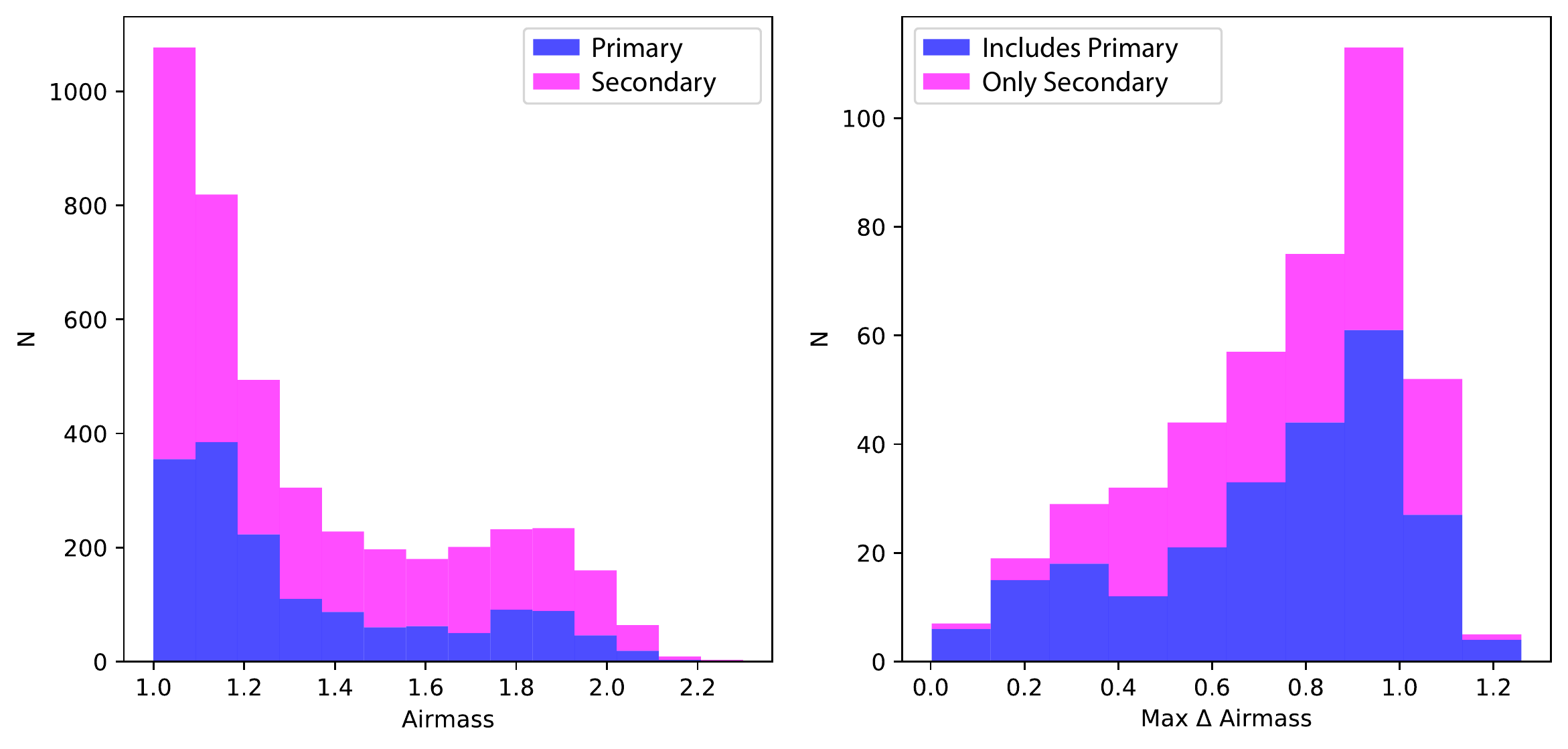}
\caption{Histograms showing the incidence of standard star observations per airmass (left) and airmass range (right). The values are color-coded by the sample --- primary (blue) or secondary (magenta) --- to which each standard star or pair of standard stars belongs. For the airmass range calculation, a pair is categorized as ``primary'' if at least one star in the pair is one of our primary standard stars. These distributions demonstrate that the airmass values and ranges are very similar for the primary (space-based CALSPEC) and secondary (SSPS, \citealt{oke1990}) standard stars.}
\label{fig:obs_airmasses}
\end{figure*}

\begin{figure*}[htbp]
\centering
\includegraphics[width=\textwidth]{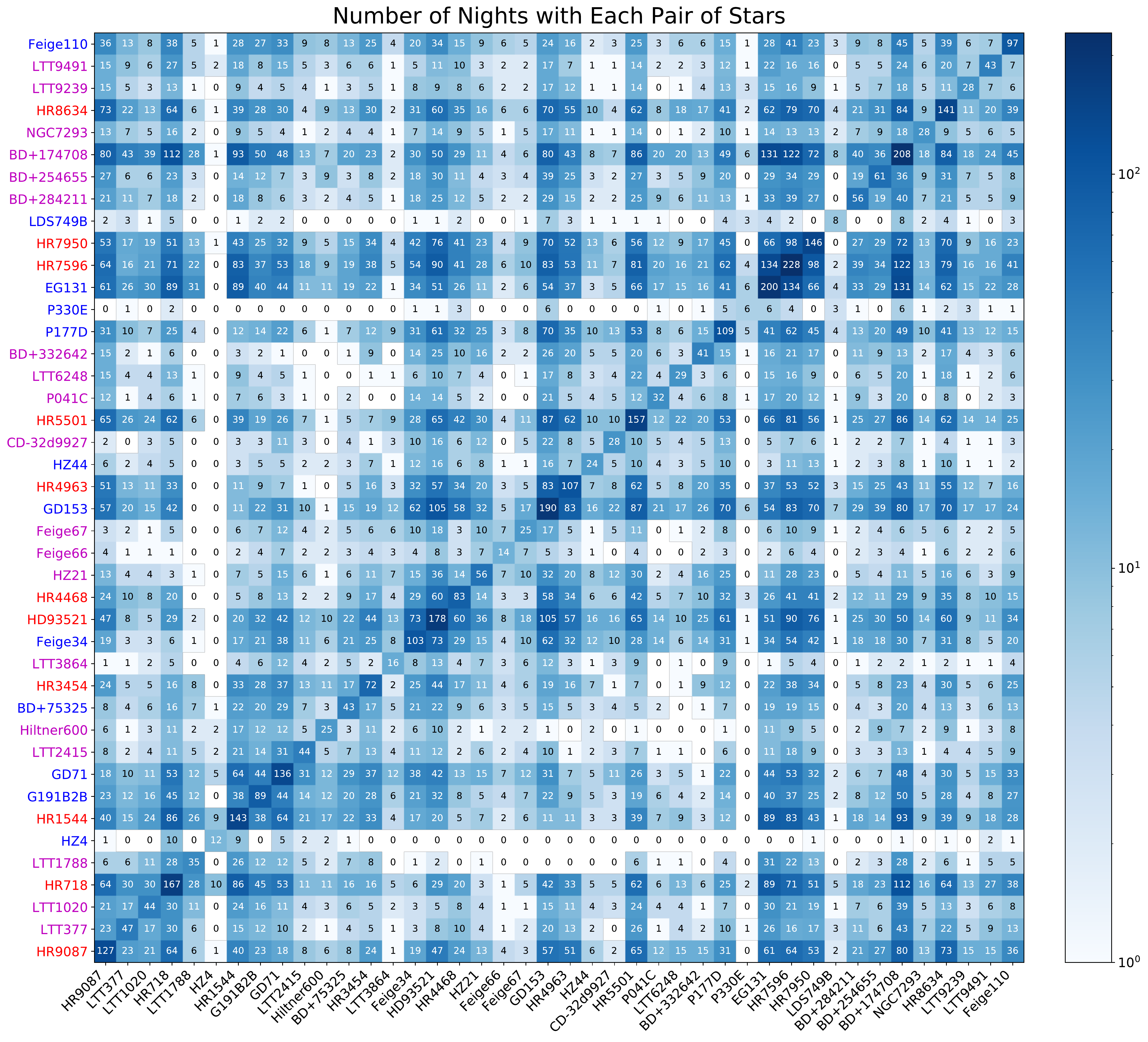}
\caption{Array showing the number of times different pairs of stars were observed on the same photometric night. The stars are ordered by right ascension and color coded (primary CALSPEC in blue, long exposure secondaries in magenta, and short-exposure secondaries in red). Four stars observed on fewer than five nights are included in our analysis, but removed from this figure for clarity. Note that the plot is symmetric. Some seasonal structure exists, but overall our standard star network is knit together quite well. Groupings of bright and faint stars result from our method for observing standard stars during twilight, with brighter stars observed when twilight is brighter and fainter stars observed when twilight is fainter or between the end and start of astronomical twilight.}
\label{fig:nobs_matrix}
\end{figure*}

The distributions of airmasses and airmass range per night are shown in Figure~\ref{fig:obs_airmasses}. These distributions reflect the combination of the {\tt{stdstar\_factory}} selection algorithm and the standard star distribution on the sky. For a large fraction of nights, a large airmass range was obtained. Nights with small airmass ranges are generally due to technical difficulties that prevented normal standard star observations or were early in the program. Figure~\ref{fig:nobs_matrix} shows which stars were observed on the same night as other standard stars. Clearly certain stars were well-placed during periods when \snfactory observed (preferentially spring, summer, and fall), or deemed more important by {\tt{stdstar\_factory}}, and so received more observations. Groupings of CALSPEC and bright (HR) stars are apparent, reflecting the high weight placed on CALSPEC stars, as well as the use of twilight observations of bright stars to improve the calibration solution.

For our primary analysis, we selected all standard-star spectra from those photometric nights (619 nights out of 1160) having at least two observations on each channel per night. We only selected observations that were part of normal scientific operations (rejecting observations that were used for engineering, such as minimizing the focus offset between the spectrograph and imaging channels). We masked the wavelengths of any spectra where the $X$ or $Y$ centroid (the centroid is a function of wavelength due to atmospheric differential refraction) was more than 6 spaxels ($2.6\asec$) from the center of the MLA. We rejected any spectra with altitude $< 25^{\circ}$ (airmass $>$~2.37). Spectra with minimum ${\mathrm{SNR}} < 15$ per wavelength bin were also removed since they would also be atypical for a standard star observation. We performed an initial robust analysis that indicated two $\sim 0.25$ magnitude outlier spectra (one outlier on the red side and one on the blue side), so we removed them. This maximal selection left \BSpectraNominal spectra on the blue side (depending on wavelength, as ADR affects the centroid cut) and \RSpectraNominal spectra on the red side of \NStandNominal standard stars over \NightsNominal photometric nights. Our primary analysis (discussed in \S\ref{sec:model}) models each wavelength independently, so it handles fractional spectra naturally.

The primary analysis is calibrated to observations of the following CALSPEC stars, each of which has full optical coverage\footnote{Specifically, we require both G430L and G750L observations with the $2\asec$ wide slit.} from the Space Telescope Imaging Spectrograph (STIS): GD~71, G~191B2B, GD~153, Feige~34, Feige~110, \BDSeventyFive, EG~131, P177D, P330E, LDS749B, HD~31128, HD~74000, HZ~4, and HZ~44.\footnote{The CALSPEC file versions we used are gd71\_stiswfcnic\_003, g191b2b\_stiswfcnic\_003, gd153\_stiswfcnic\_003, feige34\_stis\_006, feige110\_stisnic\_008, bd\_75d325\_stis\_005, gj7541a\_stis\_004, p177d\_stisnic\_008, p330e\_stiswfcnic\_003, lds749b\_stisnic\_008, hd031128\_stis\_005, hd074000\_stis\_005, hz4\_stis\_007, and hz44\_stis\_006, respectively. For a test where we calibrated directly to the fundamental white-dwarf models, we use gd71\_mod\_011, g191b2b\_mod\_011, and gd153\_mod\_11. We convert each reference spectrum from vacuum to air wavelengths to match our data.} For our network, these are our primary standards. We do not include the CALSPEC stars HR~718 ($\xi^2$~Ceti), HD~93521, or HD~165459 as primary calibrators, as these stars are too bright ($V \sim 4.3$, 7.0, and 6.9~mag, respectively) to be observed with the standard long exposures used for the other CALSPEC stars and SNe. Instead, we treat them as secondary standards.

For our recalibration of the standard stars, we use only photometric nights. As described in \S5 of \citet{buton13}, we determine whether or not a night is photometric or not using a combination of CFHT SkyProbe \citep{skyprobe2009,cuillandre2016}, the SNIFS guide star brightness (providing samples every 0.3 to 2~sec for exposures ranging from 1 to 40~minutes), and parallel observations of nearby stars obtained with the SNIFS imaging channel while the standards are observed in the spectroscopic channels. With respect to \citet{buton13}, we also improved the deglitching algorithm applied to the SkyProbe data, based in part on additional technical details, such as the fact that telescope pointing moves can affect SkyProbe frames on either side of a move in addition to those taken during a move. In addition, for the period 2011--2016 also we inspected video from the highly sensitive CFHT CloudCam, which covers the eastern half of the sky, and is very effective since cirrus predominately passes in a east-west direction. These improvements changed the status of only a few nights among those analyzed in \citet{buton13}. When employing these methods it was important to compare them in order to avoid false positive evidence for clouds since each input suffers from noise and glitches. The structure of clouds leads to attenuation, $\tau$, that follows a power-law probability distribution with $P(\tau) \sim \tau^{-1.84}$ \citep{skyprobe2009}, so with hundreds of samples between Skyprobe, the SNIFS guide stars, and cloud video, the probability is high that at some point during a non-photometric night cloud attenuation will be detectable. Thus our sensitivity is sufficient to detect essentially all non-photometric nights. Even if thin cirrus is occasionally missed, the large number of well-mixed observations of the primary calibrators and secondaries, as illustrated in Figure~\ref{fig:nobs_matrix} ensures that their fluxes are on the same scale.

In addition, a few otherwise photometric nights were excluded if partial occlusion by the dome occurred for any observation that night. This problem was evidenced by a sawtooth pattern in the guide-star signal, with jumps towards more received starlight whenever the dome corrected its position. These improvements changed the status of only a few nights among those analyzed in \citet{buton13}. The number of photometric nights on which each standard was observed is included in Table~\ref{tab:stdstars}. 

The processing of the SNIFS data is described in \citet{bacon2001, aldering2006, scalzo2010}. In brief, after bias and dark subtraction the spectrum for each spaxel is extracted from the CCD to form a data cube. The count spectrum from each spaxel in a cube is flat-fielded, corrected for per-observation dichroic shifts, and wavelength-calibrated.\footnote{This work uses wavelengths for Ar~I, Cd~II, and Hg~I as determined in the NIST Atomic Spectra Database \citep{NIST_ASD}, which are for air normalized to $P = 1013.25$~mbar and $T = 15$~C). For comparison with CALSPEC we convert its vacuum wavelengths to this system.} Note that the flat-fielding also removes the nominal spaxel-to-spaxel efficiency variations. A model point spread function, described in Appendix~\ref{app:psf}, plus uniform sky is fit at each wavelength, including allowing for atmospheric differential refraction. For standard stars this produces a spectrum, $S$ (in units of flat-fielded counts per wavelength per second), ready to be calibrated. We also take into account the shutter latency (few tens of milliseconds), which we have measured as a function of hour angle and affects the 1s exposures \citep{snfinst}.

\section{The Flux-Calibration Model}\label{sec:model}

An effective model for our analysis needed to consider/accommodate a number of factors. First, we
have heterogeneous numbers of observations of different CALSPEC stars; thus weighting each CALSPEC star equally in our calibration is not optimal. Second, \citet{bohlin15} find a scatter of \bohlinlandoltscatter when comparing Landolt and CALSPEC $UBVRI$ synthetic photometry; if any of this scatter is internal to CALSPEC then per-observation weighting will also not be optimal. Therefore, we required a model that could determine the internal dispersion of the CALSPEC system. Also, we knew from our calibration analysis presented in \citet{buton13} that there was a per-observation repeatability floor, and that it could differ for long and short exposures. We wanted a model that could determine these values, rather than having us assign them.
A common approach to this problem when performing flux calibration is to use an iterative frequentist approach \citep[cf.][]{burke2018}. A more general approach is to employ a Bayesian hierarchical model \citep[cf.][]{Narayan2019}; we opt for this approach. As described below, this model can infer the relative calibration offsets (as a function of wavelength) between the CALSPEC stars, the night-to-night dispersion in atmospheric extinction, and other parameters that a Bayesian hierarchical model is able to treat in an unbiased way.

We tried two Bayesian approaches: the first fit one model for the entire dataset, and the second fit the data for each wavelength separately (and in parallel). The primary advantages of the simultaneous model are that physical atmospheric components can be imposed, as in \citet{buton13}, and the parameters can be required to correlate or even have a strict wavelength dependence. For instance, second-order light is present at the reddest SNIFS wavelengths ($> 9500$\,\AA) for very blue stars; the brightness at blue wavelengths can be used to model second-order light as a simple transfer function to predict the brightness from this component at red wavelengths. In addition, it is possible to employ radiative transfer models, e.g., to obtain a single H$_2$O column density that determines the strength of H$_2$O at all wavelengths self-consistently (rather than using a fixed template with power-law scaling in airmass, as was done in \citealt{buton13}). In addition, there are parameters which vary only slowly with wavelength (e.g., repeatability), and their behavior is thus easier to constrain. In practice, this simultaneous model was too slow, and thus we relied on the wavelength-by-wavelength solution for the results here.\footnote{Each of 2,351 wavelengths took $\sim 3$ CPU-hours to run, and each could be run in parallel on a computing cluster. A Bayesian model that treated all wavelengths simultaneously would likely require at least as many CPU hours to converge, and it would be more difficult to efficiently spread the tasks across thousands of CPUs. We also tried a simultaneous frequentist model. The primary disadvantage of this frequentist approach is that it assumed Gaussian uncertainties and was thus not robust to the (mildly non-Gaussian, see \S\ref{sec:repeat}) tails of the residual distribution. Including non-Gaussian tails in the frequentist model made the fit convergence difficult to assess.}

\subsection{Wavelength-by-Wavelength Solution}

As described above, our primary analysis is a Bayesian hierarchical model which treats the data for every wavelength independently for computational speed. This allows the most flexibility in its uncertainty assumptions, but the lack of wavelength-wavelength interactions eliminates the ability to precisely model telluric absorption, which is non-linear with airmass by different amounts at different wavelengths. It also cannot precisely account for second-order light, and it can be more sensitive to wavelength resolution or calibration errors around strong stellar absorption features. Our Bayesian hierarchical model builds wavelength-by-wavelength models of the spectra of our standards; these models are used to determine the airmass dependence and flux zeropoint for each night. We also include in the model a Gaussian distribution (plus a separate Gaussian outlier distribution) for the repeatability of the observations in each exposure class --- short ($<\shortlong$ seconds, generally $\sim 1$s) or long. We believe that for a stable star measured with high SNR the repeatability is dominated by how well our PSF model (\S\ref{app:psf}) is typically seen to fit the observations. 
Additionally, we allow for some scatter within the system of CALSPEC primary standards, since \citet{bohlin15} found scatter ranging from \bohlinlandoltscatter when comparing synthetic and filter photometry of 11 CALSPEC stars; some of this scatter may be internal to CALSPEC. Finally, a prior is placed on the coefficients for the airmass and temperature dependence so that the small number of nights with small airmass (Figure~\ref{fig:obs_airmasses}) or instrumental temperature ranges can still be used. Note that the hierarchical model itself determines the means and standard deviations of, e.g., the atmospheric extinction and the size of the CALSPEC scatter from the ensemble of observations. The values of these hyperparameters (rather than the calibration parameters themselves) are constrained by fixed priors applied independently to each hyperparameter.

Note that we build our model in log flux, but our flux uncertainties are linear; in principle the difference can lead to a bias at low SNR, since the mean of the log is biased by $0.5/{\mathrm{SNR}}^2$ relative to the log of the mean. Since we allow for inlier and outlier distributions, we should have less bias.\footnote{For example, the log of the median of a dataset equals the median of the log of the dataset. Other measures transform differently; for the log-normal distribution, the mean shifts by $+\sigma^2/2$ compared to the mean of the log, and the mode shifts by $-\sigma^2$ compared to the mode of the log. Our robust model for each star lies between these three statistical measures, so it is plausible that our bias is bounded by $0.5/{\mathrm{SNR}}^2$.} To be conservative, we only used data with $\mathrm{SNR} > \SNRcut$, thereby limiting the bias to less than 2~mmag for any individual wavelength of any individual standard star observation. Since the SNR for most observations at most wavelengths is much higher than this for all of our stars, the net bias should be below 1~mmag (thus well below our measurement precision). By making a cut on SNR at ${\mathrm{SNR}}_{cut} = \SNRcut$ there is the potential for an Eddington-like bias \citep{Eddington1913}, going as ${\mathrm{SNR}}^{-2}\, d\, {\mathrm{ln(N(SNR >SNR_{cut}}}))/d{\mathrm{SNR}}_{cut}$ However since the population of observations with SNR falling below the limit ${\mathrm{SNR}}_{cut}$ is small, $d\,{\mathrm{ln}}({\mathrm{N(SNF > SNR}}_{cut}))/d\,{\mathrm{SNR}}_{cut} \ll 1$, this bias too can be ignored. 

The mathematical framework for implementing the Bayesian hierarchical calibration model is constructed as follows.
For the $i^{th}$ observed spectrum, $S$, of star $j$ on night $n$ with exposure-time category $t$ (i.e., long or short) and wavelength bin $l$ the model is:
\begin{equation} \label{eq:modelspectra}
    -2.5 \log_{10} (S^{\mathrm{\,mod}}_{i,l}) = m_{j,l}\, +\, k_{n,l}\, \mathrm{X}_i \, +\, b_{n,l}\, \Delta T_i\, +\, c_{n,l}\, +\, \Delta c_{t,l}(x_{i,l}, y_{i,l})\, 
\end{equation}
where $m_{j,l}$ describes the monochromatic magnitude of star $j$ (Equation~\ref{eq:standardflux}), $k_{n,l}$ is the airmass dependence (and $\mathrm{X}_i$ the airmass), $b_{n,l}$ is the nightly instrumental temperature dependence (and $\Delta T_i$ is the temperature difference between observation $i$ and the nightly mean instrumental temperature),\footnote{Allowing both the temperature and airmass dependence to vary night-to-night may seem like too many fit parameters. However, as shown in Equation~\ref{eq:hierarchy}, we infer a data-driven prior on both terms that enables calibrations of nights with sparse airmass or temperature sampling.} and $c_{n,l}$ is the flux zeropoint.\footnote{To aid with the inspection of the output and possibly help with MCMC sampling, we internally use physical fluxes in units of $10^{-15}$~erg~s$^{-1}$~cm$^{-2}$~\ang$^{-1}$ to more closely align physical units and the units of the extracted spectra $S$.} $\Delta c_{t,l} (x_{i,l}, y_{i,l})$ describes a smooth flat field inferred from stars relative to the flat field provided by the SNIFS internal continuum lamp (Equation~\ref{eq:starflat}). This term is intended to capture not only any illumination difference, but also any mean differences in the extraction of the spectra from the CCD between target and lamp spectra.
The $\Delta c$ ``star-flat'' term expands to
\begin{equation} \label{eq:starflat}
    \Delta c_{t,l}(x_{i,l}, y_{i,l}) = A_{1\ t,l} \frac{x_{i,l}}{4} + A_{2\ t,l} \frac{y_{i,l}}{4} + A_{3\ t,l} \left( \frac{x_{i,l}}{4} \right)^2 + A_{4\ t,l} \left(\frac{y_{i,l}}{4}\right)^2 + A_{5\ t,l} \frac{x_{i,l}}{4} \frac{y_{i,l}}{4}
\end{equation}
where $x_l,y_l$ are the MLA coordinates of a star at a given wavelength and defined such that at the center of the MLA, $\Delta c_{t,l}(0, 0) = 0$. We allow the star flats to differ between long and short exposures in the event that some of the star flat term is affected by the PSF.

The monochromatic magnitude of each star is given by:
\begin{equation}\label{eq:standardflux}
m_{j,l} =
\begin{cases} 
      -2.5 \log_{10} (f^{\mathrm{\,CALSPEC}}_{j,l})\, +\, \Delta m_{j,l} & \text{if CALSPEC}, \\
      -2.5 \log_{10} (f_{j,l}) & \text{if secondary}
   \end{cases}
\end{equation}

\noindent
where the wavelength-dependence of the flux for CALSPEC stars, $f^{\mathrm{\,CALSPEC}}_{j,l}$, is set relative to theoretical white dwarf models (with the gray scaling to a flux of $3.47\times10^{-9}$~erg~s$^{-1}$~cm$^{-2}$~\ang$^{-1}$ at 5556\,\ang assigned to the star Vega by \citealt{bohlin2020}). The two cases in Equation~\ref{eq:standardflux} may look similar (parameterizing the stars directly for non-CALSPEC vs. perturbations on CALSPEC for the CALSPEC stars), but for the CALSPEC stars, there is a prior around zero with an adjustable per-wavelength width given by:
\begin{equation} \label{eq:CALSPECAroundZero}
    \Delta m_{j,l} \sim \mathcal{N} (0,\, \sigma_{l}^2) \;.
\end{equation}

\noindent
This $\sigma_l$ is our estimate of the internal per-star tension inside CALSPEC. We do not require the average $\Delta m_{j,l}$ to be 0. In practice this means that there is a floor of approximately $\sigma_l/\sqrt{N_{\mathrm{CALSPEC}}}$ to how well the mean of the entire system is measured, corresponding to the measurement uncertainty from having a finite number of CALSPEC stars to calibrate to (discussed further in \S\ref{sec:rigid}).

The likelihood density from each observation is represented in the Bayesian hierarchical model as a mixture of two Gaussians

\begin{equation}\label{eq:inout}
\begin{aligned}
    -2.5\, \log_{10} (S^{\mathrm{\, obs}}_{i,l})\ \sim &\ (1 - f_{\mathrm{outl}\ t,l})\ \mathcal{N}\left(-2.5\, \log_{10}\, (S^{\mathrm{\, mod}}_{i,l}),\ \left[ \frac{2.5\, \sigma_{S\, i,l}}{\ln(10) \, S_{i,l}}\right]^2\, +\, \sigma_{\mathrm{in}\ t,l}^2 \right) + \\
     &\ f_{\mathrm{outl}\ t,l}\ \mathcal{N}\left(-2.5\, \log_{10}\, (S^{\mathrm{\, mod}}_{i,l})\, +\, \Delta m_{\mathrm{outl}\ t,l},\ \left[ \frac{2.5\, \sigma_{S\, i,l}}{\ln(10)\, S_{i,l}}\right]^2\, +\, \sigma_{\mathrm{outl}\ t,l}^2 \right) \;,
\end{aligned}
\end{equation}
where $\sigma_{\mathrm{outl}\ t,l}$ is much larger than $\sigma_{\mathrm{in}\ t,l}$. The distributions of atmospheric-extinction coefficients and temperature coefficients are also inferred
\begin{equation} \label{eq:hierarchy}
\begin{aligned}
    k_{n,l} & \sim \mathcal{N}(k_{0\ l},\, \sigma(k)^2_{l}) \\
    b_{n,l} & \sim \mathcal{N}(b_{0\ l},\, \sigma(b)^2_{l}) \;,
\end{aligned}
\end{equation}
enabling nights with small airmass or temperature ranges to be useful.

Table~\ref{tab:fitpars} provides a summary of our parameters and their priors. With tens of thousands of parameters, \numdatapoints datapoints, and non-Gaussian uncertainties, the inference poses a computational challenge. We sample from the posterior using Stan \citep{Carpenter2017} as called through the Pystan package (\doi{10.5281/zenodo.598257}). We used four chains with 3,000 iterations (1,500 warmup and 1,500 saved samples) per chain, which was almost always enough for good convergence of all standard-star $m_{j,l}$ and $\Delta m_{j,l}$ values (\citealt{GelmanRubin92}; $\hat{R} \leq 1.05$, and generally much closer to 1). For the rare runs where convergence was not achieved, we reran.

\newcommand{\nnight}{\ensuremath{N_{\mathrm{night}}}\xspace}
\newcommand{\nlamb}{\ensuremath{N_{\lambda}}\xspace}
\newcommand{\nstar}{\ensuremath{N_{\mathrm{star}}}\xspace}

\begin{deluxetable*}{rcll}[htbp]
\tablehead{
\colhead{Parameter} & & \colhead{Fixed Prior Distributions} & \colhead{Description}}
\startdata
    $k_{0\ l}$       \simNalign{0.3}{0.3^2} & Mean Atmospheric Extinction Coefficient \\
    $\sigma(k)_l$    \simNalign{0.03}{0.03^2} $\mathcal{U}(0,\ 0.2)$ & Night-to-night Dispersion in Extinction Coefficients \\
    $b_{0\ l}$       \simNalign{0}{0.1^2} $\mathcal{U}(-0.2,\ 0.2)$ & Mean Temperature Coefficient \\
    $\sigma(b)_l$\simNalign{0}{0.1^2} $\mathcal{U}(0.001,\ 0.2)$ & Night-to-night Dispersion in Temperature Coefficients \\
    $\Delta m_{j,l}$ \simNalign{0}{1^2} & Perturbations on CALSPEC Stars \\
    $\sigma_l$     \simNalign{0}{0.01^2} & Star-to-Star Dispersion of CALSPEC Perturbations \\
    $c_{n,l}$        \simNalign{0}{10^2} & Nightly Calibration \\
    $m_{j,l}$        \simNalign{0}{10^2} & $-2.5\ \log_{10}$ Star Flux \\
    $\Delta m_{\mathrm{outl}\ t,l}$ \simNalign{0}{0.1^2} & Mean Magnitude Offset of Outliers \\
    $A_{t,l\ 1--5}$   \simNalign{0}{0.1^2} & Coefficients Describing Star Flats \\
    $f_{\mathrm{outl}\ t,l}$ & $\sim$ & $\mathcal{U}(0,\ 0.2)$ & Outlier Fraction \\
    $\sigma_{\mathrm{in}\ t,l}$ & $\sim$ & $\mathcal{U}(0.001,\ 0.04)$ & Repeatability Floor \\
    $\sigma_{\mathrm{outl}\ t,l}$ & $\sim$ & $\mathcal{U}(0.04\, +\, |\Delta m_{\mathrm{outl}\ l}|/2,\ 0.8)$ & Outlier Dispersion \\
\enddata
\caption{Fixed priors (non-hierarchical) in this analysis. $i$ indexes observed spectra, $j$ indexes stars, $n$ indexes nights, $t$ indexes exposure-time category (i.e., long or short), and $l$ indexes wavelength. In general, we use weakly informative priors for these variables to roughly constrain the model to physical regions of parameter space, while allowing the data to drive the inferred parameter values.\label{tab:fitpars}} 
\end{deluxetable*}

A few minor approximations are made in our analysis: we approximate airmass as $\mathrm{X} \sim \mathrm{sec}(z)$ rather than employing the exact airmass calculation for an atmospheric shell starting above the elevation of Maunakea \citep[e.g][]{kasten1989}. For our baseline airmass range of $1 < \mathrm{X} < 2$, the peak-to-peak error when using this approximation is $\Delta \mathrm{X} \sim 0.0016$. Since our maximum extinction coefficient is $k = 0.58$, this would amount to an error on $k$ of only 0.9\,mmag/airmass. Since calibration errors propagate as differences in airmass coverage between the standard stars and supernovae, the error on the brightness of supernovae will be even less. Furthermore, we do not take Doppler effects (redshift, beaming, time dilation) into account for standard stars. Doppler effects due to the Earth's motion around the Sun can amount to more than 1\,mmag peak-to-peak even for broadband photometry \citep{rybicki79}. Doppler effects due to the motion between standard stars and the Sun are essentially static. 
For simplicity, we also assume extinction is linear with airmass for our primary analysis.
Telluric extinction nominally scales with airmass as $X^{0.6}$. But for our airmass range of $1 < \mathrm{X} < 2$, this agrees with our linear approximation to within 1.5\%. Outside the core of the $O_2$ A-band, the Maunakea telluric extinction is $k < 0.15$\,mag/airmass \citep{buton13}, so this approximation is better than $\sim1$\,mmag for our stars. We validate this approximation below.

\subsection{Model and Data Internal Consistency Checks}
\label{subsec:consistency}

To avoid any bias due to a subconscious desire to have our results conform to previous analyses (e.g., match CALSPEC with small scatter), the final calibration was kept blinded while we tested different cuts on the data and different forms for the Bayesian hierarchical model. The general approach was to determine what, if any, data selection cuts were needed, and then to try different versions of the model, alternating between these two as questions arose. For the wavelength-by-wavelength model we usually ran these tests on only a subset of wavelengths (every 20th wavelength element). This was primarily done to speed up the testing phases, but also reduced the risk of overfitting the model since so many other wavelengths remained available for validation.

For the data selection process, we examined the median residuals (as a function of wavelength and exposure time category) versus the following parameters: altitude, azimuth, hour angle, Julian day, day of year, $\chi^2$, total star flux, total sky flux, exposure time, PSF parameters (such as the seeing, $x,y$ location on the MLA, ellipticity; see Appendix~\ref{app:psf} for all of these parameters), humidity, windspeed, wind direction, temperature inside SNIFS, inside the dome, and outside, CCD flexure, FWHM values for the spaxel spectra in the cross-dispersion and wavelength directions on the CCD, CCD ADC saturation indicators, CCD temperature, and even indices indicating the observer. Of these parameters we found a trend with $\chi^2$ --- due to odd PSF shapes --- but these affected few stars and we worried that inability to detect this effect in lower SNR data might lead to a bias if a cut on $\chi^2$ were applied. Instead we performed a run in which the bright standard stars with short exposures were removed. As this did not have a significant effect on the remaining stars,\footnote{The synthetic photometry of the long-exposure standards changed with an RMS scatter star-to-star of 1--2\,mmags (depending on wavelength range) when comparing the results from our primary analysis and the short-exposure-removed analysis.} we did not implement a cut on $\chi^2$ or exposure time for our final analysis. Unsurprisingly we found that the few poorly-centered stars had larger residuals, so we tested whether rejecting stars located more than 4~spaxels from the center of the MLA affected the solution --- it did not. We found small trends with SNIFS temperature; since the temperature change inside SNIFS (which is insulated) within a night is less than a few degrees, and we expected the temperature correlation to average out for any given star, initially we did not test a model having a correction for this effect. After unblinding, we decided to make a temperature correction which became our primary analysis (discussed further in \S\ref{sec:dispersion}). There is also an indication that the repeatability for short exposures improves for windspeeds greater than 15~m/s, which corresponds to the passage of greater than half of an atmospheric turbulence cell during a 1~s exposure given the $\sim30$~m outer scale typical of Maunakea atmospheric turbulence \citep[e.g.,][]{ono2017}. Since this occurs only for a small fraction of the short-exposure standard star observations, we did not include a dependence of the repeatability on wind speed.

For the model construction process we tested several variations. One test replaced the inlier/outlier Gaussian model of Equation~\ref{eq:inout} with a Laplace probability distribution as an alternative way to allow for a heavy-tailed pull distribution. We experimented with models with and without application of star flats across the MLA (i.e., in addition to the flat-fielding performed by SNIFS internal lamps). We also separated the data into 3-year blocks, thereby by allowing each to have its own hyperparameters. This did not affect the calibration significantly, giving us confidence that little changed in the behavior of SNIFS that affects the calibration fits over the period of observations. A test was run in which a minimum airmass range of $\Delta \mathrm{X} > 0.7$ and at least \NminNights standard star observations were required, but this also did not produce much of a change on the calibration because plenty of nights remained (see Figure~\ref{fig:obs_airmasses}). We did find different calibration parameters when implementing a prior on the airmass coefficient that imposed the atmospheric extinction model of \citet{buton13}. This is due to the achromatic offset discussed in \S\ref{sec:grayoffset}. Therefore, in our final model we did not impose this constraint.

\begin{figure*}[htbp]
\centering
\includegraphics[width=\textwidth]{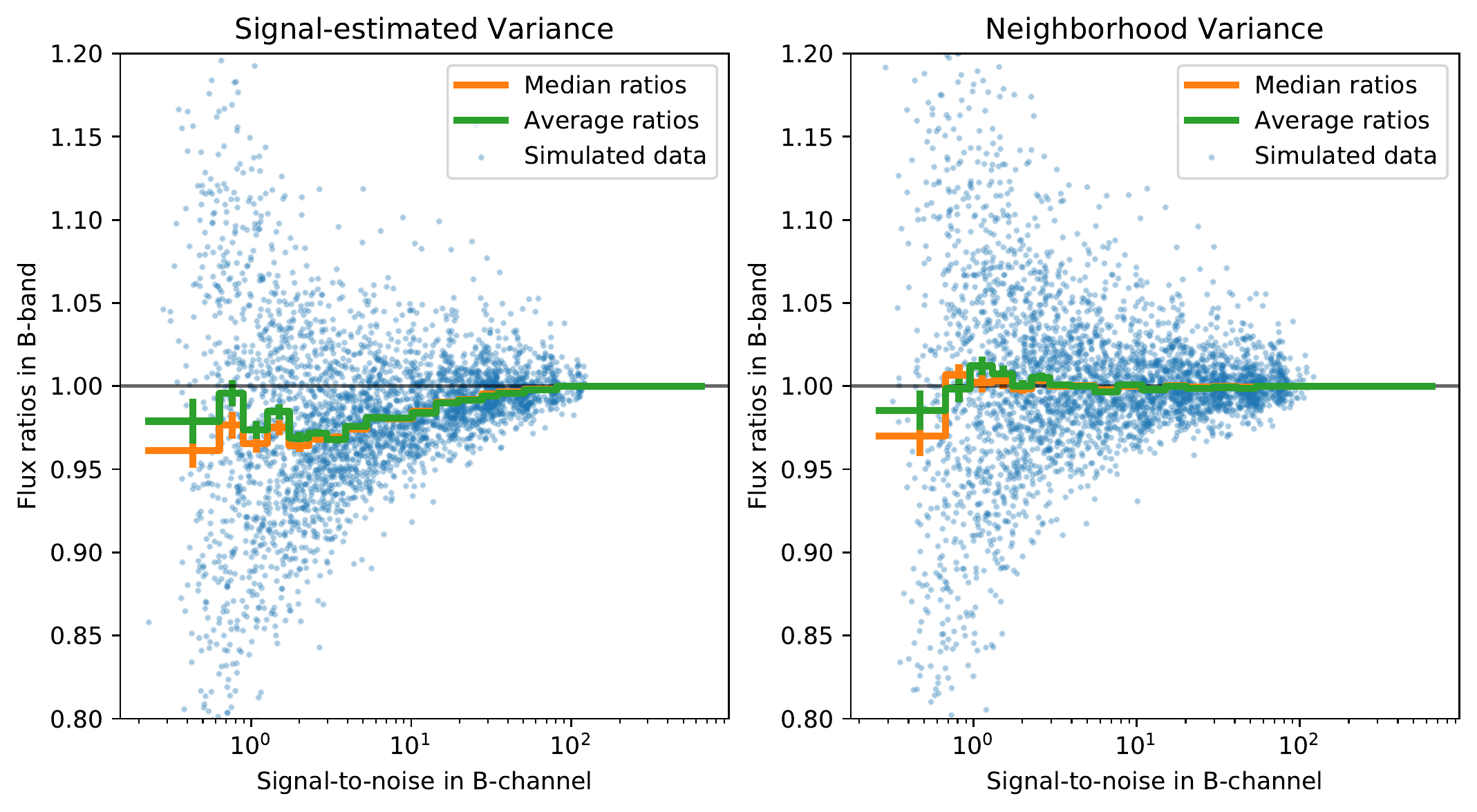}
\caption{Results of simulations to test the linearity of our {\tt{extract\_star2}} software that measures 1-dimensional spectra from SNIFS datacubes. On the left is shown a simulation in which a PSF vs wavelength is fit to the datacubes using as weights the photon and readout noise variance spectrum estimated directly from the signal. On the right is shown the cases where the initial signal-estimated variance spectrum is smoothed in wavelength, leaving out the target wavelength and the two adjacent from the kernel, thereby decorrelating the signal and the weights. Using the variance spectrum directly results in a strong bias with SNR, whereas using the smoothed variance does not. The simulations used to produce these results take high SNR standard star datacubes and add noise that simulates fainter and fainter stars. Overall, this test spans a factor of 5000$\times$ in brightness, corresponding to a range of 9.25~mag. We performed $\sim16200$ of these simulations across all spectra for 10 different standard stars in order to sample over a wide range of PSF shapes.}
\label{fig:linearity}
\end{figure*}

In parallel with testing of the Bayesian hierarchical model and selection on data parameters, we also performed internal consistency checks in other ways. For instance we tested the linearity of
the flux determinations from our weighted PSF fits by adding a range of noise to high SNR observations of a number of different stars. The right panel of Figure~\ref{fig:linearity} shows that our method is unbiased with SNR, whereas the left panel demonstrates a strong bias if the variance estimated from the data directly is used \citep{cash1979, horne1986}.

After running these tests, for our fiducial analysis we fixed the model to that described above, and decided to not make any cuts on the data parameters discussed above. After unblinding, we realized that some engineering observations and a handful of saturated exposures had been allowed into the set of observations but had not been cut. A new run excluding these observations produced the same results (within uncertainties), illustrating the robustness of our Bayesian hierarchical model fitting method.

As an analysis variant, we used telluric correction from the Line-By-Line Radiative Transfer Model \citep{clough92, clough05} retrieved through Telfit \citep{gullikson14}. We generate atmospheric models separately varying water and non-water telluric absorption, then convolve these models down to SNIFS resolution (we use a Gaussian with $\sigma = 3.7$\,\AA). For each SNIFS wavelength $l$, we build a simple model interpolation based on power-law scaling:

\begin{equation}
    k_l = \left[ a_l b^{p_l} \right]_{\mathrm{H_2 O}} + \left[ a_l b^{p_l} \right]_{\mathrm{Not\ H_2 O}} \;,
\end{equation}
where $b$ is the amount of atmospheric constituents along the line of sight, and $a_l$ and $p_l$ are separate fit parameters (as a function of wavelength) for water and non-water components. This interpolation accurately spans weak features where $p_l \sim 1$ and strong features where $p_l \sim 0.6$. Once we trained the interpolation, we computed $b$ parameters for each night, corrected the spectra, and then ran our calibration on those corrected spectra. We obtain virtually identical calibrated spectra, with largest differences over all stars and wavelengths $\lesssim 10$~mmag in the core of the A-band and $\lesssim 1$~mmag otherwise.

\section{Results} \label{sec:internalcompare}

After completing the development and testing of the Bayesian hierarchical model and data selection, we unblinded the calibration, standard star spectra, and hyperparameter values, which we now discuss. At this point, we left the comparison against external data blinded; see \S\ref{sec:compare}.

\begin{figure}[htbp]
\centering
\includegraphics[width=0.89\textwidth]{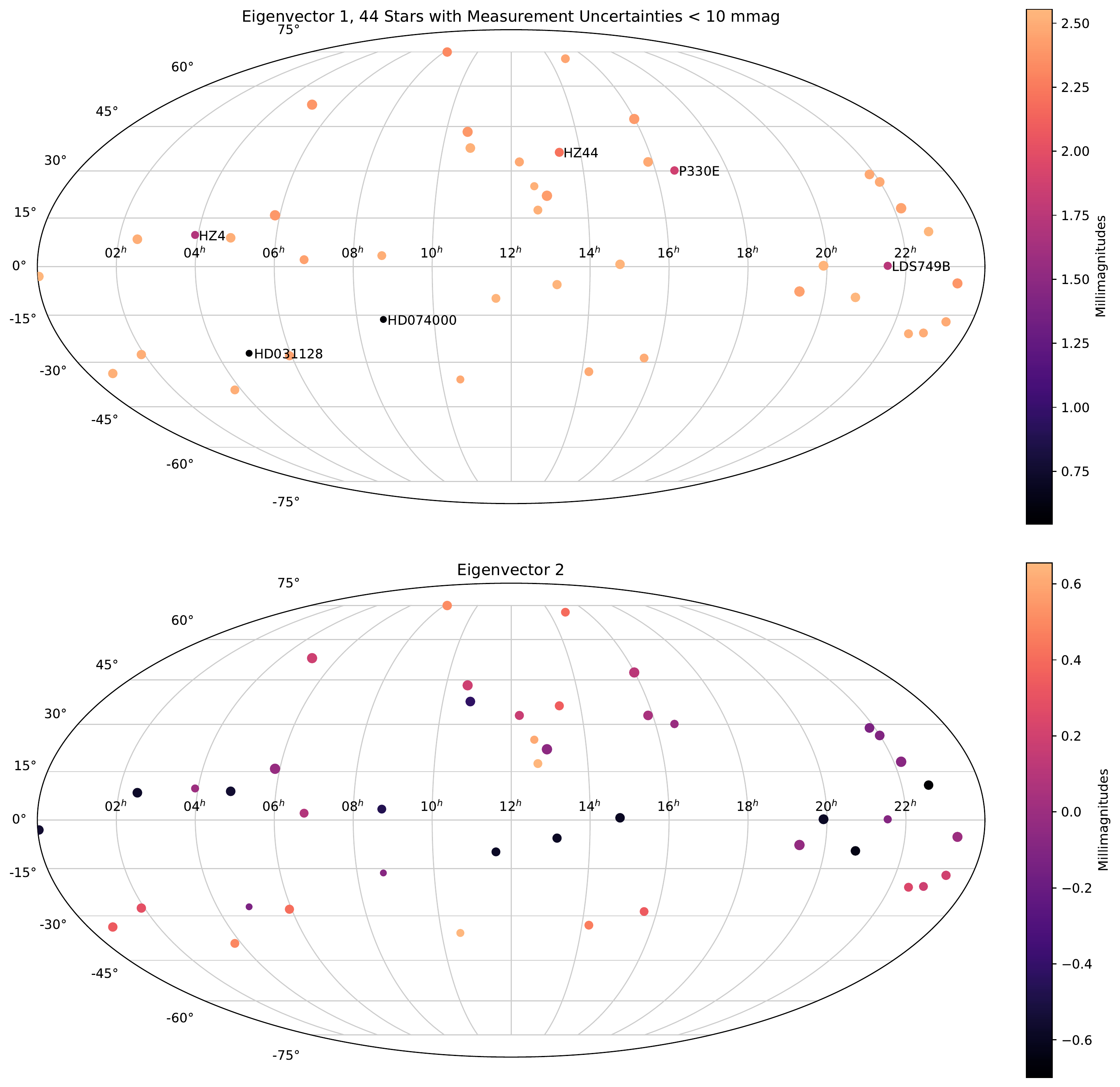}
\caption{We show eigenvectors constructed from the star-to-star covariance of the mean flux of each star averaged across 4000\AA--7000\AA\ (to improve signal to noise compared to a single wavelength).  The off-diagonal elements of the covariance matrix are decomposed into first one eigenvector (shown in the top panel) and then after removing the outer product of that vector with itself, the remaining off-diagonal covariance matrix is decomposed into the next vector (shown in the bottom panel). The linear size of the plot points indicates the inverse uncertainty of each star. The first eigenvector is nearly constant star-to-star, and the $\sim 2.5$~mmag scale indicates the high precision of the match of our network to CALSPEC. Our model naturally gives this uncertainty from the number of CALSPEC stars, the number of observations, the estimated internal consistency of our network vs. CALSPEC, the repeatability of SNIFS, and the signal-to-noise of the observations. The second vector shows very small ($\sim 1$~mmag) correlations on the sky.}
\label{fig:skyeigen}
\end{figure}

\subsection{Network Rigidity}\label{sec:rigid}

As expected from Figure~\ref{fig:stds_on_sky}, our network is rigid, as defined by the small covariance between stars. Figure~\ref{fig:skyeigen} shows the first two eigenvectors of the modeled covariance between stars in our network \citep[e.g.,][]{Padmanabhan2008}. We compute this covariance directly from the MCMC samples of the modeled $m_{j,l}$. The first eigenvector is nearly constant ($\sim 2.5$~mmag) star-to-star and represents the uncertainty of the tie of our network to CALSPEC. The second eigenvector shows very small ($\sim 1$~mmag) spatial structure.

\subsection{Airmass Dependence}\label{sec:grayoffset}

One of the diagnostics discussed above was the examination of the airmass-dependence, $k_{n,l}$. A persistent feature of our measured airmass coefficients, exhibited by those of our runs that do not enforce a physical atmosphere, is an offset of roughly 20~mmag/airmass below what a physical model \citep[as in ][]{buton13} would predict. This feature prompted us to try a number of analysis variants while the calibration results were blinded, but we found this feature to be very robust.

\begin{figure*}[htbp]
\centering
\includegraphics[width=0.8\textwidth]{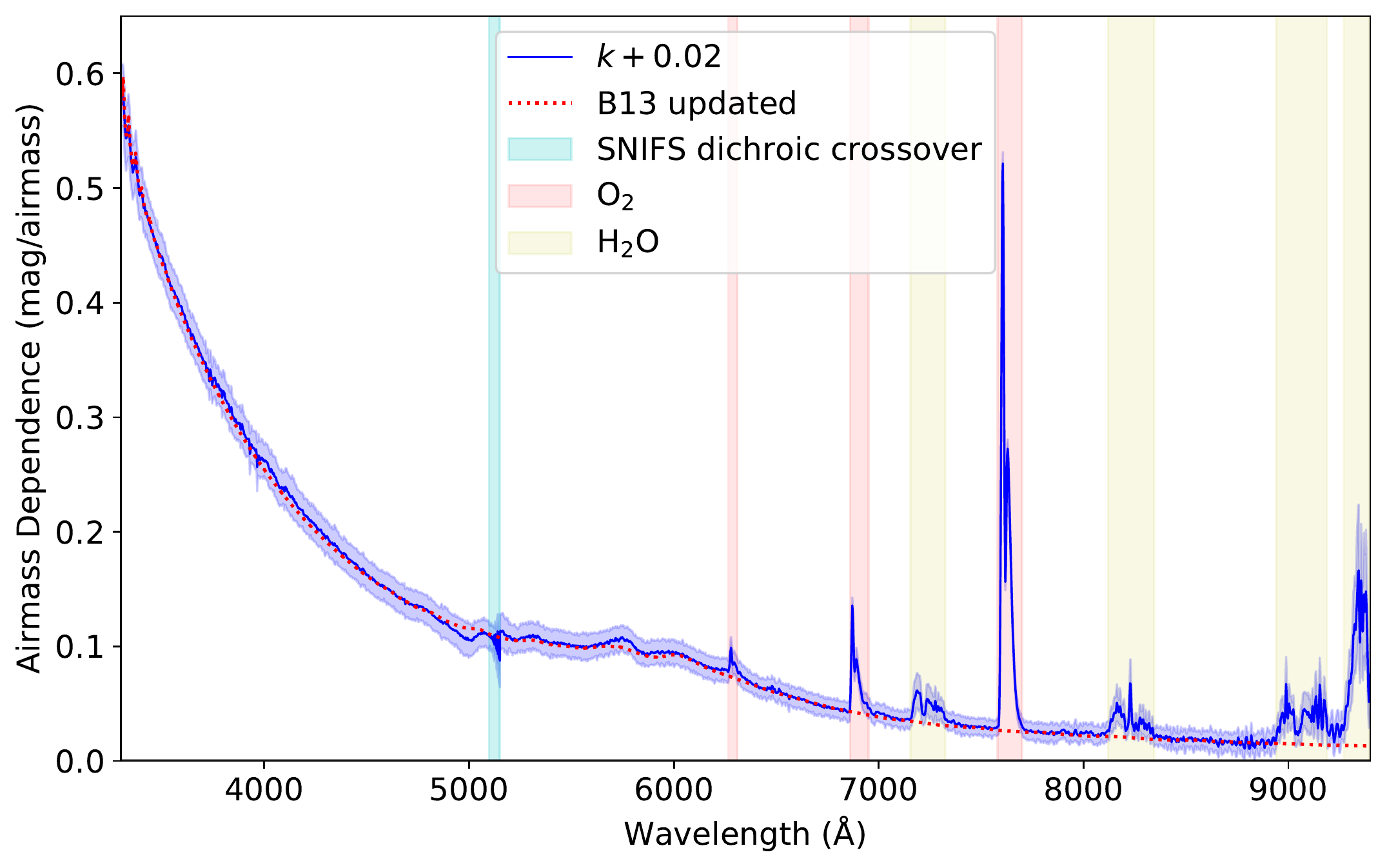}
\caption{The airmass-dependent term, $k$, plotted versus wavelength for our primary (extinction linear in airmass) analysis. For comparison purposes we have added the 20~mmag gray instrumental component discussed in \S\ref{sec:grayoffset}. The blue line is the mean $k$ and the blue band is the inferred RMS of the night-to-night variation in $k$. Major atmospheric features are marked, as is the wavelength range of the SNIFS dichroic cross-over. A physical atmospheric model, updated from \citet{buton13} to include updated ozone cross-sections from \citet{serdyuchenko2014} and a 40\% reduction in the amount of aerosol scattering, is shown as the red dashed line. (Note that \citet{buton13} modeled telluric lines in a separate step, so for their curve the telluric feaures are not included.) We see overall good agreement between our wavelength-by-wavelength model and a physical model. Note that no smoothness constraint in wavelength is imposed, so this agreement is an excellent cross-check.}
\label{fig:extinction}
\end{figure*}

One of the ways we investigated this effect was using the window spanning 8500--8800~\AA\, which is predicted to have very little extinction for the elevation of Maunakea. The physical atmospheric components \citep[e.g.][]{buton13} contribute only $\sim14$~mmag/airmass of extinction: 9~mmag/airmass due to Rayleigh scattering, zero due to ozone scattering, and with typical aerosol scattering of only $\sim5$~mmag/airmass (roughly half dust and half anthropogenic sulfates). In this window we find $k=-3\pm1$~mmag/airmass, and this quantity is found to be very robust in our various tests. We note that \citet{mccord1979} also found a low extinction of $0.005\pm0.005$~mag/airmass\footnote{\citet{mccord1979} do not provide uncertainties; we have estimated uncertainties from the airmass scans shown in their Figure~2 and then averaged the extinction measured at these two wavelengths.} at 8500 and 8800~\AA, also using the UH88, and below the component-based atmospheric prediction by about $\sim1.8\,\sigma$. We find this same offset at all wavelengths when compared to a model using nominal values for the known physical components of the atmosphere \citep[cf.][]{buton13}.\footnote{A similar effect is even seen in another Maunakea dataset (CFHT MegaCam); \citet{betoule13} finds generally higher airmass extinction coefficients (by up to 20~mmag) for large-aperture photometry (their Table~3) relative to their nominal aperture photometry, which is performed with an image-quality-dependent radius. If the average PSF profile were not changing with airmass, then the (on average) larger aperture radii used on images taken at higher airmass with worse image quality would not capture more light.}

Examining the atmospheric extinction coefficients in the SNIFS imaging channel data (taken in parallel with the spectrophotometric observations) provided a more conclusive test. For simplicity, we used a \citet{moffat1969} PSF for the imaging photometry and found a correlation between the Moffat $\beta$ parameter and airmass. Fitting for a different $\beta$ for each observation of each star results in the expected extinction coefficients, while fixing $\beta$ results in unphysical extinction coefficients (this test confirms that the impact on the exctinction coefficients is achromatic to $\lesssim 0.01$~mag over the wavelength range of $V$, $r$, and $i$). We ran a further test of the PSF, separating the data into observations with seeing above and below 0\farcs9. However, both runs showed very similar atmospheric extinctions in the 8500--8800~\AA\ window.

However, we note that for flux calibration of new objects this offset has a small effect since the extinction solution is simply being used as a convenient functional form for interpolating the calibration with airmass and the primary standards and secondary standards have similar distributions in airmass (Figure~\ref{fig:obs_airmasses}).

\begin{figure}[htbp]
\centering
\includegraphics[width=0.7\columnwidth]{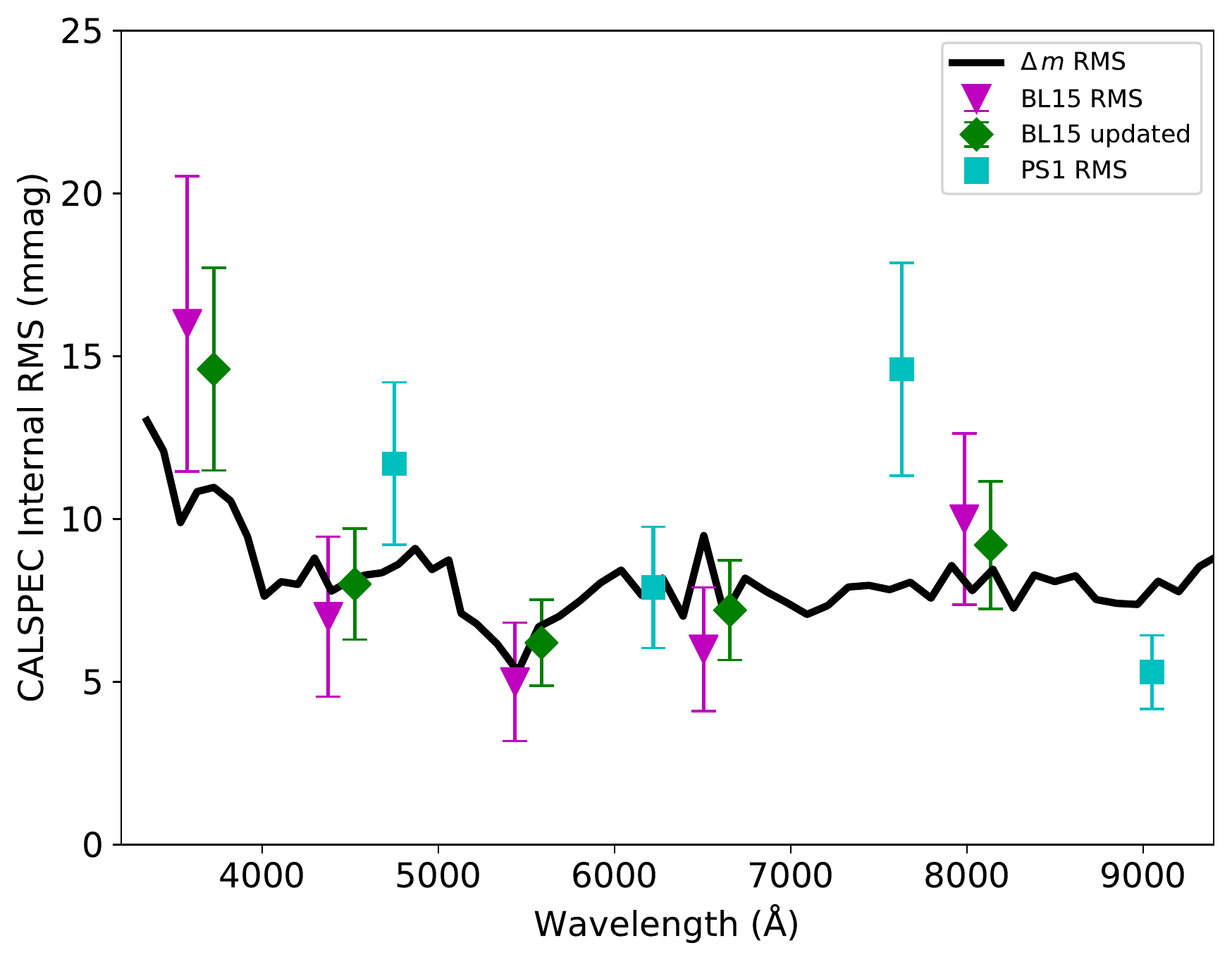}
\caption{The dispersion of the $\Delta m$ values ($\sigma_l$ from Equation~\ref{eq:CALSPECAroundZero}) for our \NPrimary CALSPEC standards (black curve). The calculation is performed in bins of 40 wavelength samples in order to lower the uncertainty on the individual $\Delta m$ values for stars with fewer (STIS and/or SNIFS) observations. Overplotted are the dispersions when comparing filter photometry for the $UBVRI$ filters \citep[][ BL15 magenta diamonds]{bohlin15}. We have updated the $UBVRI$ intrinsic dispersion, including subtraction of the quoted filter photometry measurement uncertainty, to include newer space-based CALSPEC standard stars (green squares). We also plot a similar comparison we have done for Pan-STARRS1 in the $griz$ filters. These points have been offset slightly in wavelength for clarity. This demonstrates that the dispersion of $\Delta\, m$ values determined from our Bayesian hierarchical standardization model are consistent with other external checks of CALSPEC.}
\label{fig:calspec_rms}
\end{figure}

\subsection{CALSPEC Dispersion}
\label{sec:dispersion}

After completing a full run we looked at the per-wavelength offsets, $\Delta m_{j,l}$ for the space-based CALSPEC stars. Figure~\ref{fig:calspec_rms} shows the modeled dispersion, $\sigma_l$, versus wavelength. The median $\sigma_l$ is \internalscatter, while the smallest dispersion is approximately 6~mmag around 5300~\AA. Overplotted are the dispersions measured from filter photometry, in $UBVRI$ by \citet{bohlin15}, and in $UBVRI$ and $griz$ (as in Table~2 of \citet{scolnic15}) by us\footnote{See \S\ref{subsec:phot_comp} for technical details.}. Our updated dispersions reflect the addition of new CALSPEC stars since the publication of the previous dispersions. Note that in this case it was possible to remove the contribution from the quoted filter photometry measurement uncertainties. Following \citet{bohlin15}, our primary comparison is for stars having photometry from \citet{landolt2007,landolt2009,bohlin15}. This excludes the filter photometry for EG~131, HD~31128, and HD~74000\footnote{Limited to $UBV$ photometry for the latter two stars in any case.}, which otherwise drives up the filter photometry dispersion substantially (see Figure~\ref{fig:filtercomp}), especially in the $U$ and $I$ bands. These measures are only a check on the internal consistency between $\Delta\, m_{j,l}$ and CALSPEC in our case, or between filter photometry and CALSPEC, but do suggest that there is real dispersion within the space-based CALSPEC system at the level that we have measured.

\begin{figure}[htbp]
\centering
\includegraphics[scale=0.4]{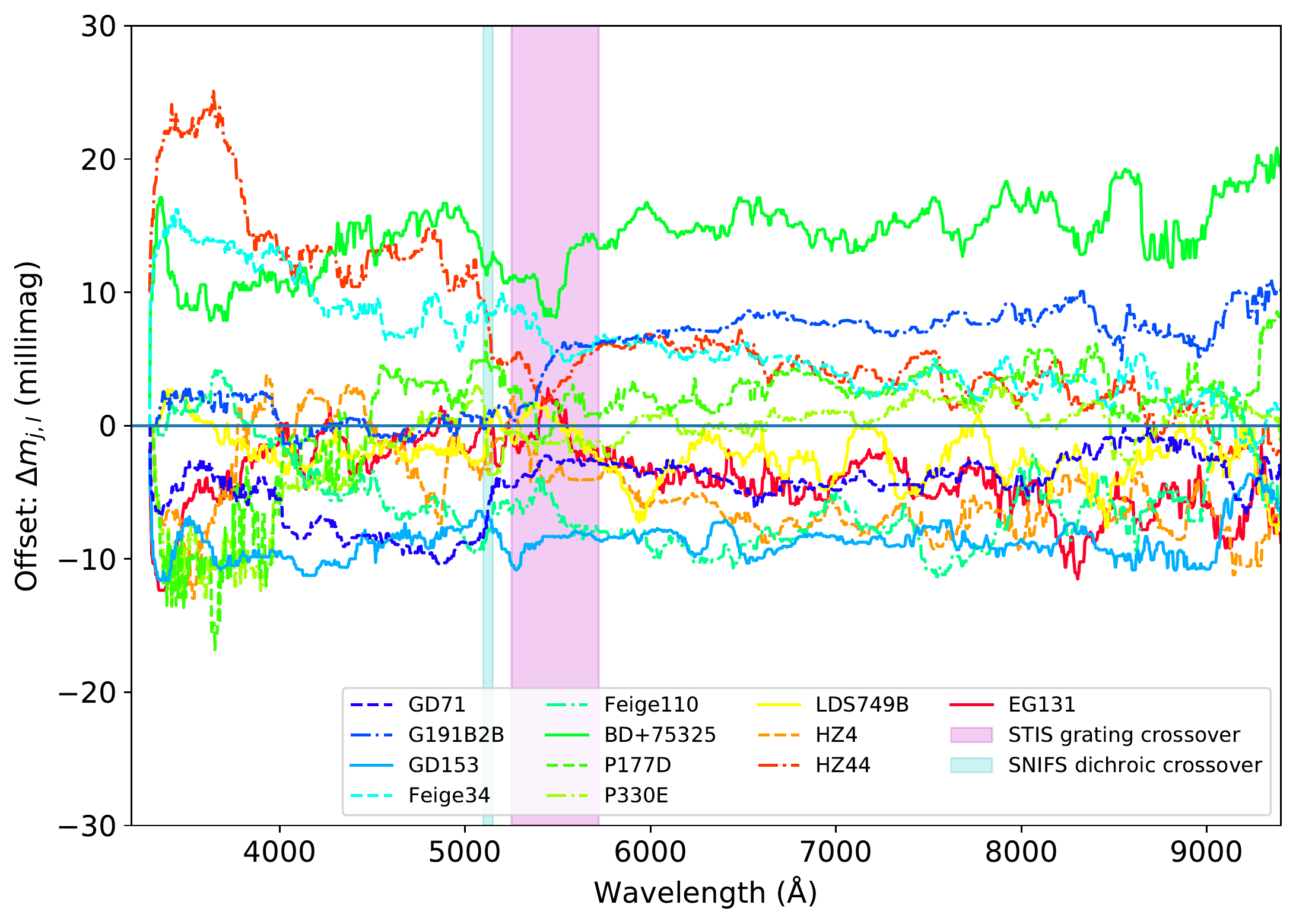}
\includegraphics[scale=0.4]{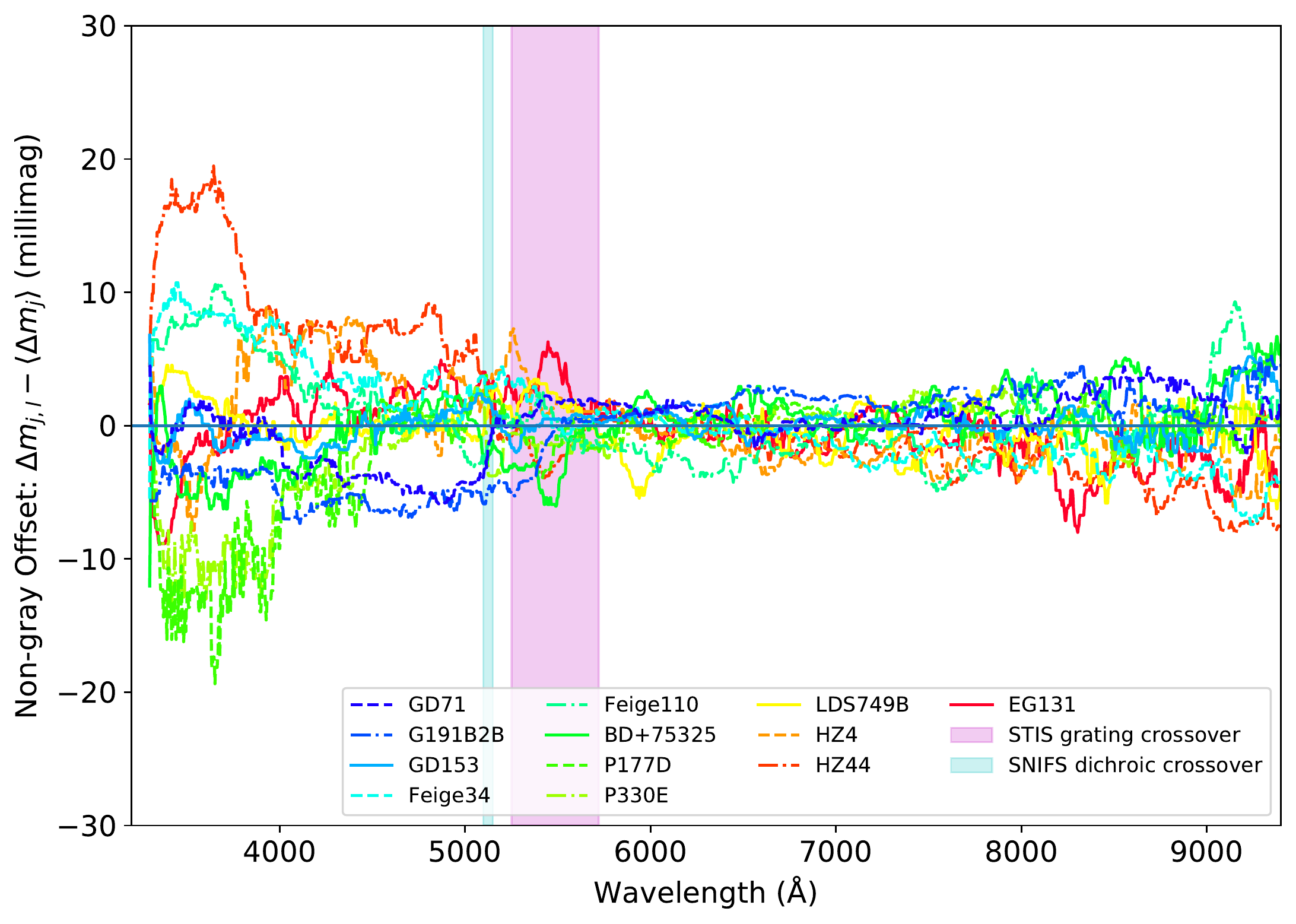}
\caption{The $\Delta m_l$ values for each star showing the offset with respect to CALSPEC. On the left we show the full $\Delta m_l$ for each star, making plain that most of the power is in the form of constant-in-wavelength offsets. On the right we have removed the constant term in order to highlight the chromatic components. The vertical blue-shaded region shows the SNIFS dichroic crossover wavelength range; there does not appear to be much structure between the blue and red sides of SNIFS. The vertical magenta-shaded band is the region where the CALSPEC \STIS observations change grating coverage. The values have been median smoothed in wavelength for clarity. Only primary calibrators observed on more than one night are shown, since otherwise the per-observation repeatability dominates, forcing the $\Delta\, m_l$ to fall back on the prior, $\sigma_l$.}
\label{fig:deltaf_per_star}
\end{figure}

Since we calculate $\Delta m_l$ for each CALSPEC star, we can examine these as well. Figure~\ref{fig:deltaf_per_star} shows these versus wavelength, labeled by star name. While the $\Delta m_l$ values are calculated for each SNIFS wavelength bin, we have median-smoothed the values in wavelength to enhance the signal-to-noise while preserving any jumps in the curves. The wavelength-combined RMS of the differences is only \calspecdispersion. (This is slightly less than the median of the internal scatter of \internalscatter due to the influence of the prior, $\sigma_l$, and the difference in how the stars are weighted between to two types of measurement.) The largest absolute mean offset is \bdseventyfivemmag~mmag for \BDSeventyFive. As noted in \S\ref{sec:data}, there are suggestions in the literature that this star might be variable. Similar to \citet{bohlin15}, the ensemble average of the CALSPEC stars does not seem to be exactly centered on the three fundamental white dwarfs (which should set the \STIS calibration for the other CALSPEC stars). Without an explanation for the scatter we observe comparing \STIS and SNIFS, the reason for this is not clear. 

Most of the per-star $\Delta m_l$ values consist of offsets, therefore, next we remove the mean offsets in order to examine the chromatic component. This shows excellent chromatic agreement redward of $\sim5700$\,\ang. Blueward of this there are spectral tilts. The vertical blue-shaded region shows the SNIFS dichroic crossover wavelength range; there does not appear to be much structure between the blue and red sides of SNIFS. However, at slightly longer wavelengths than the SNIFS dichroic crossover is the crossover between the \STIS G430L and G750L gratings (vertical magenta-shaded band); a few stars appear to have jumps there. We examined the case with the most structure in $\Delta\, m_l$, HZ~44, to see whether our spectrum or the CALSPEC spectrum appeared more realistic, for instance, having a smoother continuum. Unfortunately this star has a dense forest of absorption lines in this wavelength region that precludes any strong statements in this regard.

\begin{figure}[htbp]
    \centering
    \includegraphics[width = 0.75\textwidth]{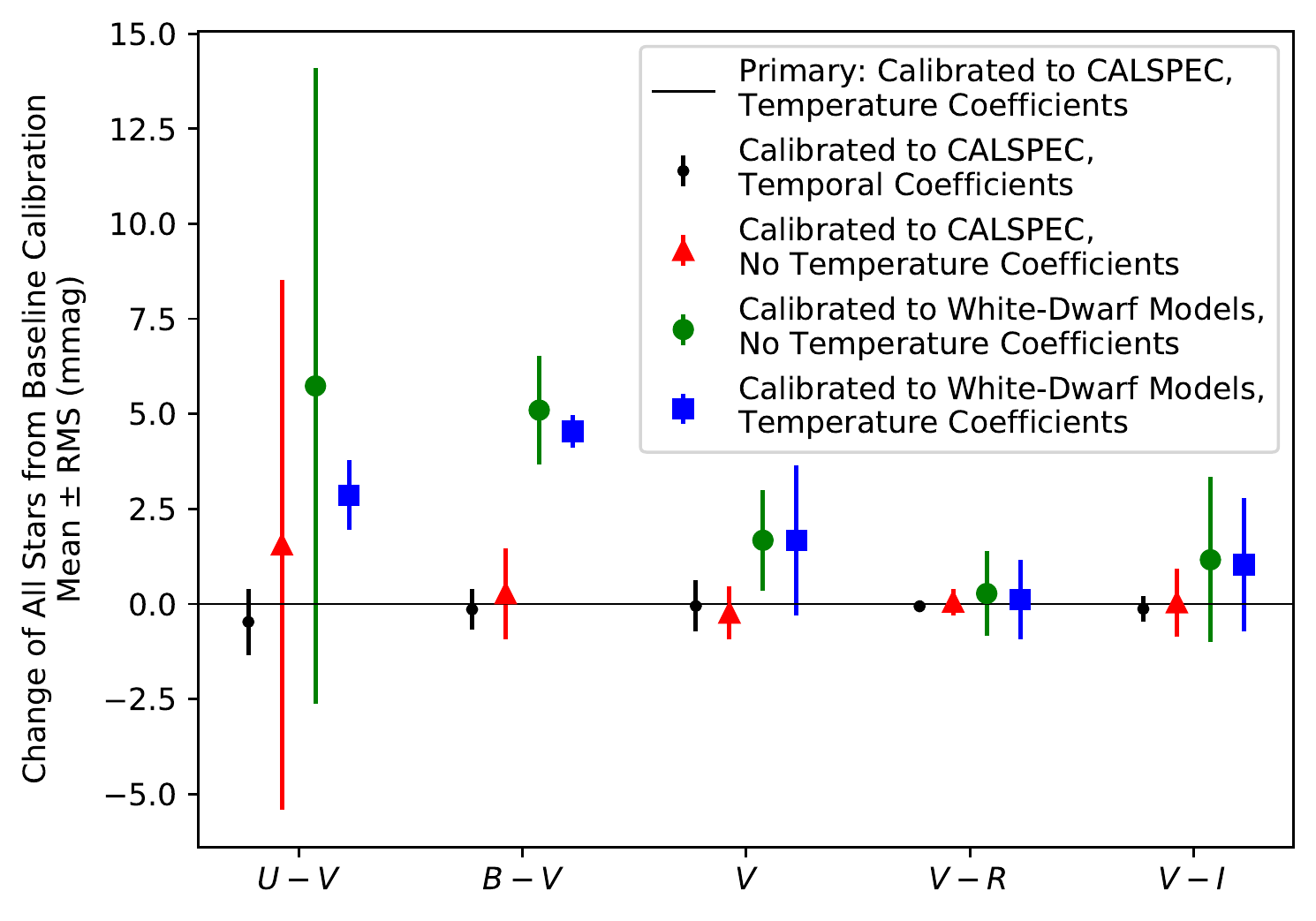}
    \caption{The mean and dispersion (not the uncertainty on the mean) of the per-star changes in synthetic photometry for different analysis variants compared to our primary calibration to \NPrimary CALSPEC stars, with the inclusion of nightly SNIFS temperature coefficients. The small black points indicate the changes if we regress on time of night instead of temperature (i.e., replace $\Delta T$ with $\Delta$date in Equation~\ref{eq:inout}). This has essentially no effect on the recovered standard star fluxes (changes are $\sim 1$~mmag), indicating that there is no evidence of changes in the SNIFS calibration during each night that are not driven by temperature changes. The red triangles show our original unblinded calibration: calibrated to CALSPEC but with no nightly SNIFS temperature coefficients. The mean star essentially does not change (indicating good mixing of primary and secondary stars through each night), but scatter of several mmags star-to-star is seen in the $U$ band. Next, the green points show the results from calibrating to the three fundamental CALSPEC white-dwarf models directly (with STIS observations not used at all) with no nightly SNIFS temperature coefficients.  We see a shift of $\sim 5$~mmags in the $U$ and $B$ bands, with the same several mmag scatter star-to-star. Finally, the blue squares show the three-white-dwarf calibration, but with temperature coefficients. The $U$ and $B$ offsets decrease somewhat, and the star-to-star scatter drops to 1--2~mmags. This indicates that some of the original tension we saw between the three-white-dwarf calibration and the CALSPEC calibration is due to when in the night the white-dwarfs tended to be observed. Importantly, it also indicates that our network is rigid, and that changing the primary calibration stars moves the entire network together (at least if nightly temporal or temperature coefficients are used). As for the remaining difference between the CALSPEC observations and models for the three WD fundamental calibrators, it is in line with the disagreement shown around the Balmer jump in Figure~9 of \cite{bohlin2020}.}
    \label{fig:synthcomparison}
\end{figure}

After unblinding the original version of Figure~\ref{fig:deltaf_per_star}, we were perplexed by small, but statistically significant $\sim 10$~mmag, wavelength-dependent offsets between the three fundamental CALSPEC white dwarfs that varied over the SNIFS B channel. We searched for possible variables that might not average down over many observations of the three white dwarfs. We noticed that GD~153 is generally observed in the first half of the night, while G~191B2B and GD~71 are generally observed in the second half (as the SNfactory did not usually run during the winter). Thus, the instrumental temperature of SNIFS was generally higher (by a median of 0.5$^{\circ}$C) when observing GD153 than for the other two stars. With the high precision of our dataset, we decided to take this into account, even though its importance was only realized after unblinding (as noted in \S\ref{subsec:consistency}, we had observed the temperature trend while blinded, but incorrectly believed it would average out). This effect may be due to small wavelength-dependent changes in focus with temperature due to CaF$_2$ in the optics preceding the SNIFS microlens array. The net changes are shown in Figure~\ref{fig:synthcomparison}, where the principal effect is to greatly improve the consistency of the $U$-$V$ color when calibrating only to the three fundamental white-dwarf models rather than the STIS observations of the CALSPEC network, which we now discuss.

\subsection{Calibrating to CALSPEC STIS observations versus calibrating to WD models directly}
\label{sec:wdmodels}
While our use of the calibrated CALSPEC spectra provides a large sample of primary standards, we also considered calibration directly using only the calculated stellar atmosphere models of the three fundamental CALSPEC white dwarfs. This would be the optimal choice if most of the scatter we observe against our CALSPEC primaries is caused by internal tension between the STIS observations of these stars. We expected several differences in doing so. First, as shown in Figure~9 of \cite{bohlin2020}, the models and the STIS observations differ by up to 1\% around the Balmer jump, so there may be increased uncertainty in this region. In addition, since the three WDs are concentrated in the northern portion of the northern winter sky, using only these three standards could somewhat weaken the robustness of our network. Finally, this variant decreases the number of primary calibrators from \NPrimary to 3, and hence increases the statistical uncertainty on the mean of the network (as Figure~\ref{fig:deltaf_per_star} shows, the three white dwarfs do not seem to show a dispersion $\sqrt{14/3}=2.2\times$ smaller than the other CALSPEC stars).

Figure~\ref{fig:synthcomparison} shows this variant, relative to our primary calibration, as the square blue symbols. As expected from the dispersion in the blue relative to the red seen in Figure~\ref{fig:deltaf_per_star}, there is a clear offset for the colors $U$-$V$ and $B$-$V$. Even so, the discrepancy in the mean remains below 5~mmag. For the remaining colors, the differences are less than 2~mmag. Our results confirm that our network is indeed rigid (which \S\ref{sec:rigid} also discusses) with a small ($\sim$~2~mmag) star-to-star RMS when also controlling for instrumental temperature variations.

\subsection{Other Global Parameters}\label{sec:repeat}

\begin{figure}[htbp]
\centering
\includegraphics[width=\columnwidth]{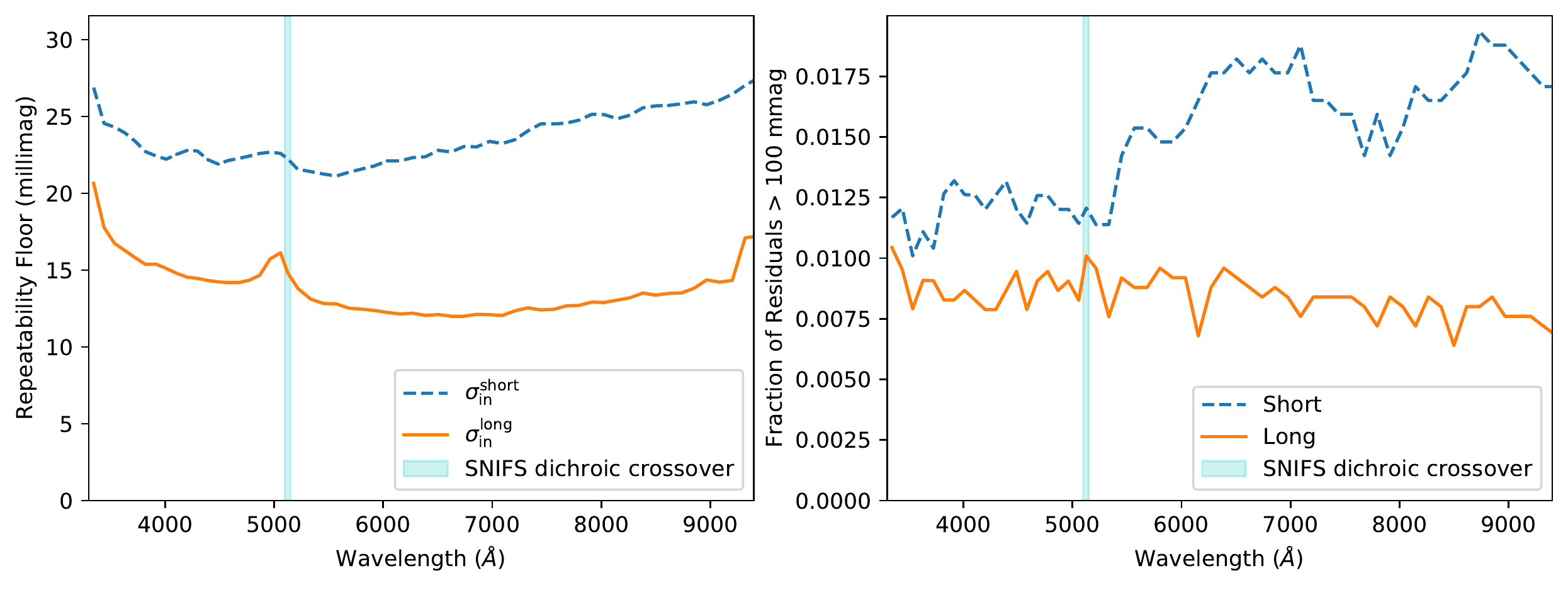}
\caption{The inferred repeatability floor (left) and observed outlier fractions (right) versus wavelength from our robust Bayesian hierarchical calibration model. To improve the signal-to-noise, we show these results binned in 40-wavelength bins. The repeatability is obviously much better for long exposures than short exposures, and the outliers occur more frequently for short exposures as well. See \S\ref{sec:repeat} for a detailed discussion.
}
\label{fig:repeat}
\end{figure}

Next we examine the values found for several other parameters of our model.  The repeatability of measurements of the same star could depend on a number of factors such as PSF knowledge, atmospheric transparency, instrument stability, flat-fielding errors, shutter timing, etc. Figure~\ref{fig:repeat} shows our repeatability, the standard deviation, $\sigma_{in}$, of the inlier population, as a function of wavelength for both long and short exposures. For each exposure category there is somewhat worse repeatability around the SNIFS dichroic crossover wavelength region. Intra-night atmospheric transparency variations must be subdominant since, as Figure~\ref{fig:extinction} shows, the atmosphere is nearly transparent near $\sim 8800$\,\ang yet the repeatability at this wavelength is not any lower than at other nearby wavelengths. PSF variations are the most likely cause of the repeatability limit, especially since short exposures have much larger values of $\sigma_{in}$ and their PSFs are seen to have much more structure. The larger atmospheric refraction at bluer wavelengths is well-known to lead to poorer seeing, and this includes the potential for more structured PSFs; this could explain the trend to higher $\sigma_{in}$ at bluer wavelengths. With SNIFS we rely on an analytic PSF (see Appendix~\ref{app:psf}) whereas imaging surveys have many stars per field allowing a potentially more detailed characterization of PSF structure. Even so, our repeatability is comparable to that found for PS1; \citet{schlafly2012} quote repeatabilities of 11, 10, 11, 12, and 16\,mmag in the PS1 $griz$ filters, while the updated analysis of \citet{magnier2019} finds repeatabilities of 14, 14, 15, 15, 18\,mmag. The Dark Energy Survey obtained somewhat better repeatabilities of 7.3, 6.1, 5.9, 7.3, 7.8\,mmag \citep{burke2018}. The assignment of repeatability to PSF modeling is re-enforced by the 2--3~mmag repeatability achieved using large-aperture photometry \citep[e.g.,][]{bernstein2018} and the sub-mmag achieved with defocused stars using the SNIFS imaging channel \citep{mann2011}, or the space-based repeatabilities of 2-4~mmag for STIS and 4.5~mmag for the WFC3 IR grism
\citep{bohlin2000,bohlin2019}. 

Figure~\ref{fig:repeat} also shows the outlier (residuals $> 100$~mmag) fraction versus wavelength for both long and short exposures; we see evidence that 1--2 percent of our observations are outliers (not well described by the uncertainties in the data and the Gaussian repeatability floor). The wavelength-by-wavelength solution does not know which SNIFS wavelength is being processed, yet it clearly finds a higher fraction of outliers on the SNIFS red channel for short-exposure standard stars. We believe this arises from the combination of better intrinsic seeing at longer wavelengths coupled with the highly-structured PSF that can occur for short exposures. 
For long exposures, the outlier fraction is fairly independent of channel, and is comparable to the level of outliers found, e.g., for DES by \citet{burke2018}. Recall though, that our model does not make cuts or assign a given spectrum entirely to the inlier or outlier population. Rather, as Equation~\ref{eq:inout} shows, each spectrum from each channel has a finite probability to be in either population. Overall, the values of these hyperparameters are consistent with, or better than, our expectations from our frequentist analysis in \citet{buton13}.

\begin{figure}[htbp]
\centering
\includegraphics[width=0.7 \columnwidth]{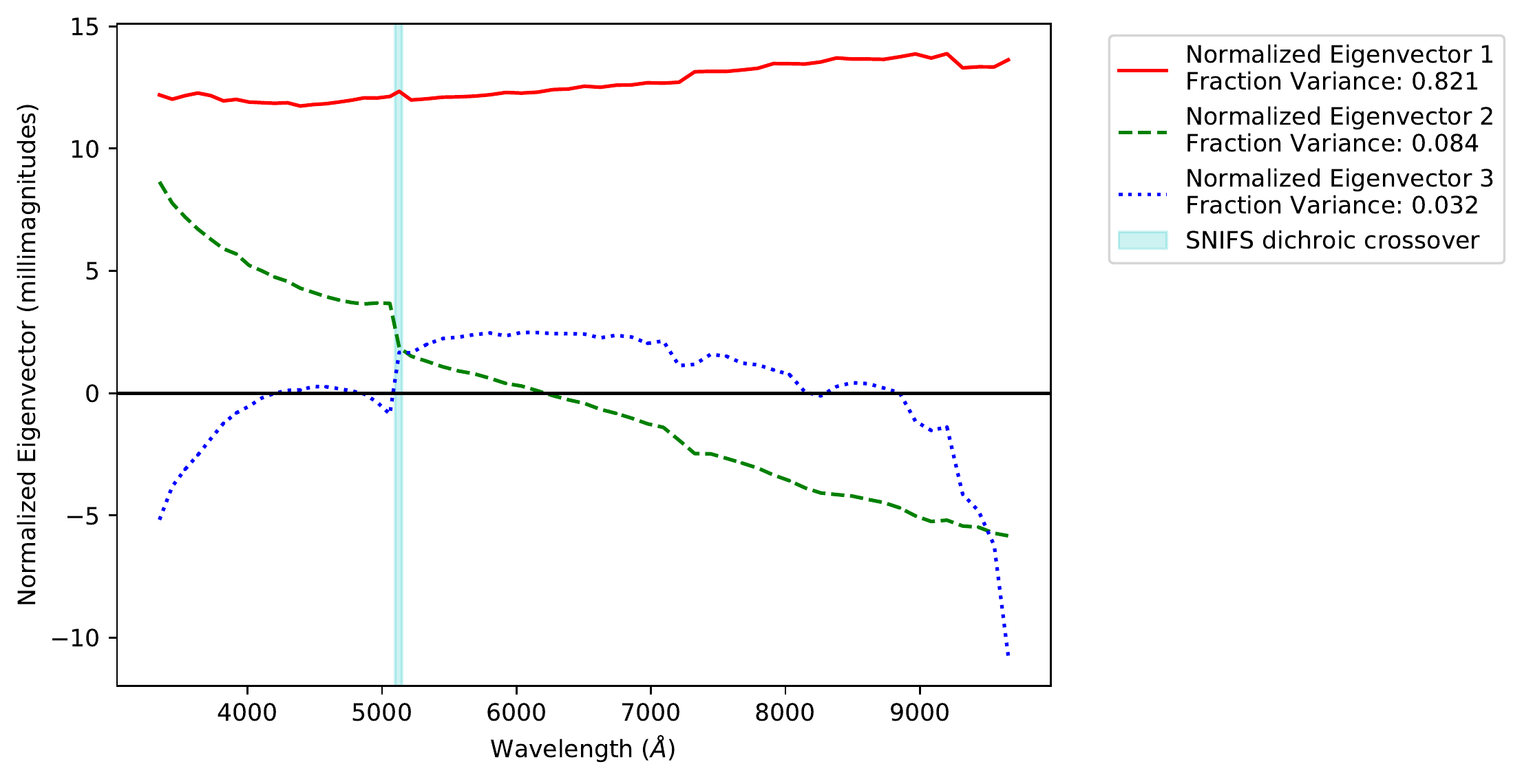}
\caption{{This figure shows a robust decomposition of the per-spectrum residuals (as a function of wavelength) into eigenvectors to investigate the repeatability floor seen in Figure~\ref{fig:repeat}.} We show the first three eigenvectors here; these explain more than 90\% of the variance. Each eigenvector is multiplied by the square root of the eigenvalue, so that the amount of dispersion is shown (in mmag). The first eigenvector is mostly gray and explains about 5/6ths of the variance. The second eigenvector introduces a small tilt with wavelength with a peak-to-peak size of $\sim$ 14\,mmags. The third eigenvector shows curvature and a small amount of H$_2$O variation, with a peak-to-peak size of $\sim$ 12\,mmags.}
\label{fig:eigenvectors}
\end{figure}

Figure~\ref{fig:eigenvectors} shows a robust principal component decomposition of our per-spectrum residuals using \texttt{SkiKit-learn MinCovDet} \citep{rousseeuw84, rousseeuw99, scikit-learn}. To reduce the wavelength-to-wavelength noise in the eigenvectors, we work in bins of 40 wavelengths. Most of the residual variation is approximately achromatic, but some tilt and curvature also is present. Specificially, the dominant, largely achromatic, eigenvector of the residuals describes 82\% of the variance; it is likely due to PSF differences with respect to our PSF model. The next eigenvector of the residuals (8\% of the variance) is nearly monotonically chromatic. The third component (3\% of the variance) varies in shape most strongly for wavelengths near the ends of the full spectral range. Both the second and third eigenvectors as show weak features around the telluric H$_2$O features. So it seems possible that the combination of the second and third eigenvectors of the residuals might be due to fluctuations in extinction. It is notable that the region around the dichroic is very weak, reinforcing the discussion in \S\ref{sec:dispersion} that the chromatic jumps seen for some CALSPEC stars are not due to SNIFS.

\subsection{Leave-One-Out Tests}\label{sec:LOO}

\begin{figure}[htbp]
\centering
\includegraphics[width=0.9\textwidth]{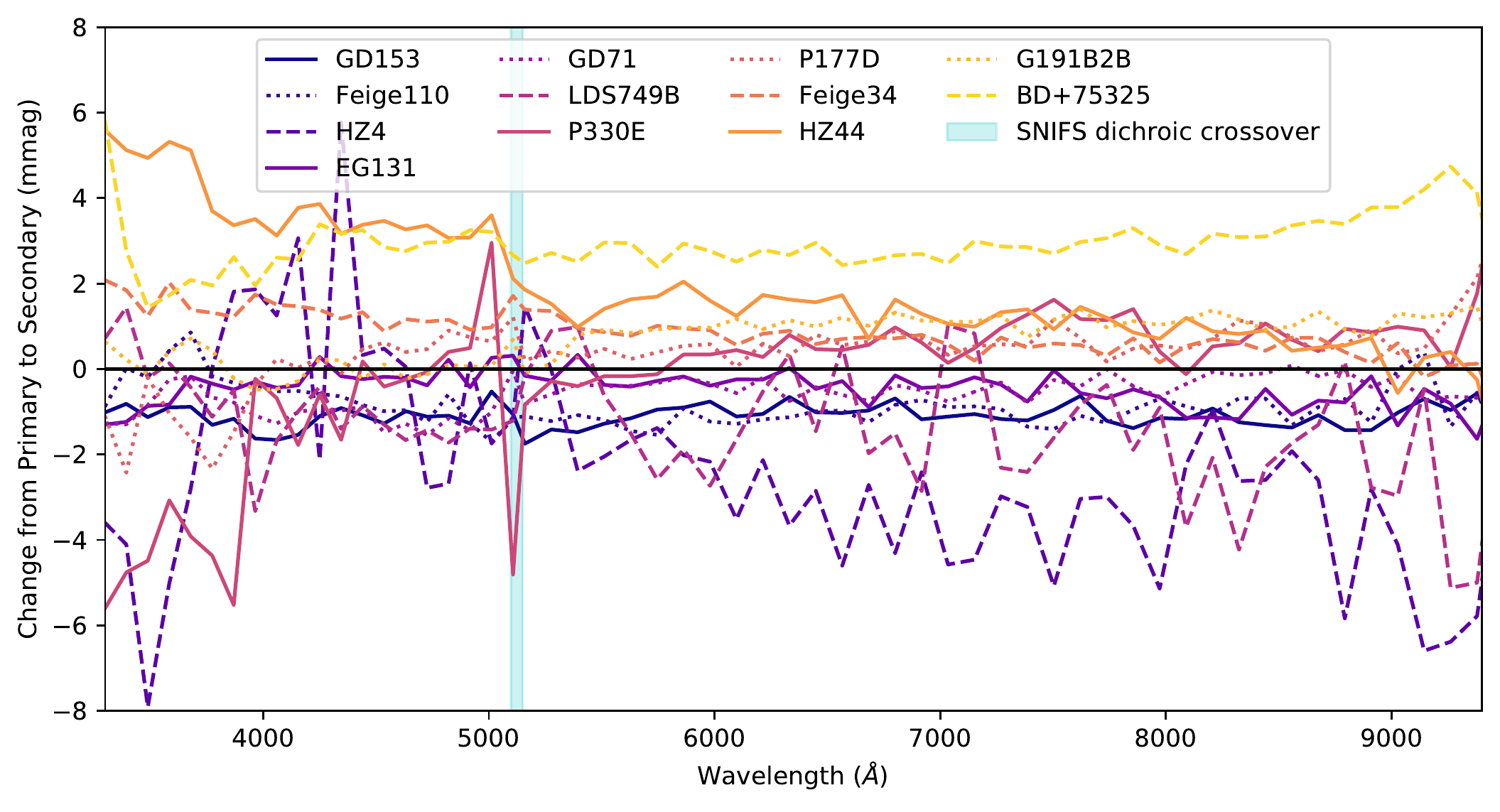}
\caption{The results of our leave-one-out test. The calibration model has been rerun for each primary standard star, moving that star to a secondary standard. We plot the change (in mmag) from our calibration of the star when it is a primary to our calibration when it is a secondary. For better signal-to-noise, we again bin in wavelength. Overall, our network is robust to the loss of any one primary standard star. We do not show HD~31128 and HD~74000 because they have few measurements and thus their measurement uncertainties are dominated by repeatability.}
\label{fig:LOO}
\end{figure}

In order to estimate the external accuracy of our recalibrated standard star network we carried out a ``leave-one-out'' test, in which each primary calibrator was removed in turn from the primary list (i.e., we imposed no constraint that the left-out standard should be similar to its CALSPEC value) and the calibration recalculated. The results are shown in Figure~\ref{fig:LOO}. Here the comparison is made to the spectra of the primary calibrators with their $\Delta m_l$ terms applied in order to separate this mean effect, which is already estimated by the Bayesian hierarchical model, from the differential effect of removing a primary standard. The average change to the calibration is $\sim 2.6$~mmag, and can be as small as $\sim 1.2$~mmag. This demonstrates that the zeropoint of our standard star network is robustly tied to these CALSPEC stars.

\section{Comparisons to non-CALSPEC External Data}\label{sec:compare}

Many of our standard stars have extensive external data beyond that from CALSPEC. In \S\ref{subsec:spec_comp} we examine how well our spectral recalibration of the SSPS stars agrees with expectations due to known factors. In \S\ref{subsec:phot_comp} we compare synthesized photometry of our recalibrated standard stars with photoelectric photometry from the literature. In both cases we find very good agreement with expectations from the literature.

\subsection{Spectral Comparison for Stars from the Southern Spectrophotometric Standards Compilation}
\label{subsec:spec_comp}

As noted above, our bright standard stars and our fainter southern stars are taken from the lists of \citet{hamuy92,hamuy94}. Up to now the SNfactory has used the original full-resolution spectra obtained by \citet{hamuy92,hamuy94}, corrected for telluric absorption by us, since these provide $\sim3\times$ better sampling than the published SSPS tables. We can expect a number of differences between our recalibration of these stars onto the CALSPEC system relative to the original calibration that was employed. To begin with, originally these stars were zeropointed to the flux of Vega as given by \citet{hayes1985}, using magnitudes for the bright secondary standard stars relative to Vega given by \citet{taylor1984} but then adjusted by \citet{hamuy92,hamuy94} to agree with the then-existing $V$-band photometry (including their own). Several of the \citet{taylor1984} flux points were rejected by \citet{hamuy92,hamuy94} due to inconsistencies, leaving some large gaps in wavelength coverage, e.g., across the Balmer jump and in the range 8376--9834~\AA, over which the response of their new observations was interpolated\footnote{\citet{stritzinger2005} have since recalibrated five of these stars using stellar models to interpolate the original calibration.}\footnote{\citet{bessell1999} has made several alternative improvements in the calibration of these stars.}. Furthermore, as these were wide-slit observations, we have found that wavelength zeropoint errors of several \ang can occur for stars mis-centered in the slit; these offsets can differ between the blue and red spectrograph setups used by \citet{hamuy92,hamuy94}. Finally, the original spectral resolution of the SSPS data is $\sim16$\,\ang; the resolution of our recalibrated spectra is about 4$\times$ higher.

Therefore, we can expect differences due to the mismatch between \citet{hayes1985} and CALSPEC Vega, larger differences where the response of the original system was poorly constrained, larger residuals near strong stellar absorption lines due to wavelength shifts and resolution differences\footnote{Since these differences are measurable from the original spectra, we have already corrected for them in the reference spectra we have used, e.g., in \citet{buton13}.}, and possible additional mean and random offsets of order 10~mmag.

\begin{figure}[htb]
\centering
\includegraphics[width=0.8\columnwidth]{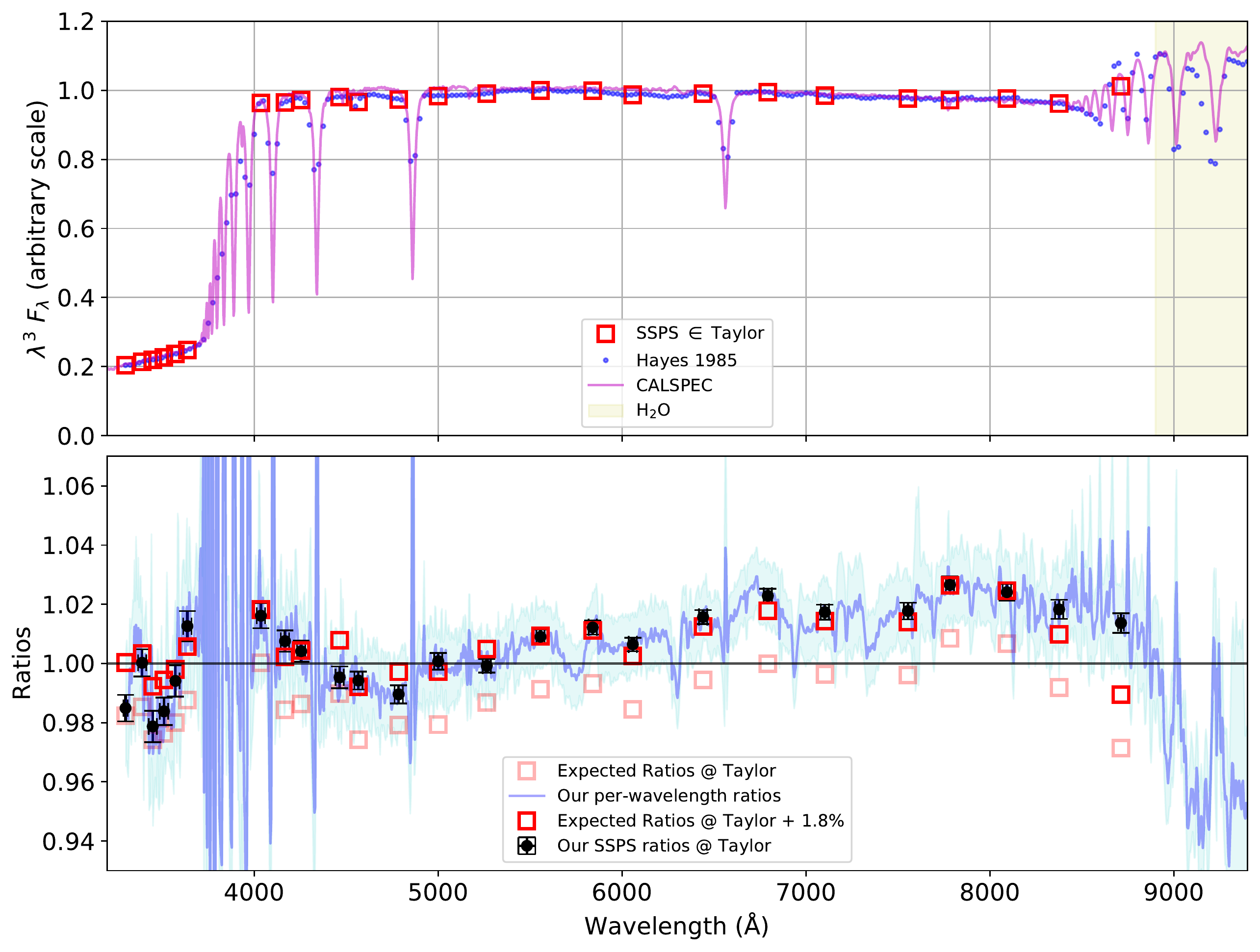}
\caption{Comparison of the expected and measured corrections to place the SSPS sample on the CALSPEC system. The upper panel shows the CALSPEC spectrum of Vega with the \citet{hayes1985} flux values overlaid. The red squares represent the wavelength bins, originally defined by \citet{taylor1984}, used to establish the calibration for the SSPS sample by \citet{hamuy92,hamuy94}. This illustrates the difficulties around the Balmer and Paschen absorption lines that needed to be avoided for the original SSPS calibration. (Note that the \citet{taylor1984} bin around 8700\,\ang was not used for SSPS calibration.) The lower panel shows the expected ratio of \citet{hayes1985} to CALSPEC (light red squares), our mean measured ratio between the original SSPS spectra and our recalibration (light blue line, with a cyan band representing the standard deviation among our sample of the SSPS stars), and our ratios at the SSPS/\citet{taylor1984} wavelengths (black points, with uncertainties in mean along the ratio axis and the width range along the wavelength axis). Our ratios are higher than the prediction, i.e., the original SSPS spectra gave fluxes higher than for the CALSPEC system, by around 1.8\%.}
\label{fig:calib_ratio}
\end{figure}

Figure~\ref{fig:calib_ratio} shows a comparison of the changes in calibration that we find here, compared to those expected from the \citet{hayes1985} and \citet{taylor1984} calibration of Vega used by \citet{hamuy92, hamuy94} versus the CALSPEC spectrum of Vega from \citet{bohlin2020}\footnote{Specifically, {\tt alpha\_lyr\_stis\_010.fits}.}. The top panel overlays the SSPS flux calibration windows from Taylor onto the CALSPEC spectrum of Vega over the spectral range of interest here. The \citet{hayes1985} flux points are also shown.\footnote{The (heavily smoothed) ratio over all the \citet{hayes1985} flux points is shown in Figure~2 of \citet{bohlin2004} and Figure~7 of \citet{bohlin14}, illustrating the problems surrounding the Balmer and Paschen series absorption lines that \citet{hamuy92,hamuy94} tried to avoid when selecting which \citet{taylor1984} points to use.} The lower panel compares the ratios of our recalibration of SSPS to that expected from the known differences in calibration methods. The solid black and red squares compare our derived recalibration ratios to those expected, at the \citet{taylor1984} wavelength bins used for SSPS, but after shifting the flux ratio by an achromatic normalization factor of 1.8\%. The light blue curve shows the calibration ratio at full spectral resolution, and the cyan band shows the standard deviation across all of the recalibrated SSPS stars. Most of the very high frequency differences surround strong stellar absorption lines, and arise from small wavelength shifts and resolution differences, as anticipated. The large dip redward of the 9000\,\ang is due to incomplete correction for H$_2$O in the SSPS, and is expected for reasons other than using the \citet{hayes1985} plus \citet{taylor1984} versus CALSPEC flux calibration as reference. 

There are a few \citet{taylor1984} bins for which our recalibration differs from what was expected. These include two of the six \citet{taylor1984} bins blueward of the Balmer jump; here we find that our spectra for the Morgan-Keenan A-type stars in our SSPS sample exhibit continua that are very linear in flux versus wavelength --- a behavior that would be hard to mimic accidentally, and which CALSPEC shows for its Vega spectrum. This leads us to believe that our recalibrations here are sound, and that the original SSPS had some structure here in addition to the differences attributable to the reference flux calibration that was used. There is also a strong difference for the reddest \citet{taylor1984} bin, likely due to issues around the Pashcen lines in the \citet{hayes1985} calibration, previously noted by \citet{bohlin2004}.

Overall, we conclude that the smooth trends in our recalibration are those expected from the SSPS versus CALSPEC calibrations, whereas the high-frequency variations arise from the better wavelength calibration, wavelength resolution, S/N, model-insensitivity, and robustness of our SNIFS spectrophotometric calibration.

\subsection{Comparison to Filter Photometry}
\label{subsec:phot_comp}

Our spectrophotometry allows us to synthesize magnitudes on any photometric system, in principle making such comparisons straightforward. Ideally we would like to compare our new spectrophotometric calibration with a homogeneous external source, such as a set of filter photometry on a common system. However, our brightness range $3<V<15$ is problematic for all of the existing homogeneous all-sky surveys: SDSS and PS1 are saturated for all but a few of our fainest stars, \GAIA exhibits non-linearity, and has wavelength coverage slightly too broad compared to our spectra; Hipparcos saturates for our brightest stars and does not extend to our faintest stars. We have already calibrated to CALSPEC to the extent possible, but this covers only a third of our stars. This situation forces us to compare subsets of our standard star network to different photometry sources, which we now do. The most complete coverage for comparison with our standards comes from filter photometry on the Johnson-Kron-Cousins system obtained from a variety of observers spanning several decades. 

\begin{figure}[htbp]
\centering
\includegraphics[width=0.8\columnwidth]{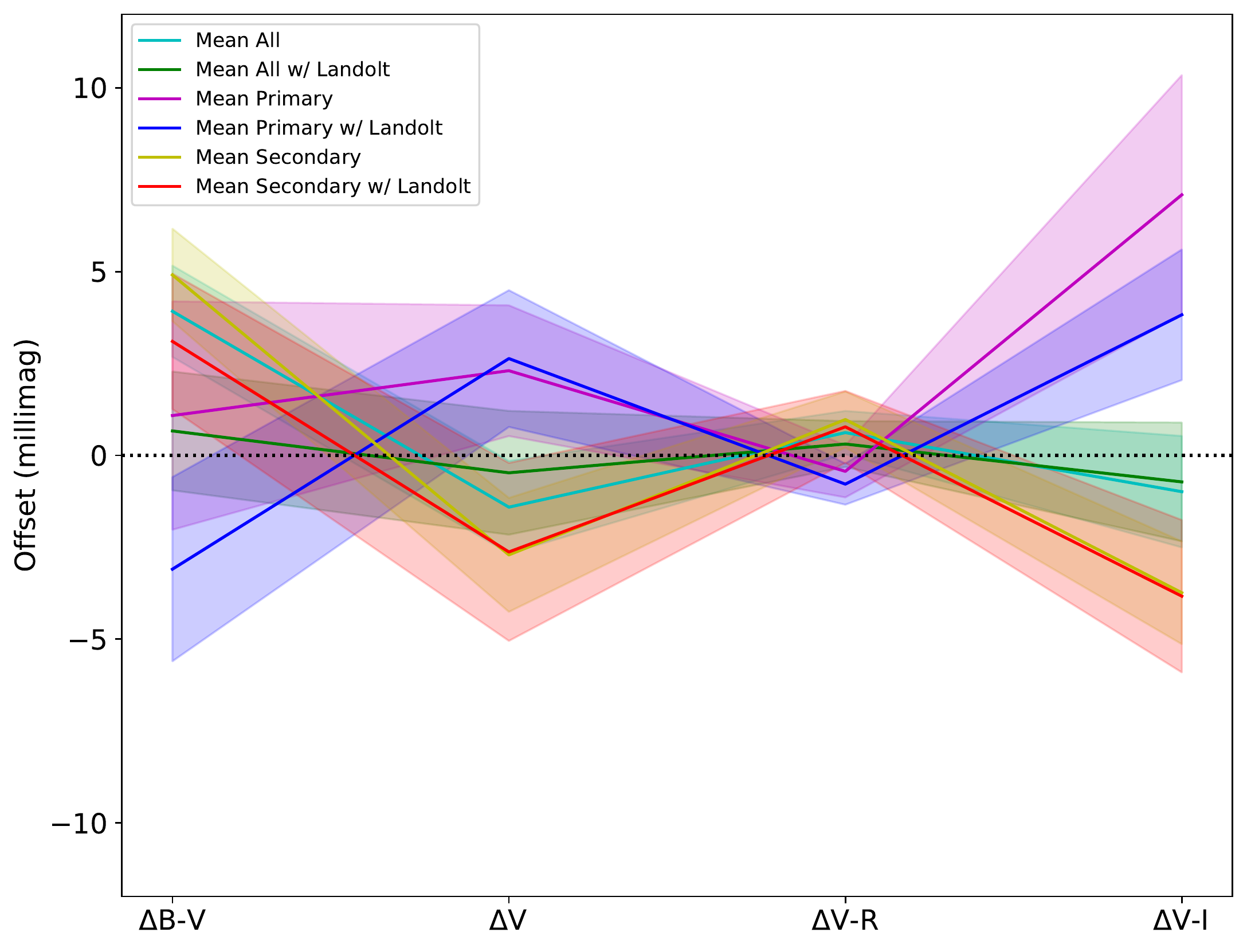}
\caption{The mean offset between synthetic photometry and literature filter photometry for six different groupings of our standard stars. We plot the residuals for each of $B-V$, $V$, $V-R$, $V-I$, in which the filter photometry was originally analyzed and reported. The shaded bands, when evaluated at the discrete photometric indices, give the error on the means. We see generally good agreement --- within a few mmag --- between our flux-calibrated spectra and filter photometry. The agreement in $V-R$ is especially impressive. The means of the different subsets agree within their uncertainties, except for $V-I$, where the primary and secondary calibrators are in tension by $2.8\,\sigma$.}
\label{fig:filtercomp}
\end{figure}

We begin by collecting filter photometry on the $UBVRI$ system from the literature. For the CALSPEC stars in common with \citet{bohlin15}, we use the same sources of filter photometry, namely, their paper and \citet{landolt2007}. Filter photometry of the SSPS standard stars was presented in \citet{hamuy92}, \citet{landolt1992} and \citet{bessell1999}. For the bright HR stars, both \citet{hamuy92} and \citet{bessell1999} rely heavily on older photometry from the SAAO group \citep{cousins1971,cousins1980,cousins1984,kilkenny1989}. For standard stars not in these two sets, we have collected $UBVRI$ photometry from \citet{K62, EG65, P73, G74, C78, D82, MM97, Koen2010}. Since almost all of the filter photometry measurements were reduced as $V$ along with color indices, we analyze the data in this same way, rather than as per-band magnitudes, so that the measurement uncertainties remain uncorrelated. Since the uncertainties of our spectra and the CALSPEC spectra are strongly correlated across wavelength, a normalization plus colors (instead of independent bands) is also the best way to express our synthetic photometry. Because the companion stars of \BDTwentyEight and Hiltner~600 are included in the photoelectric photometry apertures, but not in the SNIFS measurements, they are excluded from this comparison.

We calculate two sets of synthetic magnitudes; the first in $UBVRI$ using the CALSPEC spectra in order to obtain initial zeropoints for the $UBVRI$ system on the new CALSPEC system. The second, in $BVRI$, using our recalibrated spectra; for these $U$ is omitted because the SNIFS spectra miss between 0.05 and 0.17~mag over the color range $-0.37 <B\mathrm{-}V < +0.65$ from the blue side of the $U$ band. Synthetic magnitudes are calculated by integrating over our spectra using \sncosmo \citep{sncosmo} with filter bandpasses defined by \citet{bessell2012}.\footnote{We do not shift the \citet{bessell2012}, in contrast to \citet{bohlin15}, since there did not seem to be a significant improvement in doing so.} The \citet{bessell2012} $R-$ and $I$-band filter transmission curves omit telluric absorption; here we include telluric absorption \citep{hinkle2003} typical of observatories, like CTIO, KPNO and SAAO where most of the filter photometry was obtained.

While Table~5 of \citet{bohlin15} provided the zeropoints for the $UBVRI$ system for the $f_{\lambda}$ spectra of the version of CALSPEC then in use, the newest revision of CALSPEC presented by \citet{bohlin2020} has significant changes. Using the CALSPEC spectra for the same 15 stars as in \citet{bohlin15} having full $UBVRI$ coverage, we find zeropoints of \zpV, \zpBmV, \zpUmB, \zpVmR, and \zpVmI\,mag
for $V$, $B$-$V$, $U$-$B$, $V$-$R$ and $V$-$I$, respectively.

While this comparison uses 15 stars in $UBVRI$, by using our spectra for all of the stars in our sample having filter photometry, a much larger sample can be created. Although $U$-band must be dropped in this approach, this sample contains 60, 63, 53, and 52 stars with $V$, $B$-$V$, $V$-$R$ and $V$-$I$ photometry, respectively. In addition to this parent sample containing all available photometry in all available bands, we create several subsets. Two subsets are based on whether or not a star is a primary or secondary standard star; this is of interest because here we present new standardization for the secondaries. Each of these two subsets are split further based on whether or not the source of filter photometry is by Landolt and collaborators; this set is of interest because the Landolt system is pervasive, and likely the most homogeneous, and for those reasons form the basis for the analysis in \citet{bohlin15}. 

\begin{figure}[htbp]
\centering
\includegraphics[width=0.48\textwidth]{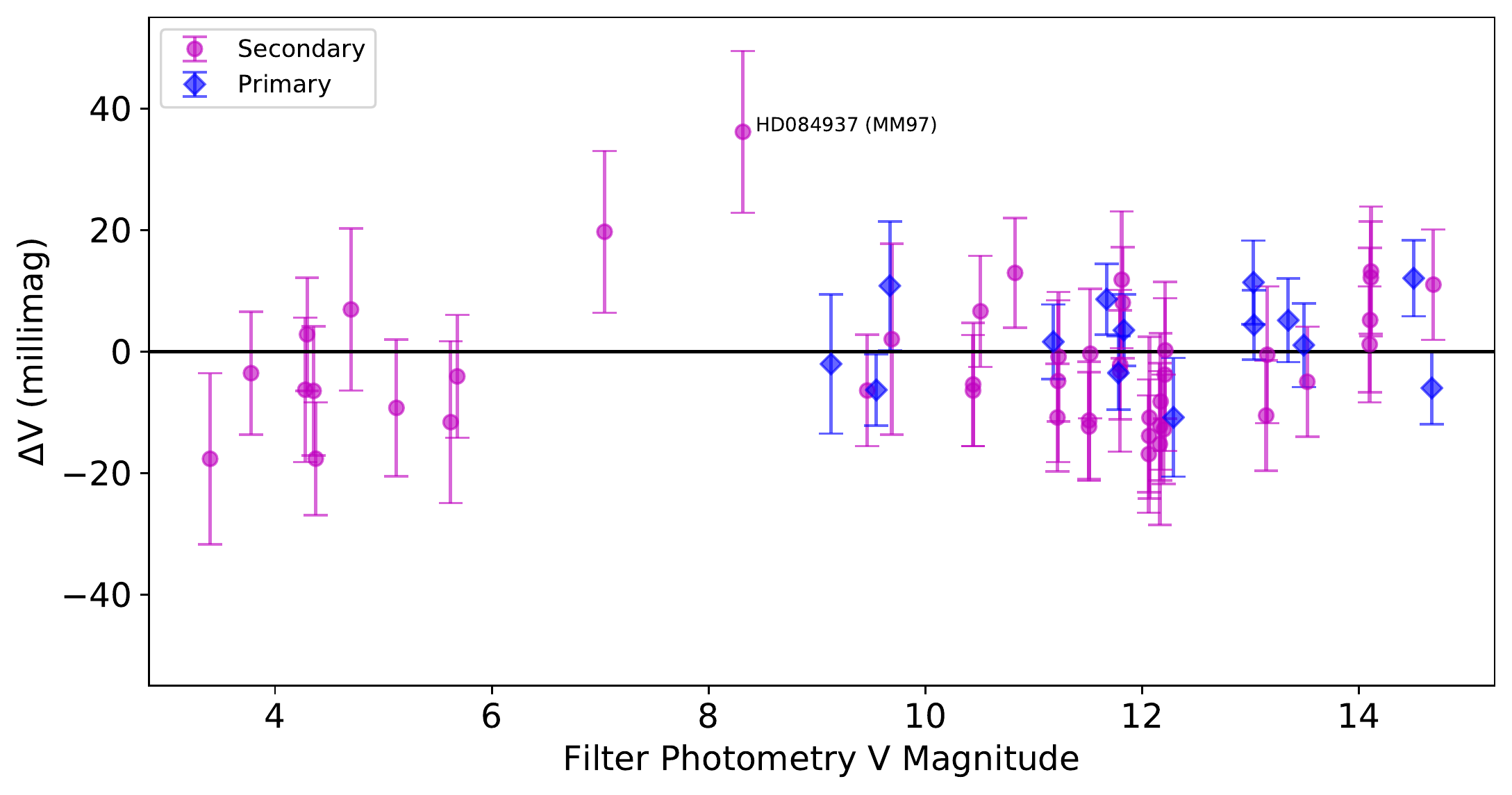}
\includegraphics[width=0.48\textwidth]{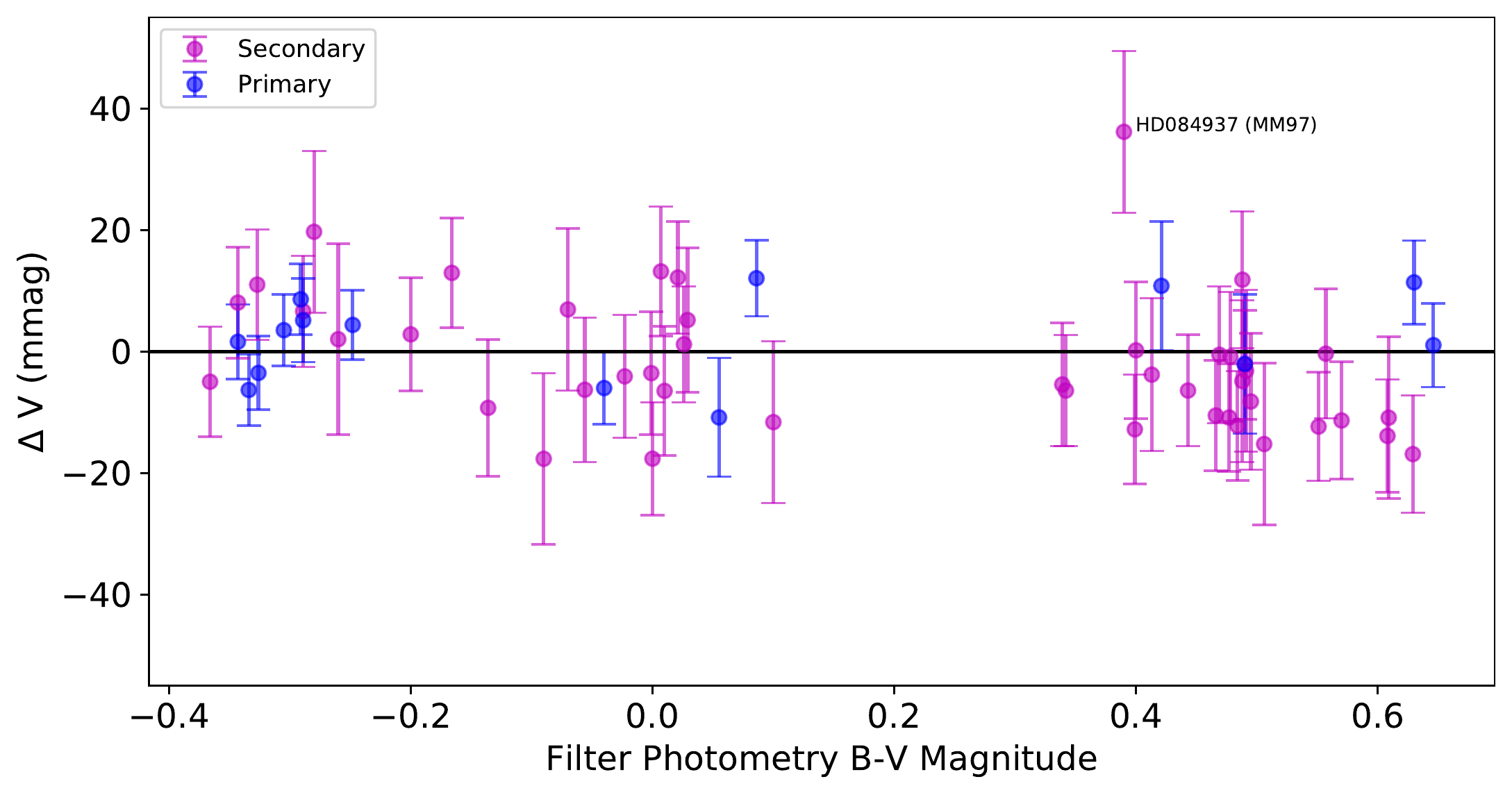}
\caption{Comparison of synthetic photometry of our standard star spectra. Left: $V$-band residuals versus $V$-band magnitude. Here synthetic magnitudes have been subtracted from filter photometry magnitudes. Stars are color-coded by whether they are in the primary (blue) or secondary (magenta) samples. Note that some stars have several independent sources of filter photometry from the literature, hence those stars can overlap in their V-band magnitudes on the scale of this plot. The stars deviating the most from the mean are labeled with their name and source of photometry. These outliers have photometry from our most heterogeneous literature sources (often reporting on only a single star in our sample), so are not a serious concern. The results show that our standard star network is linear (relative to filter photometry) over a span of 12 magnitudes. Right: $V$-band residuals versus $B-V$ This shows that for blue stars our primary and secondary standards agree well. For intermediate colors, there is enhanced scatter for both the primary and secondary standards. For redder colors there appears to be a $\sim9$~mmag offset between most of the primary standards and secondary standards. This offset appears to be the primary driver of the small $V$-band offset we measure for our standard star network, and is more apparent because we have added a large number of red stars, e.g., in comparison to \citet{bohlin15}.}
\label{fig:dVV}
\end{figure}

We obtain the best agreement with the full set of filter photometry by adjusting the zeropoints relative to those given above to \ourzpV, \ourzpBmV, \ourzpVmR, and \ourzpVmI\,mag
in  $V$, $B$-$V$, $V$-$R$ and $V$-$I$, respectively.  These values were chosen to split the difference between the primary and secondary subsets with Landolt photometry. They are close to the means for the full sample having Landolt photometry, but offset slightly because the sample of secondaries with Landolt photometry is larger (16 versus 11 stars). The means and errors on the means for all of our samples are shown in Figure~\ref{fig:filtercomp}. We note that our agreement in $V$-$R$ for all the stars in our sample is exceptional, with a dispersion of only \VmRrms\,mmag for the subset of \NLandoltAll stars with Landolt $V$-$R$. There is mild tension for the $V$-band zeropoints and $V$-$I$ color relative to synthetic photometry for stars between our CALSPEC-referenced primary standards and our secondary standards. This difference persists whether or not telluric absorption is included in the filter transmission function. The differences across all filters for those with Landolt photometry, in the sense primary minus secondary, are \fmsV, \fmsBmV, \fmsVmR, and \fmsVmI\,mmag
in $V$, $B$-$V$, $V$-$R$ and $V$-$I$, respectively.

By comparison, the filter photometry of the SSPS tertiary stars was found to agree with synthetic photometry of the original spectra within offsets in the means of $+6$, $-2$, and $+3$~mmag and RMS values of 11, 6, 18~mmag, respectively in $B$, $V$, and $R$ \citep{hamuy92, hamuy94}. The subset of those with the most homogeneous filter photometry, from \citet{landolt1992}, were found to have mean offsets of $+13$, $+9$ and $+8$~mmag and RMS values of 9, 7 and 11~mmag, respectively, in $B$, $V$ and $I$ in \citet{hamuy92, hamuy94}. Our recalibrated spectra exhibit much better agreement with filter photometry.

With $BVRI$ filter photometry and synthetic magnitudes in hand, we can also look for trends with brightness and color. The left panel of Figure~\ref{fig:dVV} shows differences between $V$-band filter and synthetic photometry versus $V$~mag, demonstrating that our standard star network is linear relative to filter photometry over a span of 12 magnitudes. The plotted uncertainties include the published measurement uncertainties for the filter photometry, our internal dispersion (see \S\ref{sec:dispersion}), plus an extra dispersion of 5~mmag needed to obtain $\chi^2_\nu = 1$ meant to account for heterogeneity between the various sources of filter photometry. The only significant standout is HD084937, which has few SNIFS spectra and limited filter photometry from \citet{MM97}. The figure suggests an offset of around $-5 \pm 4$~mmag for the brightest stars; such offsets are not uncommon, e.g., \citet{hamuy92} and \citet{landolt1992} when comparing their filter photometry for fainter stars. The right panel of Figure~\ref{fig:dVV} shows the V-band differences versus $B$-$V$ color; here one sees that for blue colors the primary and secondary stars agree well and for intermediate color there is somewhat larger scatter. But for redder stars there is some disagreement. Because our sample of secondary stars includes many more red stars than our primary sample, an offset appears. Even so, the evidence for a systematic effect is small, being at the $\sim2\,\sigma$ level.

\section{Conclusions}\label{sec:conclusion}

\begin{figure}[htbp]
\centering
\includegraphics[width=0.95\textwidth]{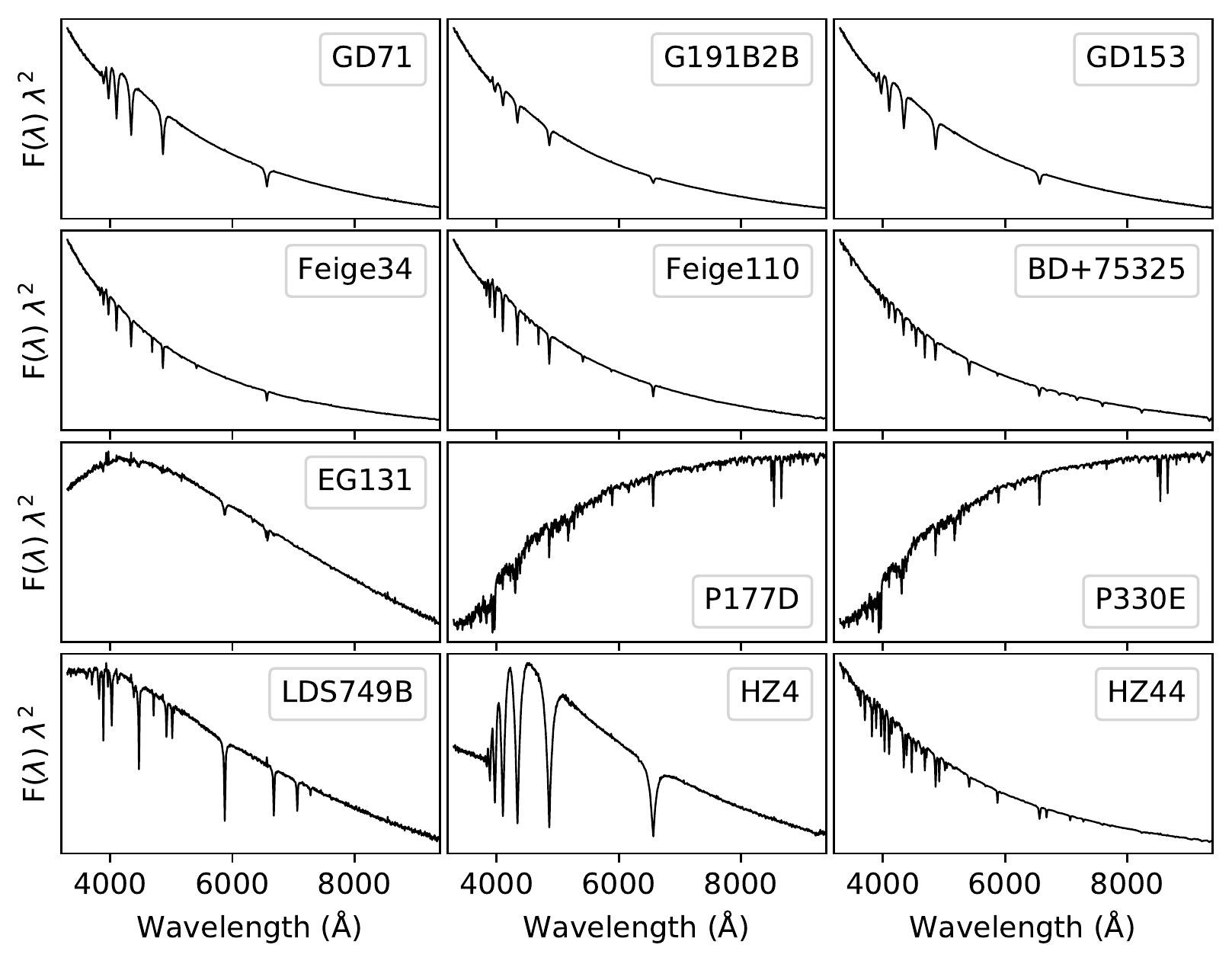}
\caption{Our spectra of space-based CALSPEC stars that were used as primary calibrators and having observations on more than one night. Fluxes are linear in F($\lambda$)$\lambda^2$ in order to balance the range of spectral slopes across the ensemble of standard stars, and the flux labeling is suppressed for clarity.}
\label{fig:calspec_ngt1}
\end{figure}

\begin{figure}[htbp]
\centering
\includegraphics[width=0.95\textwidth]{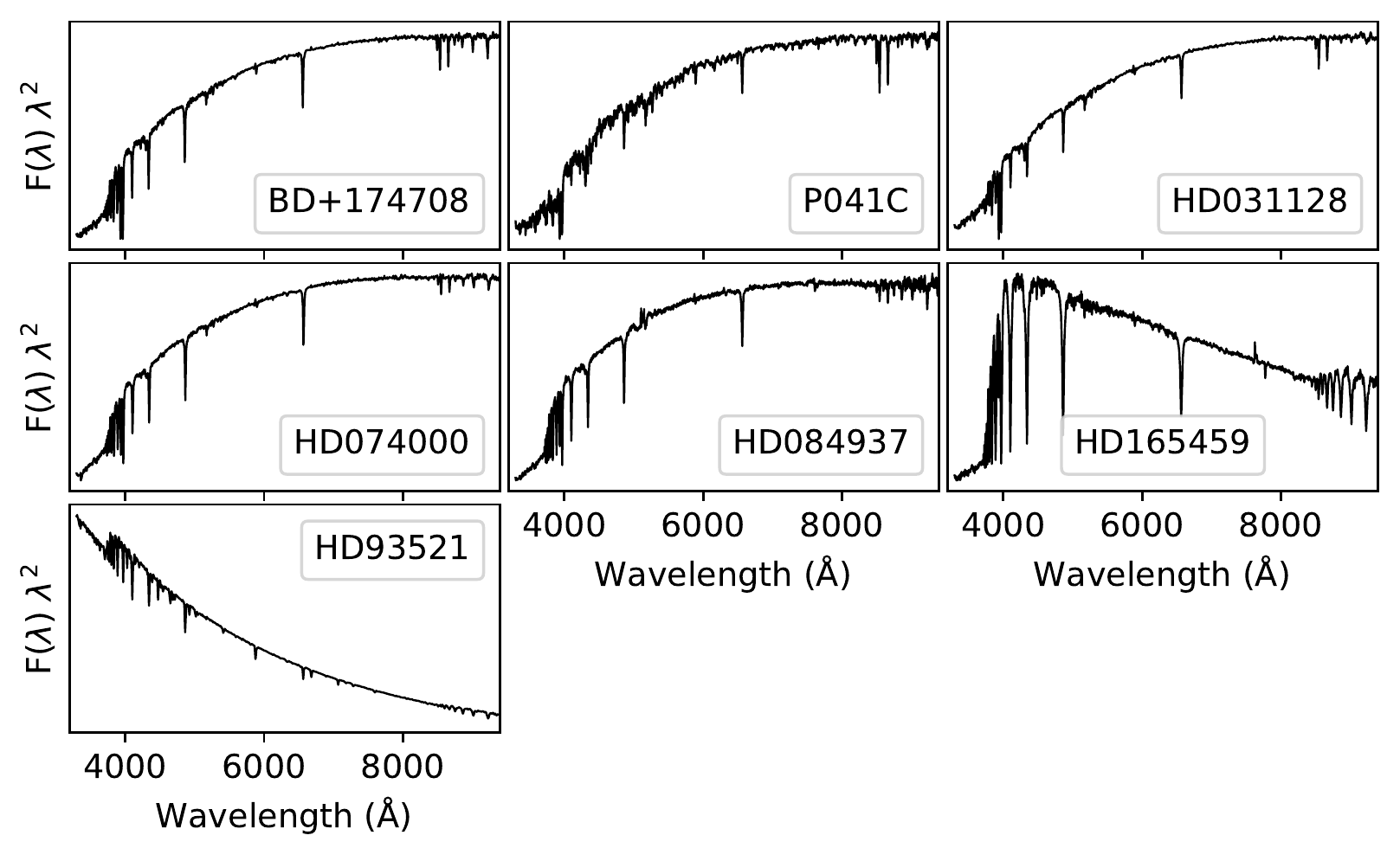}
\caption{Our spectra of space-based CALSPEC stars that were not used as primary calibrators or having observations on two or fewer nights. The presentation follows that of Figure~\ref{fig:calspec_ngt1}.}
\label{fig:calspec_supp}
\end{figure}

\begin{figure}[htbp]
\centering
\includegraphics[width=0.92\textwidth]{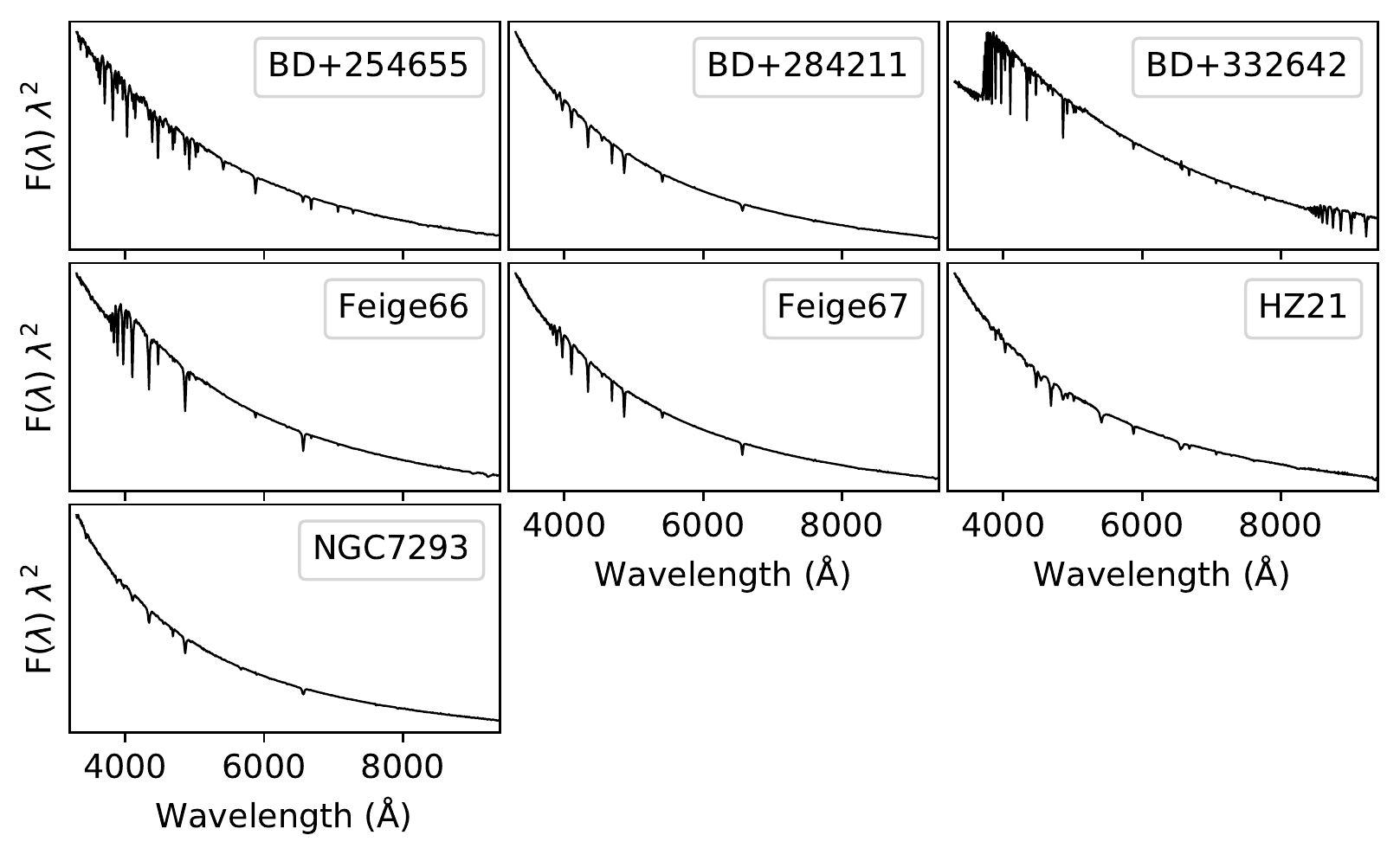}
\caption{Our spectra for stars from \citet{oke1990} that were used as secondary calibrators. The presentation follows that of Figure~\ref{fig:calspec_ngt1}.}
\label{fig:calspec_oke}
\end{figure}

\begin{figure}[htbp]
\centering
\includegraphics[width=0.92\textwidth]{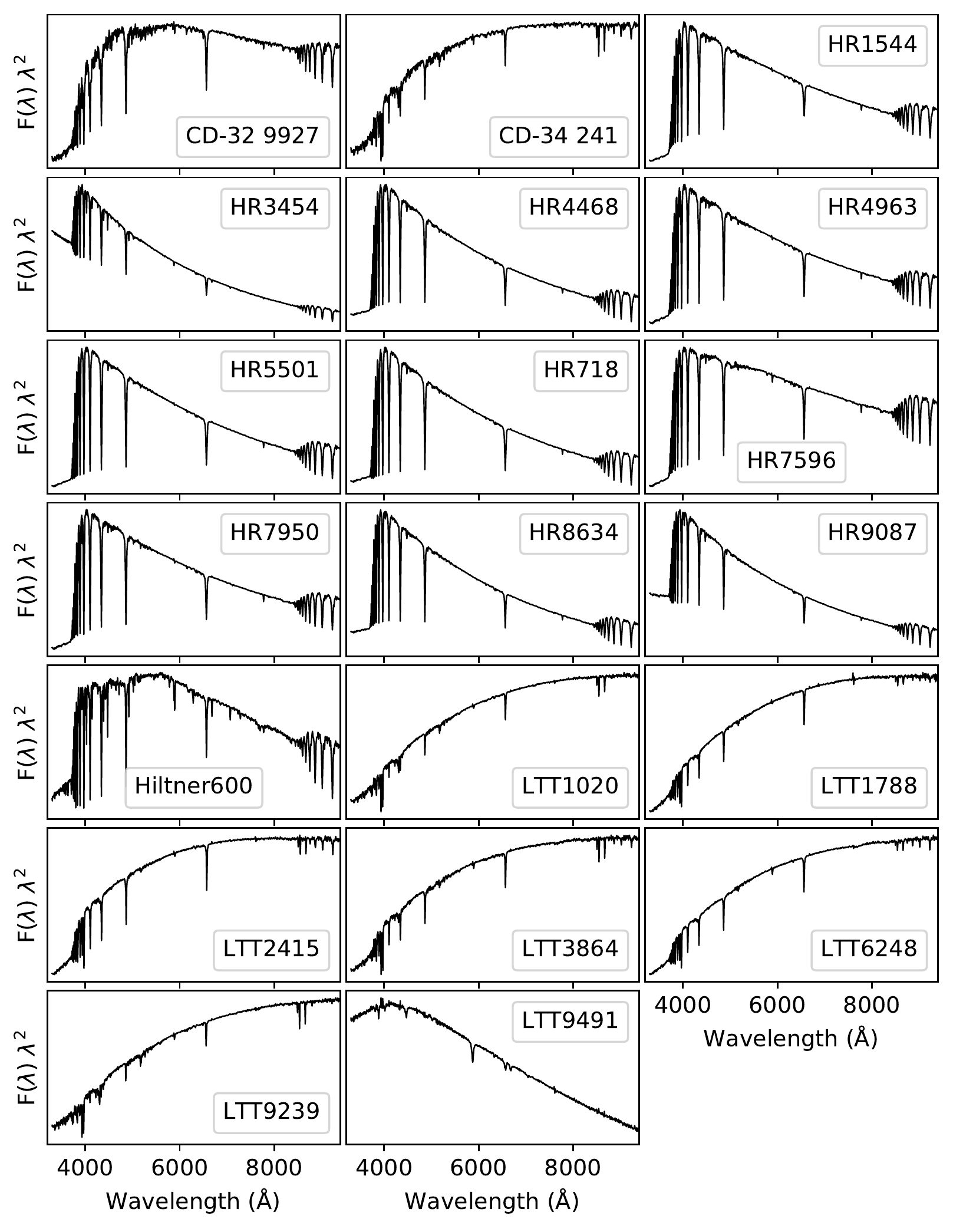}
\caption{Our spectra for stars from the SSPS sample of \citet{hamuy92,hamuy94}. The presentation follows that of Figure~\ref{fig:calspec_ngt1}.}
\label{fig:calspec_ssps}
\end{figure}

This work presents the results from a large sample of optical spectrophotometry of \NStandNominal stars from the SNIFS instrument on the UH 2.2m telescope. We present a Bayesian hierarchical model that intercalibrates the whole network of observations, placing all stars on the CALSPEC system with an accuracy of a few mmag. Among other factors, this model accounts for distributions of inliers and outliers, instrumental repeatability, and tensions between primary CALSPEC calibrators. Figures~\ref{fig:calspec_ngt1}, \ref{fig:calspec_supp}, \ref{fig:calspec_oke}, and \ref{fig:calspec_ssps} show our final calibrated spectra, their high signal-to-noise, and exceptionally smooth continuum regions. As our system response and atmosphere model have no enforced smoothness wavelength-to-wavelength, the smoothness seen in these plots constitutes another crosscheck on our analysis.

We characterize the residuals of the system, finding 1--2\% outliers and long-exposure repeatability of \repeatibility, depending on wavelength. Most of the residuals are gray (wavelength-independent).

While blinded, we perform a series of crosschecks on the analysis, including subsets of the data and searching for correlations with observing conditions and instrumental parameters. After unblinding, we find good linearity against filter photometry over 12 magnitudes. These standard stars are being used to calibrate the SNfactory SNe~Ia, which in turn will be used to measure dark energy parameters. \ConcludingSentence Another use of these recalibrated standard stars will be to place archival data that used these stars onto the space-based CALSPEC system. Our measured mean spectra are available at [SNfactory link; Zenodo link].

\clearpage

\acknowledgments{
The authors wish to recognize and acknowledge the very significant cultural role and reverence that the summit of Maunakea has always had within the indigenous Hawaiian community.  We are most fortunate to have the opportunity to conduct observations from this mountain. We thank the anonymous referee for careful and useful feedback. We thank the technical staff of the University of Hawaii 2.2~m telescope, and Dan Birchall for observing assistance. We thank Jean-Charles Cuillandre and Billy Mahoney for useful discussions of the SkyProbe data, Ralph Bohlin for providing photometry of \BDSeventeen in digital form, and Marc Betoule for clarifying the SNLS extinction results in \citet{betoule13}. This work was supported in part by the Director, Office of Science, Office of High Energy Physics of the U.S. Department of Energy under Contract No.~DE-AC02-05CH11231. Support in France was provided by CNRS/IN2P3, CNRS/INSU, and PNC and French state funds managed by the National Research Agency within the Investissements d'Avenir program under grant reference numbers ANR-10-LABX-0066, ANR-11-IDEX-0004-02 and ANR-11-IDEX-0007. Additional support comes from the European Research Council (ERC) under the European Union's Horizon 2020 research and innovation program (grant agreement No 759194-USNAC). Support in Germany was provided by DFG through TRR33 "The Dark Universe" and by DLR through grants FKZ 50OR1503 and FKZ 50OR1602. In China the support was provided from Tsinghua University 985 grant and NSFC grant No 11173017. The technical support and advanced computing resources from University of Hawaii Information Technology Services - Cyberinfrastructure, funded in part by the National Science Foundation MRI award \# 1920304, are gratefully acknowledged. Some results were obtained using resources and support from the National Energy Research Scientific Computing Center, supported by the Director, Office of Science, Office of Advanced Scientific Computing Research of the U.S. Department of Energy under Contract No. DE-AC02-05CH11231. We thank the Gordon \& Betty Moore Foundation for their support. Additional support was provided by NASA under the Astrophysics Data Analysis Program grant 15-ADAP15-0256 (PI: Aldering). This work has made use of data from the European Space Agency (ESA) mission \GAIA (\url{https://www.cosmos.esa.int/gaia}), processed by the \GAIA Data Processing and Analysis Consortium (DPAC, \url{https://www.cosmos.esa.int/web/gaia/dpac/consortium}). Funding for the DPAC
has been provided by national institutions, in particular the institutions participating in the \GAIA Multilateral Agreement.}

\facilities{MKO, UH2.2m (SNIFS), HST, \GAIA} 

\software{
astropy \citep{Astropy},
LBLRTM \citep{clough92, clough05}
Matplotlib \citep{matplotlib}, 
Numpy \citep{numpy}, 
pystan (\doi{10.5281/zenodo.598257}),
scikit-learn \citep{scikit-learn}, 
SciPy \citep{scipy},
\sncosmo \citep{sncosmo},
Stan \citep{Carpenter2017},
Telfit \citep{gullikson14}
}

\appendix

\section{Analytic Point Spread Functions}\label{app:psf}

The SNIFS spectroscopic channel only contains observations from the currently observed target, and it is therefore not possible to follow the standard approach of building an empirical PSF model from other objects observed in the field with the same channel. In principle, the SNIFS imaging channel could be used to constrain the PSF, but this functionality has not yet been implemented. This then requires an analytic description of the PSF that can be fit to individual observations from the spectroscopic channel and that is internally consistent --- that is, reporting the correct flux across the wide range of observing conditions, e.g., seeing, centering within the micro-lens array (MLA), signal-to-noise ratio, etc. experienced with real data. We investigated two analytic models. First, we considered a PSF that is the sum of Moffat \citep{moffat1969} and Gaussian profiles that we refer to as the ``\PSFMG'' \citep{buton13}. Second, we considered the convolution of several instrumental components along with an atmospheric term inspired by Kolmogorov turbulence, implemented in Fourier space, that we refer to as the ``\PSFF''.

\begin{figure}
    \centering
    \includegraphics[width = 0.9\textwidth]{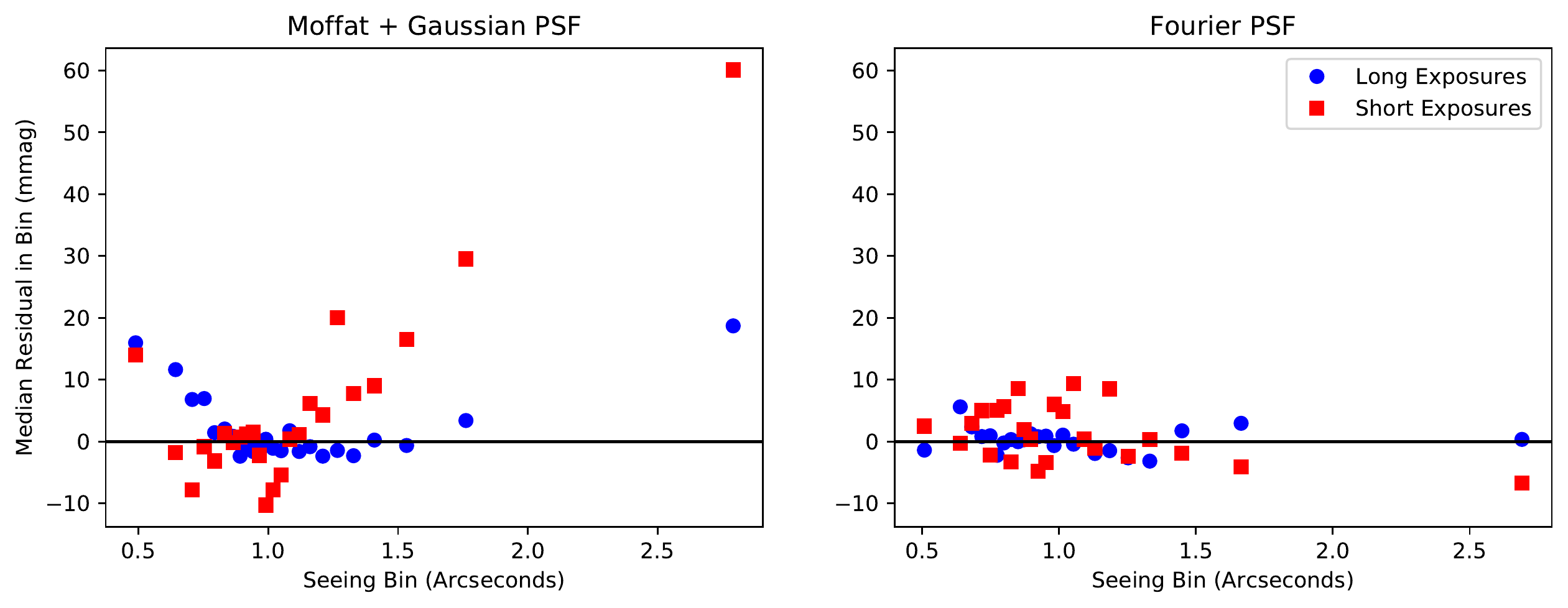}
    \caption{Median residuals vs. seeing in equal-percentile bins for spectra extracted with both PSFs, separated by long and short exposures. The \PSFMG PSF (left panel) shows large systematic residuals for both very good and very bad seeing, and these trends are different between long and short exposures. Encouragingly, the \PSFF PSF does better on all fronts, having much smaller residuals (even for very bad seeing), and having similar behavior for both long and short exposures. On the basis of results like this, we chose the \PSFF PSF for our primary results. We note that, for long-exposure standard stars, the two PSFs give consistent calibration to within mmags, as discussed in the text.}
    \label{fig:resid_vs_seeing}
\end{figure}

We have performed two separate calibration runs, both using the same observations separately extracted with each PSF. Although 10--20~mmag differences between the results for the PSFs are seen for the short-exposure ($\sim 1$s) standards, the long-exposure standards calibrated with each match to an RMS of 2--4~mmags, depending on wavelength. We use the \PSFF for our primary results, as it has smaller residuals with seeing (Figure~\ref{fig:resid_vs_seeing}).

\subsection{The Moffat + Gaussian PSF}

The \PSFMG\ PSF model is composed of the sum of Moffat and Gaussian profiles:
\begin{equation}
\PSFMG(\tilde{r},\lambda)\ =\ \mathcal{A_{MG}}(\lambda) \times  \left[ \eta(\alpha)\ {\rm{exp}}\left(-\frac{\tilde{r}^2}{2\,\sigma(\alpha)^2}\right)\ + \left(1 + \left(\frac{\tilde{r}}{\alpha(\lambda)}\right)^2\right)^{-\beta(\alpha)}\right]
\label{eq:psfmg}
\end{equation}
where the normalization is given by
\begin{equation}
\mathcal{A_{MG}}(\lambda) = \frac{\sqrt{\epsilon - \zeta^2}}{ \pi \left(2\, \eta(\alpha)\   \sigma(\alpha)^2\ +\ {\alpha(\lambda)^2}/{(\beta(\alpha) -1)}\right)}
\end{equation}

The elliptical radius $\tilde{r}$ is defined as
\begin{equation}
\begin{split}
\tilde{r}^2 \ &=\ \left[x -x_*\left(\lambda\right)\right]^2\ \\
&+\ \epsilon\times \left[\left(y - y_*(\lambda\right)\right]^2\ \\
&+\ 2\zeta\times \left[x -x_*\left(\lambda\right)\right]\times \left[y -y_*\left(\lambda\right)\right]
\end{split}
\end{equation}
with respect to the position of the source, $x_*(\lambda), y_*(\lambda)$. $\epsilon$ and $\zeta$ encode the ellipticity and orientation of the PSF with respect to the axes of the MLA. We assume that the Gaussian and Moffat shape parameters can be described as linear functions of a single underlying parameter $\alpha$:
\begin{equation}
\begin{split}
\beta(\alpha) &= \beta_0\ +\ \beta_1 \times \alpha \\
\sigma(\alpha) &= \sigma_0\ +\ \sigma_1 \times \alpha \\
\eta(\alpha) &= \eta_0\ +\ \eta_1 \times \alpha
\end{split}
\end{equation}
We determine the values of the coefficients $\beta_0$, $\beta_1$, $\sigma_0$, $\sigma_1$, $\eta_0$, and $\eta_1$ from fits to large numbers of high signal-to-noise observations of standard stars. We find that we require separate sets of coefficients for short exposures (formally $<\shortlong$ seconds, but generally $\sim 1$s) compared to longer exposures, and determine the coefficients for these two subsets of observations separately.

Atmospheric effects are chromatic, and we expect both the position of the source on the MLA and the width of the PSF to vary with wavelength. The variation in the central position of the source on the MLA is due to atmospheric differential refraction, and can be written as:
\begin{equation}
\label{eq:adr}
\begin{split}
x_*(\lambda) = x_0 + \frac{1}{2} \left(\frac{1}{n^2(\lambda)} - \frac{1}{n^2(\lambda_0)} \right) \times \delta \sin(\theta) \\
y_*(\lambda) = y_0 - \frac{1}{2} \left(\frac{1}{n^2(\lambda)} - \frac{1}{n^2(\lambda_0)} \right) \times \delta \cos(\theta) \\
\end{split}
\end{equation}
where $x_0$ and $y_0$ are the position of the source at a reference wavelength $\lambda_0$ of 5000~\AA. The index of refraction $n(\lambda)$ is calculated using an updated version of the Edl\'en equation \citep{edlen1966, stone2001}. In principle, $\delta$ is the tangent of the zenith angle and $\theta$ is the parallactic angle, but in practice we allow these parameters to vary from their nominal values as part of the PSF fitting procedure.

We parameterize the chromatic variation of the PSF width $\alpha(\lambda)$ using the following equation:
\begin{equation}
\alpha(\lambda) = \alpha_2 \left( \frac{\lambda}{\lambda_0} \right)^{\alpha_1 + \alpha_0 (\lambda / \lambda_0 - 1)}
\end{equation}
where $\alpha_0$, $\alpha_1$ and $\alpha_2$ are all free parameters. We assume that the ellipticity of the PSF is dominated by guiding errors, so we do not include a chromatic term for the ellipticity parameters $\epsilon$ or $\zeta$.  The final PSF model has nine free parameters that must be fit for each exposure: $x_0$, $y_0$, $\delta$, $\theta$, $\alpha_0$, $\alpha_1$, $\alpha_2$, $\epsilon$, and $\zeta$. Finally, to account for the pixelization of the PSF, we evaluate it on a three-times subsampled grid and sum the subsamples corresponding to each pixel.

\subsection{The Fourier PSF}

The \PSFF\ PSF model is composed of the convolution of atmospheric, instrumental, and tracking terms. We implement this PSF in Fourier space so that the convolutions 
are simply 
the product of the different terms, and then take the inverse Fourier transform of the final Fourier-space PSF model $\PSFFtilde(k_x, k_y, \lambda)$ to obtain the real-space model $\PSFF(x, y, \lambda)$. The Fourier-space model can be written as the product of the following terms:
\begin{equation}
\PSFFtilde(k_x, k_y, \lambda) = \mathcal{\tilde{P}}_{\mathrm{atmospheric}}(k_x, k_y, \lambda) \times \mathcal{\tilde{P}}_{\mathrm{instrumental}}(k_x, k_y) \times \mathcal{\tilde{P}}_{\mathrm{tracking}}(k_x, k_y)  \times \mathcal{\tilde{P}}_{\mathrm{ADR}}(k_x, k_y, \lambda) \times \mathcal{\tilde{P}}_{\mathrm{pixel}}(k_x, k_y)
\end{equation}

We model the atmospheric component of the PSF as:
\begin{equation}
\mathcal{\tilde{P}}_{\mathrm{atmospheric}}(k_x, k_y, \lambda) = \exp\left(-\left(k w(\lambda)\right)^\tau\right)
\end{equation}
where $k = \sqrt{k_x^2 + k_y^2}$ and $w$ is a parameter measuring the width of the PSF. For Kolmogorov turbulence, $\tau$ is 5/3, although in practice we find that our observed PSFs prefer a slightly lower value of $\tau$.

We find that the instrumental response can be modeled as the convolution of a narrow Gaussian core with a function having extended wings:
\begin{equation}
\mathcal{\tilde{P}}_{\mathrm{instrumental}}(k_x, k_y) = \exp\left(-\frac{1}{2} (k_x^2\sigma_{\mathrm{inst.},x}^2 + k_y^2\sigma_{\mathrm{inst.},y}^2)\right) \times \exp\left(-\left(k w_{\mathrm{inst.}}\right)^{\tau_{\mathrm{inst.}}}\right)
\end{equation}
where the parameters $\sigma_{\mathrm{inst.},x}$, $\sigma_{\mathrm{inst.},y}$, $w_{\mathrm{inst.}}$, and $\tau_{\mathrm{inst.}}$ control the shape and widths of these profiles.

In this PSF model, we assume that all non-instrumental forms of ellipticity are due to tracking errors. We model these tracking errors as the convolution of the PSF with what is effectively a 1D Gaussian in the direction of the tracking error:
\begin{equation}
\mathcal{\tilde{P}}_{\mathrm{tracking}}(k_x, k_y) = \exp\left(-\frac{1}{2} (k_x^2\sigma_{\mathrm{tracking},x}^2 + k_y^2\sigma_{\mathrm{tracking},y}^2 + 2 \rho k_x k_y \sigma_{\mathrm{tracking},x} \sigma_{\mathrm{tracking},y})\right)
\end{equation}
where $\sigma_{\mathrm{tracking},x}$ and $\sigma_{\mathrm{tracking},y}$ set the width and direction of the tracking uncertainty. We set the ellipticity $\rho$ to 0.99 to enforce that the tracking uncertainty is effectively a 1D contribution to the final PSF. In practice, we find that the guiding errors generally occur in right ascension, but can have a strong component in declination if wind shake is strong. The recovered direction is almost always aligned nearly perfectly with either the right ascension or declination axis.

The ADR term $\mathcal{\tilde{P}}_{\mathrm{ADR}}(k_x, k_y, \lambda)$ uses the same ADR model that was used for the Gaussian + Moffat PSF (Equation~\ref{eq:adr}). The positional offset from ADR is implemented as a convolution with a delta function at the given offset:
\begin{equation}
\mathcal{\tilde{P}}_{\mathrm{ADR}}(k_x, k_y, \lambda) = \exp\left(-i (k_x x_*(\lambda) + k_y y_*(\lambda))\right)
\end{equation}

Finally, convolution with a pixel can be done analytically for a PSF in Fourier space:
\begin{equation}
\mathcal{\tilde{P}}_{\mathrm{pixel}}(k_x, k_y) = \sinc\left(\frac{k_x}{2 \pi}\right) \times \sinc\left(\frac{k_y}{2 \pi}\right)
\end{equation}

To evaluate this PSF model, we first evaluate all of the previously described terms on a grid of $k_x$ and $k_y$ and then use an inverse fast Fourier transform to obtain the real-space PSF. The normalization of this PSF is set by the $(k_x=0, k_y=0)$ pixel which is 1 in all of the different PSF components. With this implementation, this normalization corresponds to the sum of the real-space pixels rather than the sum of the full PSF directly, so it is essential to evaluate the PSF on a real-space pixel grid that is large enough to encompass the full PSF. To ensure that the full PSF is contained in our real-space pixel grid, we evaluate this PSF with a border of 15 real-space pixels around the target real-space image. To mitigate aliasing artifacts, we also subsample the real-space pixels by a factor of two.

As for the Gaussian + Moffat PSF, the central position of the source on the MLA varies with wavelength due to ADR, as described in Equation~\ref{eq:adr}. We parametrize the chromatic variation of the PSF width $w(\lambda)$ as:
\begin{equation}
w(\lambda) = w_0 \left(\frac{\lambda}{\lambda_0}\right)^\gamma
\end{equation}
where $w_0$ and $\gamma$ are free parameters. In practice, we do not observe significant wavelength dependence of the instrumental or tracking PSF components so we do not include those in the model.

We fit this PSF model, with all of the previously-described parameters unconstrained, to large numbers of high signal-to-noise observations of standard stars.  We find that a single set of values for the parameters $\tau$, $\sigma_{\mathrm{inst.},x}$, $\sigma_{\mathrm{inst.},y}$, $w_{\mathrm{inst.}}$, and $\tau_{\mathrm{inst.}}$ is sufficient to generate accurate PSF models for our entire dataset, and we fix those parameters to values determined from these fits. Unlike the Moffat + Gaussian PSF, we do not find a need for separate sets of parameters for short and long exposures. The final PSF model has eight free parameters that must be fit for each exposure: $x_0$, $y_0$, $\delta$, $\theta$, $w_0$, $\gamma$, $\sigma_{\mathrm{tracking},x}$, and $\sigma_{\mathrm{tracking},y}$.

No analytic model can capture all details of real PSFs. As just one example, exposures of only 1~sec are used for brightest standards and in that time very few independent atmospheric turbulence phase distortion cells are sampled, resulting in PSFs that are not as smooth as those in much longer exposures. We have directly verified this type of variation from video taken of bright stars. PSF structure not captured by the PSF model will increase the $\chi^2$ achieved by the PSF fit. If such structure is fixed, $\chi^2$ for standard stars (where readout and sky noise are negligible) will increase linearly with SNR. The pull of such structure will also change with SNR, possibly leading to a SNR-dependent bias. We have investigated this question and find that each star has a wide range of $\chi^2$ values at similar SNR, indicating that mismatches in PSF structure are not fixed but rather randomized across observations. This is consistent with the typical PSF fit residuals, and with the lack of bias for standard stars across observing conditions.

\section{Is \BDSeventeen A Useful Standard Star?}
\label{app:bdseventeen}

\BDSeventeen has been used as a spectrophotometric standard star for decades, however measurements by both \citet{bohlin15} and \citet{marinoni16} have suggested that it may be variable. Here we first review whether there is any {\it a priori} reason why \BDSeventeen might be a photometric variable, based on what is known about this extensively-studied system. Then we examine our \BDSeventeen photometric-night observations of \BDSeventeen spanning 12 years for indications of variability, along with additional photometry time series from \citet{bohlin15} and Hipparcos.

\BDSeventeen first sparked interest as a nearby (119~pc; \citet{GAIA2018} halo star, featuring a low metallicity of $\mathrm{[Fe/H]} \sim -1.6$ and a high proper motion. It was monitored over the course of 4~hrs with a sensitivity of 4~mmag by \citet{mcmillan1976} and showed no short-term variability. \citet{okegunn83} presented spectrophotometry of this star, ushering in its use as a popular spectrophotometric standard. It was subsequently adopted by SDSS has one of its three fundamental standard stars \citep{fukugita1996} and incorporated into CALSPEC \citep{bohlin15}. \BDSeventeen is an F8 subdwarf \citep{mishenina2000}, a type that has a comparatively low variability fraction \citep[e.g.][]{eyer2019}. However, it was eventually determined to be a single-line spectroscopic binary \citep{latham1988,latham2002} with a period of $219.19 \pm 0.12$ days

There are also reports from speckle imaging obtained circa 1986 and 1990 of a companion separated by 0.21\asec \citep{lu1987,balega1994}, amounting to a projected separation of 25~AU at the distance of \BDSeventeen. The \citet{lu1987} observations inferred approximately equal luminosities for the primary and secondary. Using the modern parallax and assuming a circular orbit (as in \citealt{lu1987}) implies a period of 80~yrs. This is much different than the period found by \citet{latham2002}, and since only a single set of lines was detected, the orbit of this purported companion would need to possess a small inclination to not have revealed the companion. Our examination of images of \BDSeventeen from \HST, e.g, using the Advanced Camera for Surveys High Resolution Camera observations from 2002 having stellar $\mathrm{FWHM} \sim 0.05$\asec in F330W, or $\sim 0.07$\asec in F775W, do not show a companion at this separation. Similarly, speckle images circa 1993 \citep{balega1994} and 2007 \citep{ras2008} do not detect a companion \citep{ras2008}. Hipparcos (sensitive to separations greater than 0.1\asec) did not report detection of a double star. Therefore, we suspect that the early reports of a companion based on speckle imaging may not be reliable; henceforth we focus on the companion detected via radial velocities.

The presence of a companion raises the possibility of variability due to phase- and seeing-dependent contaminating light from a lower-mass companion, variability of the companion, or residual effects from the post-main-sequence evolution of an initially more massive companion. The mass function of $0.00207 \pm 0.00024$ from the radial velocity analysis of \citet{latham2002} along with a mass of $\sim0.91$~\msun (estimated by \citealt{ramirez2006} from measurement of the surface gravity) can be used to constrain the possible companion configurations. For one, the minimum companion mass is $\sim0.12$~\msun --- roughly that of an M-type subdwarf, possessing a bolometric luminosity less than 0.2\% that of the primary. The maximum angular separation is constrained to be less than 8~milli-arcsec for non-degenerate companions, so the system would be unresolved even by \HST. Over a wide range of potential companion masses the separation is at most a few AU. Thus, if the companion were initially more massive, such a small orbital separation would have resulted in contact, possibly including mass transfer, between the stars during the red-giant phase of the more massive star. Initial-final mass relations for degenerate stars \citep{kovetz2009} predict that a companion with a main sequence mass slightly greater than than of \BDSeventeen would be only $\sim0.6$~\msun today. Significant mass transfer from an initially larger companion would leave a non-degenerate, non-main-sequence companion. If instead the companion is a lower-mass main-sequence star, its mass would need to be less than $\sim 0.7$~\msun in order to be faint enough to avoid the detection of its own set of spectral absorption features, e.g., in the study of \citet{latham2002} or the numerous detailed metallicity studies of this star \citep[e.g.,][]{ramirez2006}. A lower current-mass companion is statistically preferred due to the higher probability of larger inclinations. Roughly speaking, a configuration with a degenerate companion covers roughly 40\% of the inclination probability while a configuration featuring a lower mass companion covers the rest, but with some overlap between these configuration in the 0.6--0.7~\msun range. Invoking eclipses as the source of variability requires the companion mass to be at its minimum. Since the luminosity of the companion would be small at optical wavelengths in this case, a transit of the secondary has the larger effect. Such transits would result in dimming by less than 2\%\footnote{Using the \GAIA DR2 radius of 1.09~\rsun for the primary and the mass-radius relation for late-type subdwarfs \citep{subdwarf2018}}, and have a duration of only $\sim 0.44$~days; the random chance of observing such a transit while taking a standard star observation is around 0.2\%.

\begin{figure*}[htbp]
\centering
\includegraphics[width=0.49\textwidth]{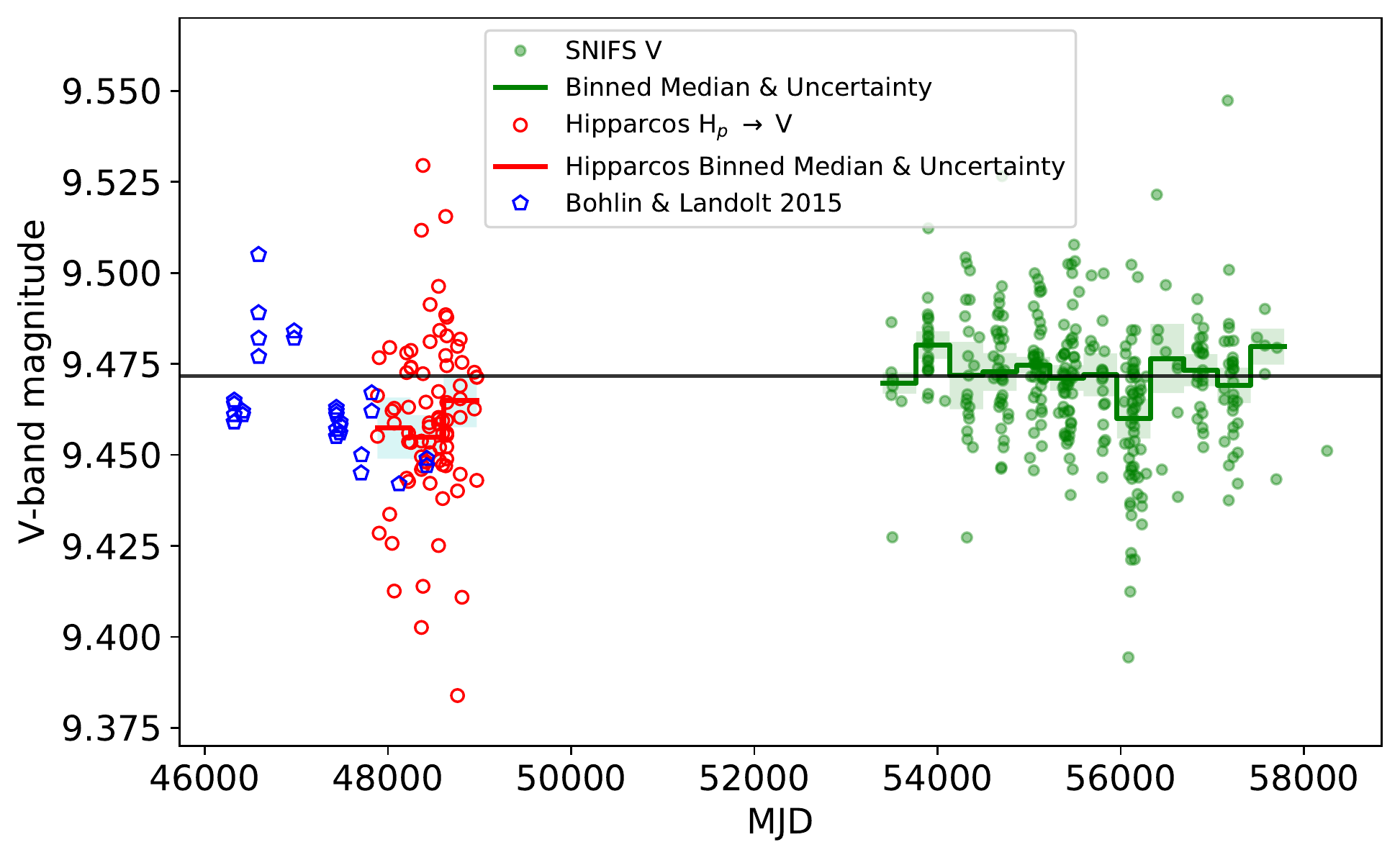}
\includegraphics[width=0.49\textwidth]{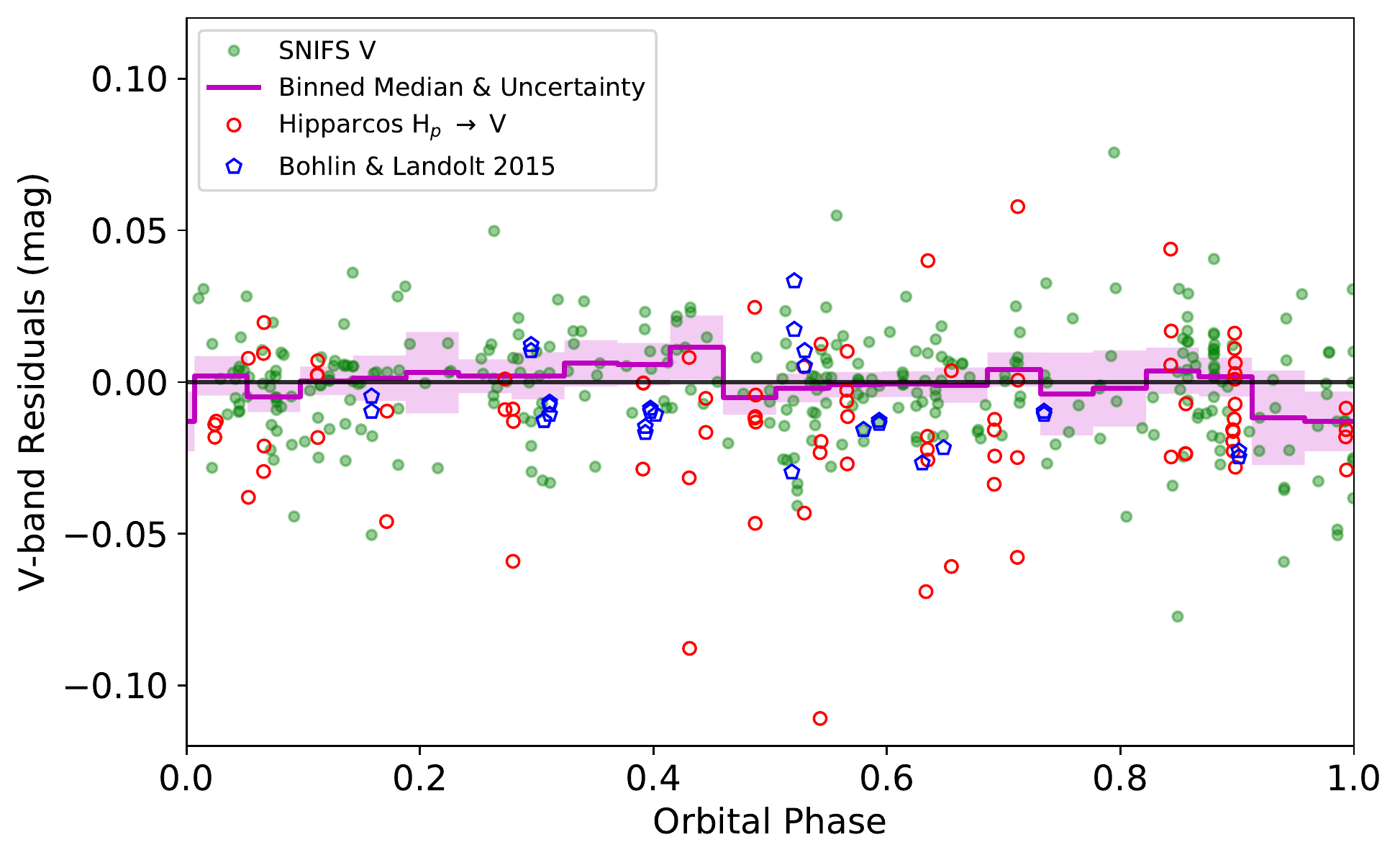}
\caption{Photometry of \BDSeventeen. Left: Photometry versus Julian date from the SNfactory (filled green circles), Hipparcos \citep{Perryman1997}, converted from $Hp$ magnitudes to $V$ using the formula of \citet{bessell2000} (open red circles), and from \citet{bohlin15} (open blue pentagons). The brightening trend in the \citet{bohlin15} photometry is readily apparent. It is difficult to definitively ascertain whether that brightening is supported by the Hipparcos data or not. But our new data show that such a brightening did not continue. Right: The same photometry versus the orbital phase determined by \citet{latham2002}. The phase zeropoint is arbitrary. For both figures, for the SNfactory data we show the binned medians and then shade the range of the robust error on the median. For the left figure we also show the binned medians and robust error on the median for the Hipparcos data.}
\label{fig:bd17photometry}
\end{figure*}

Turning now to the photometry, in Figure~\ref{fig:bd17photometry}a we plot the \BDSeventeen  V-band magnitude versus time from \citet{bohlin15}, Hipparcos, and SNfactory. A linear fit to the SNfactory data alone indicates a slope over 12~yrs of $0.9 \pm 0.3$~mmag/yr. Note that for this determination, \BDSeventeen was removed from the flux calibration solution so that any variation in \BDSeventeen could not be absorbed into the calibration. This result is significantly smaller than the trend of roughly $8 \pm 1$~mmag/yr reported by \citet{bohlin15}. As a check on our measurement uncertainty we measured this slope for a stable and well-observed star, EG131, finding a slope of only $0.7 \pm 0.4$~mmag/yr. This demonstrates that our $2.9\,\sigma$ measurement of a very gradual brightening of \BDSeventeen, while likely real, excludes a linear trend of the size reported by \citet{bohlin15}. In the degenerate companion scenario, \BDSeventeen might be slowly fading as it recovers from an ancient interaction with its companion, but we find a very slow brightening of \BDSeventeen, which would require some other mechanism.

The full photometric time series from \GAIA is not yet available, but we did examine its RMS, as for the other standard stars (\S\ref{sec:variables}). \GAIA \citep{GAIA2016, GAIA2018, gaia_edr3} monitored \BDSeventeen over 189 epochs, spanning 25~July 2014 to 28~May 2017, and from the \GAIA G-band flux, error on the flux, and number of observations, we can infer that the RMS of these observations is only 3.8~mmag over this period. 

\citet{marinoni16} monitored \BDSeventeen in V-band with respect to two comparison stars over the course of 54~minutes, a much shorter timescale than the orbital period or the variability seen by \citet{bohlin15}. We find that the RMS of the \citet{marinoni16} measurements is only 6~mmag, whereas their typical measurement uncertainty is 7~mmag. The Spearman correlation coefficient between brightness and HJD is $\rho=0.26$, having a probability of $p = 0.10$. So, these observations do not present compelling evidence of variability, a result consistent with the lack of variability within a night found by \citet{mcmillan1976}. \citet{marinoni16} comment that their photometry spanning 7~years does show variability of $\sim 30$~mmag; a value much higher than the we find from SNfactory data or found by \GAIA.

In the subdwarf/planet scenario discussed above, any effects should be synchronized with the orbital period. Therefore, we look for the latter type of periodic variability by phasing our observations with a period of 219.19 days, referenced to the measured periastron date of 47129.9~MJD. Figure~\ref{fig:bd17photometry}b shows this result for the SNfactory data, as well as for the \citet{bohlin15} and Hipparcos data. Here we see no evidence for any phase dependence, though our sampling is too sparse to rule out an eclipse shorter than a few weeks. A companion could also modify the color of \BDSeventeen, but we see no evidence for this either. But there is still the potential that the level of variability detectable by the photometry available to date depends on details of when samples were obtained. The full \GAIA time series may shed further light on this question.

We conclude that there is evidence that \BDSeventeen is brightening, but apparently at a level significantly less than seen by \citet{bohlin15} or \citet{marinoni16}. The trend is so small, and at $0.9 \pm 0.3$~mmag/yr better constrained than for many other standard stars in routine use, that we conclude that \BDSeventeen remains a valuable standard star. As in this paper and with the CALSPEC system, it should be pooled with many other standard stars in order to deweight small levels of variability.

This study highlights the need for high-resolution spectroscopic and imaging monitoring of existing and potential standard stars to reduce the fraction possessing hidden companions, which could compromise photometric stability. Space-based monitoring such as from Kepler \citep[e.g.][]{hermes2017}, \GAIA, and {\it TESS} can also monitor the photometric stability directly.

\section{Physical Model of the Maunakea Atmosphere}\label{app:MKatm}

Here we present our measured atmospheric extinction curve, its night-to-night dispersion, and the error on the mean. While not used to infer the standard stars presented in this work, we employ the atmospheric extinction constituents --- Rayleigh scattering, aerosol scattering and absorption, ozone absorption --- along with the achromatic offset component, to aid in understanding our measured values. As O$_2$ and H$_2$O are non-linear with airmass, we discuss their model separately in Appendix~\ref{app:nonlinearairmass}. After removing the impact of these tellurics, we fit the linear-in-airmass model shown in Appendix~\ref{app:linearairmass}.

\subsection{Non-Linear-in-Airmass Model} \label{app:nonlinearairmass}

As discussed in \S\ref{subsec:consistency}, we fit Line-By-Line Radiative Transfer Models \citep{clough92, clough05} based on scaling the water and non-water tellurics (with separate scaling parameters for each night) after convolving them down to SNIFS resolution using a Gaussian with $\sigma = 3.7$\,\ang. Table~\ref{tab:tellurictable} presents this model evaluated at airmass 1, 1.5, and 2 over the wavelengths 6,000\,\ang to 11,000\,\ang (for the user's convenience, running as red as the silicon detector cutoff wavelength); this table shows both the median and the RMS over all nights.

\begin{deluxetable*}{c|cc|cc|cc}[htbp]
\tablehead{
\colhead{Wavelength} &
\multicolumn{2}{c}{$\mathrm{X}=1$} & 
\multicolumn{2}{c}{$\mathrm{X}=1.5$} & 
\multicolumn{2}{c}{$\mathrm{X}=2$} \\[-0.5em]
\colhead{} & 
\colhead{Extinction} & 
\colhead{RMS} &
\colhead{Extinction} & 
\colhead{RMS} &
\colhead{Extinction} & 
\colhead{RMS}}
\startdata
    6000.0 & 0.0004 & 0.0003 & 0.0006 & 0.0005 & 0.0007 & 0.0007\\ 
6002.0 & 0.0003 & 0.0003 & 0.0005 & 0.0004 & 0.0007 & 0.0006\\ 
6004.0 & 0.0003 & 0.0002 & 0.0004 & 0.0004 & 0.0006 & 0.0005\\ 
6006.0 & 0.0002 & 0.0002 & 0.0003 & 0.0003 & 0.0005 & 0.0004\\ 
6008.0 & 0.0002 & 0.0002 & 0.0003 & 0.0003 & 0.0004 & 0.0003\\ 
6010.0 & 0.0002 & 0.0002 & 0.0003 & 0.0002 & 0.0003 & 0.0003\\ 
6012.0 & 0.0002 & 0.0001 & 0.0002 & 0.0002 & 0.0003 & 0.0003\\ 
6014.0 & 0.0002 & 0.0001 & 0.0002 & 0.0002 & 0.0003 & 0.0003\\ 
6016.0 & 0.0002 & 0.0001 & 0.0002 & 0.0002 & 0.0003 & 0.0003\\ 
6018.0 & 0.0001 & 0.0001 & 0.0002 & 0.0002 & 0.0003 & 0.0002\\ 
\enddata
\caption{Atmospheric extinction in magnitudes from our radiative-transfer model (convolved to SNIFS resolution), evaluated at three different airmass values. We show the nightly median and the RMS over all nights. This table is published in its entirety in the machine-readable format. A portion is shown here for guidance regarding its form and content.}
\label{tab:tellurictable}
\end{deluxetable*}

\subsection{Linear-in-Airmass Model} \label{app:linearairmass}

\begin{figure}[htbp]
    \centering
    \includegraphics [width = 0.95\textwidth]{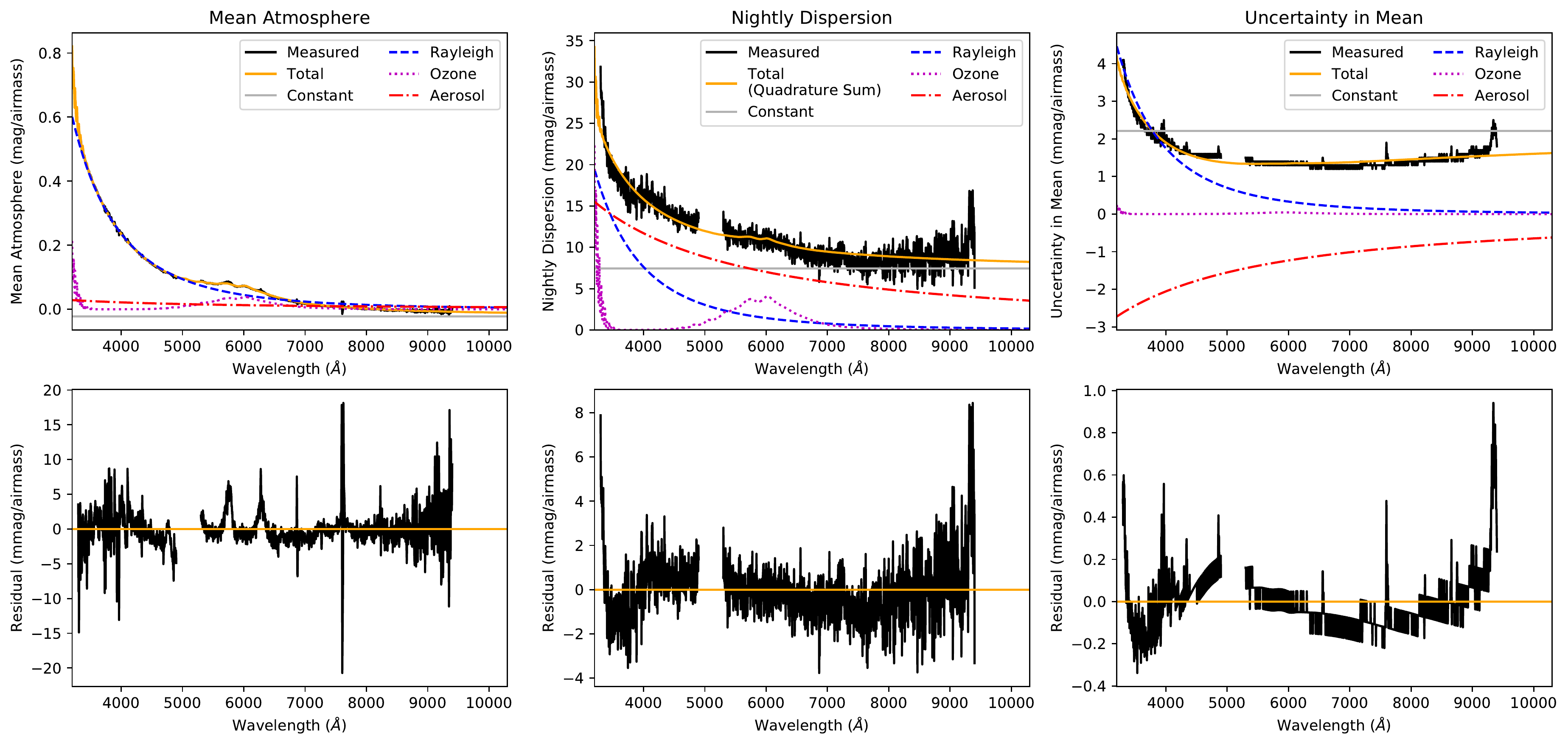}
    \caption{The {\bf left column} shows the mean atmosphere model ($k_{0\ l}$ from Equation~\ref{eq:hierarchy}), {\bf middle column} shows the night-to-night dispersion ($ \sigma(k)_{l}$ from Equation~\ref{eq:hierarchy}), and the {\bf right panel} shows the uncertainty in $k_{0\ l}$ (rounded to 0.1~mmag). {\bf Upper panels} show the measured values with their physical decompositions, while the {\bf bottom panels} show the residuals after the model is subtracted from the measurements. In the left residual panel a pair of small discrepancies at the peaks of the ozone Chappuis band is apparent; these are not apparent in the dispersion, suggesting a small error in the \cite{serdyuchenko2014} ozone template. The residual panels show evidence of a slight convolved-template mismatch at the $O_2$ A-band, but this seems to be static (a glitch visible in the mean, but not in the dispersion). At all other wavelengths, the physical decomposition matches our measured atmosphere to better than a few mmag/airmass.}
    \label{fig:atmdecomp}
\end{figure}

After removing telluric absorption from the input spectra, we rerun the inference of the full model and save this new set of parameters. We fit the remaining linear-in-airmass physical components to the mean atmospheric extinction coefficients ($k_{0\ l}$ from Equation~\ref{eq:hierarchy}), the night-to-night dispersion around the mean ($ \sigma(k)_{l}$ from Equation~\ref{eq:hierarchy}), and the uncertainty of the mean ($k_{0\ l}$). Figure~\ref{fig:atmdecomp} shows our results and Table~\ref{tab:atmdecomp} shows these models in a machine-readable format.

The results of the decomposition into physical components, shown in the left-most pair of panels in Figure~\ref{fig:atmdecomp} exhibits very good agreement with our per-wavelength measurements. This confirms the efficacy of this approach to determining atmospheric extinction --- the approach we took in B13. Small discrepancies occur at the peaks of the ozone Chappuis band and in the $O_2$ A-band. As there is no increase in the nightly dispersion at these wavelengths, these represent small but real differences between our observations and the physical component templates we have used in this Appendix. The quadrature decomposition of the nightly dispersion, shown in the middle pair of panels in Figure~\ref{fig:atmdecomp} is interesting, as it suggests that the achromatic offset that we found necessary to include is likely to have a scatter of only 8~mmag, relative to its mean value of 20~mmag. This is evidence that the effect is persistent, and not due to, e.g., a mix of nights with and without the effect. The nightly dispersion decomposition also highlights aerosol scattering as the most variable atmospheric constituent. But due to the elevation of Maunakea, the effect of its variability is still small. Finally, the right-most pair of panels showing the error on the mean, and its linear decomposition, illustrates the impressively small uncertainty on our mean extinction curve. 

\begin{deluxetable*}{c|ccc|ccc}[htbp]
\tablehead{
\colhead{Wavelength} &
\multicolumn{3}{c}{Measured Extinction} &
\multicolumn{3}{c}{Modeled Physical Extinction} 
\\[-0.5em]
\colhead{(\AA)} &
\colhead{Mean} & 
\colhead{Nightly} & 
\colhead{Uncertainty} &
\colhead{Mean} & 
\colhead{Nightly} & 
\colhead{Uncertainty} 
\\[-0.5em]
\colhead{} & 
\colhead{} & 
\colhead{Dispersion} &
\colhead{on Mean} & 
\colhead{} &
\colhead{Dispersion} & 
\colhead{on Mean}}
\startdata
    3298.68 & 0.5571 & 0.0318 & 0.0040 & 0.5760 & 0.0228 & 0.0013\\ 
3301.06 & 0.5606 & 0.0295 & 0.0041 & 0.5842 & 0.0229 & 0.0013\\ 
3303.44 & 0.5639 & 0.0292 & 0.0039 & 0.5952 & 0.0230 & 0.0013\\ 
3305.82 & 0.5637 & 0.0286 & 0.0039 & 0.5893 & 0.0229 & 0.0013\\ 
3308.20 & 0.5646 & 0.0281 & 0.0039 & 0.5857 & 0.0228 & 0.0013\\ 
3310.58 & 0.5623 & 0.0289 & 0.0039 & 0.5995 & 0.0231 & 0.0013\\ 
3312.96 & 0.5581 & 0.0280 & 0.0041 & 0.5897 & 0.0229 & 0.0013\\ 
3315.34 & 0.5486 & 0.0279 & 0.0039 & 0.5769 & 0.0226 & 0.0013\\ 
3317.72 & 0.5415 & 0.0287 & 0.0040 & 0.5667 & 0.0224 & 0.0013\\ 
3320.10 & 0.5328 & 0.0273 & 0.0040 & 0.5588 & 0.0223 & 0.0012\\ 
\enddata
\caption{Atmospheric extinction, in magnitudes per airmass. The {\bf left group of columns} show the measured values and the {\bf right group of columns} show the best-fit physical model (without the constant-in-wavelength component that appears to be linked to PSF variation and not the atmosphere). This table is published in its entirety in the machine-readable format. A portion is shown here for guidance regarding its form and content.}
\label{tab:atmdecomp}
\end{deluxetable*}

\end{document}